\documentclass[a4paper, 11pt, abstracton, titlepage, oneside]{scrartcl}
\makeatletter

\usepackage[T1]{fontenc}
\usepackage[utf8]{inputenc}
\usepackage[onehalfspacing]{setspace}
\usepackage[headsepline]{scrlayer-scrpage}
\clearscrheadfoot 
\usepackage{layout}
\usepackage[english]{babel}
\usepackage[left=2cm,right=2cm,top=2.5cm,bottom=2.5cm]{geometry}

\usepackage{algorithm}
\usepackage[noend]{algpseudocode}
\usepackage{adjustbox}
\usepackage{eurosym}
\usepackage{lscape}
\usepackage{nicefrac}
\usepackage{array}
\usepackage{multicol}
\usepackage{paralist}
\usepackage{authblk}
\usepackage{graphicx}
\usepackage{booktabs}
\usepackage{placeins}
\usepackage{amsmath}
\usepackage{mathtools}
\usepackage{bm}
\usepackage{amssymb}
\usepackage{color}
\usepackage{changepage} 
\usepackage{supertabular}
\usepackage[acronym]{glossaries}
\usepackage{pgfplots}
\usepackage[normal]{threeparttable}
\usepackage{ulem}
\usepackage{natbib}
\usepackage[titletoc,title]{appendix}
\usepackage[hyphens]{url} 
\usepackage{ellipsis}
\usepackage{breqn} 
\let\counterwithin\relax

\usepackage{chngcntr}  
\usepackage{enumerate}
\usepackage{theorem}
\usepackage{etoolbox}
\usepackage{multirow}
\usepackage{colortbl}
\usepackage{subcaption}
\newcommand\fnote[1]{\captionsetup{font=small}\caption*{#1}}
\usepackage{tikz}
\usetikzlibrary{arrows.meta}
\usepackage{relsize}
\usepackage{stackengine}
\usepackage{wrapfig}
\usepackage{lscape}
\usepackage{epstopdf}
\usepackage{rotating}
\patchcmd{\endabstract}{\null}{}{}{} 

\usepgfplotslibrary{groupplots}
\usepgfplotslibrary{dateplot}
\usepackage[utf8]{inputenc}
\ifdefined\DeclareUnicodeCharacter
\DeclareUnicodeCharacter{2212}{-}
\fi

\usepackage{array}
\newcolumntype{x}[1]{>{\centering\arraybackslash\hspace{0pt}}p{#1}}

\usepackage{amsthm}
\newtheorem{prop}{Proposition}

\newtheorem{res}{Result}
\let\vec\mathbf

\ifdefined\mathleft
\else
\newcommand{\mathleft}{}
\fi	
\ifdefined\mathright
\else
\newcommand{\mathright}{}
\fi
\ifdefined\mathcenter
\else
\newcommand{\mathcenter}{}
\fi

\DeclareMathOperator*{\argmin}{arg\,min}

\makeatletter
\providecommand\phantomcaption{\caption@refstepcounter\@captype}
\makeatother

\newglossary{infoslist}{1i}{1o}{Information variants}
\makeglossaries
\newacronym{cs}{CS}{Charging Station}
\newacronym{ev}{EV}{Electric Vehicle}
\newacronym{sdp}{SDP}{Stochastic Dynamic Programming}
\newacronym{mdp}{MDP}{Markov Decision Process}

\newglossaryentry{low_avail}{name={\textit{low-25}},
	description={Low-density scenario},
	type=infoslist}
\newglossaryentry{high_avail}{name={\textit{high-60}},
	description={Low-density scenario},
	type=infoslist}
\newglossaryentry{0}{name={DEC-MYO},
	description={Low-density scenario},
	type=infoslist}
\newglossaryentry{1}{name={DEC},
	description={Low-density scenario},
	type=infoslist}
\newglossaryentry{2}{name={DEC-I},
	description={Low-density scenario},
	type=infoslist}
\newglossaryentry{3}{name={DEC-O},
	description={Low-density scenario},
	type=infoslist}
\newglossaryentry{4}{name={DEC-IO},
	description={Low-density scenario},
	type=infoslist}
\newglossaryentry{5}{name={CEN},
	description={Low-density scenario},
	type=infoslist}
\newglossaryentry{6}{name={CEN-LAB},
	description={Low-density scenario},
	type=infoslist}
\newglossaryentry{7}{name={CEN-MYO},
	description={Low-density scenario},
	type=infoslist}
\newglossaryentry{8}{name={DEC-I-dyn},
	description={Low-density scenario},
	type=infoslist}
\newglossaryentry{9}{name={DEC-O-dyn},
	description={Low-density scenario},
	type=infoslist}
\newglossaryentry{10}{name={DEC-IO-dyn},
	description={Low-density scenario},
	type=infoslist}
\newglossaryentry{HLHC}{name={HLH-c},
	description={collaborative hierarhical label heuristic},
		type=infoslist}
\newglossaryentry{HLH}{name={HLH},
	description={hierarchical label heuristic},
	type=infoslist}
\newglossaryentry{RO}{name={RO},
	description={rollout},
	type=infoslist}
\newglossaryentry{LHRO}{name={LH-RO},
	description={rolled-out label},
	type=infoslist}
\newglossaryentry{sc1}{name={\textit{SC1}},
	description={rolled-out label},
	type=infoslist}
\newglossaryentry{sc2}{name={\textit{SC2}},
	description={rolled-out label},
	type=infoslist}
\newglossaryentry{sc3}{name={\textit{SC3}},
	description={rolled-out label},
	type=infoslist}

\definecolor{lightblue}{rgb}{0.60784,0.76078,0.90196}
\definecolor{darkblue}{rgb}{0.26667,0.44706,0.76863}
\definecolor{lightgreen}{rgb}{0.66275,0.81569,0.55686}
\definecolor{darkgreen}{rgb}{0.43922,0.67843,0.27843}
\definecolor{orange}{rgb}{0.92941,0.49020,0.19216}
\definecolor{yellow}{rgb}{1.00000,0.75294,0.00000}
\definecolor{grey}{rgb}{0.64706,0.64706,0.64706}
\definecolor{purple}{rgb}{0.51373,0.23529,0.04706}

\DeclareOldFontCommand{\rm}{\normalfont\rmfamily}{\mathrm}
\DeclareOldFontCommand{\sf}{\normalfont\sffamily}{\mathsf}
\DeclareOldFontCommand{\tt}{\normalfont\ttfamily}{\mathtt}
\DeclareOldFontCommand{\bf}{\normalfont\bfseries}{\mathbf}
\DeclareOldFontCommand{\it}{\normalfont\itshape}{\mathit}
\DeclareOldFontCommand{\sl}{\normalfont\slshape}{\@nomath\sl}
\DeclareOldFontCommand{\sc}{\normalfont\scshape}{\@nomath\sc}
\makeatother
\counterwithin*{equation}{section}

\bibpunct[, ]{(}{)}{,}{a}{}{,}%
\newcolumntype{L}[1]{>{\raggedright\let\newline\\\arraybackslash\hspace{0pt}}p{#1}}
\newcolumntype{C}[1]{>{\centering\let\newline\\\arraybackslash\hspace{0pt}}p{#1}}
\newcolumntype{R}[1]{>{\raggedleft\let\newline\\\arraybackslash\hspace{0pt}}p{#1}}
\addtokomafont{captionlabel}{\footnotesize\bfseries} 
\addtokomafont{caption}{\footnotesize\bfseries} 

\AtBeginDocument{%
	\abovedisplayskip=2.5pt plus 0pt minus 0pt
	\abovedisplayshortskip=2.5pt plus 0pt minus 0pt
	\belowdisplayskip=2.5pt plus 0pt minus 0pt
	\belowdisplayshortskip=2.5pt plus 0pt minus 0pt
}

\newsavebox{\abstractbox}
\renewenvironment{abstract}
{\begin{lrbox}{0}\begin{minipage}{\textwidth}
			\begin{center}\normalfont\sectfont\abstractname\end{center}\quotation}
		{\endquotation\end{minipage}\end{lrbox}%
	\global\setbox\abstractbox=\box0 }

\makeatletter
\expandafter\patchcmd\csname\string\maketitle\endcsname
{\vskip\z@\@plus3fill}
{\vskip\z@\@plus2fill\box\abstractbox\vskip\z@\@plus1fill}
{}{}
\makeatother

\makeatletter
\def\BState{\State\hskip-\ALG@thistlm}
\makeatother

\counterwithin*{equation}{section}

\newenvironment{myfont}{\fontfamily{ccr}\selectfont}{\par}
\DeclareTextFontCommand{\textmyfont}{\myfont}

\begin{document}

\definecolor{lightblue}{rgb}{0.60784,0.76078,0.90196}
\definecolor{darkblue}{rgb}{0.26667,0.44706,0.76863}
\definecolor{lightgreen}{rgb}{0.66275,0.81569,0.55686}
\definecolor{darkgreen}{rgb}{0.43922,0.67843,0.27843}
\definecolor{orange}{rgb}{0.92941,0.49020,0.19216}
\definecolor{yellow}{rgb}{1.00000,0.75294,0.00000}
\definecolor{grey}{rgb}{0.64706,0.64706,0.64706}
\definecolor{purple}{rgb}{0.51373,0.23529,0.04706}

\newcommand{\orange}[1]{\textcolor{orange}{#1}}

\newacronym{cs}{CS}{charging station}
\newacronym{dp}{DP}{dynamic programming}
\newacronym{abk:eu}{EU}{European Union}
\newacronym{abk:ecv}{ECV}{electric commercial vehicle}
\newacronym{ev}{EV}{electric vehicle}
\newacronym{abk:ghg}{GHG}{greenhouse gas}
\newacronym{abk:icev}{ICEV}{internal combustion engine vehicles}
\newacronym{mdp}{MDP}{Markov decision process}
\newacronym{abk:rcspp}{RCSPP}{resource constraint shortest path problem}
\newacronym{abk:ref}{REF}{resource extension function}
\newacronym{sdp}{SDP}{stochastic dynamic programming}
\newacronym{Soc}{SoC}{state of charge}
\newacronym{scps}{SCPS}{stochastic charge pole search}


\title{Coordinated Charging Station Search in Stochastic Environments: A Multi-Agent Approach}

\author[1,2]{\normalsize Marianne Guillet}
\author[1,3]{\normalsize Maximilian Schiffer}
\affil{\small 
	TUM School of Management, Technical University of Munich, 80333 Munich, Germany
	
	\textsuperscript{2} TomTom Location Technology Germany GmbH, 12435 Berlin, Germany
	
	
	\textsuperscript{3} Munich Data Science Institute, Technical University of Munich, 80333 Munich, Germany,
					
	\scriptsize
	marianne.guillet@tomtom.com,
	schiffer@tum.de}

\date{}

\lehead{\pagemark}
\rohead{\pagemark}

\begin{abstract}
	Range and charge anxiety remain essential barriers to a faster electric vehicle market diffusion. To this end, quickly and reliably finding suitable charging stations may foster an electric vehicle uptake by mitigating drivers' anxieties. Here, existing commercial services help drivers to find available stations based on real-time availability data but struggle with data inaccuracy, e.g., due to conventional vehicles blocking the access to public charging stations. In this context, recent works have studied stochastic search methods to account for availability uncertainty in order to minimize a driver’s detour until reaching an available charging station. 
	So far, both practical and theoretical approaches ignore driver coordination enabled by charging requests centralization or sharing of data, e.g., sharing  observations of charging stations' availability or visit intentions between drivers. Against this background, we study coordinated stochastic search algorithms, which help to reduce station visit conflicts and improve the drivers' charging experience.
	We model a multi-agent stochastic charging station search problem as a finite-horizon Markov decision process and introduce an online solution framework applicable to static and dynamic policies. In contrast to static policies, dynamic policies account for information updates during policy planning and execution. 
	We present a hierarchical implementation of a single-agent heuristic for decentralized decision making and a rollout algorithm for centralized decision making. 
	Extensive numerical studies show that compared to an uncoordinated setting, a decentralized setting with visit-intentions sharing decreases the system cost by 26\%, which is nearly as good as the 28\% cost decrease achieved in a centralized setting, and saves up to 23\% of a driver's search time while increasing her search reliability.
\begin{singlespace}
{\small\noindent\\
\smallskip}
\end{singlespace}
\end{abstract}

\maketitle	

\section{Introduction}
	
	 Electric vehicles (EV) can play a crucial role in decarbonizing the transportation sector, given a rapid energy transition and continuous battery technology improvement \citep{SchullerStuart2018}. 
	 To sustain the global electric vehicle market's growth of the past two years \citep[][]{Deloitte2020}, it is necessary to mitigate the remaining barriers to private EV adoption.
	 In addition to the well-known range anxiety, a new phenomenon referred to as charge anxiety, caused by unreliable and insufficient public charging infrastructure, remains an essential barrier, particularly in cities \citep{Myersdorf2020}.
	Besides increased price transparency, a seamless charging experience may reduce these anxieties if drivers can easily find and use an available charging station \citep{McKinsey2020}.

	In this context, increasing public infrastructure coverage and improving interoperability between charging service providers is necessary to facilitate easy and reliable access to public charging stations but requires a long planning horizon \citep{Volkswagen2019}. 
	Accordingly, complementary short-term solutions are necessary to alleviate deficiencies in the currently undersized charging infrastructure. In theory, cooperative charging strategies based on vehicle-to-vehicle communication can allow for fairer charging capacity allocation \citep{YouYangEtAl2016} but are not implemented yet due to immature technology and uncertain economic-value \citep{Lauinger2017}.
	 In practice, existing map-based services help EV drivers to locate available charging stations based on real-time charging station availability data, but fail to provide a reliable charging station search experience: stations reported as available can be unusable due to inaccurate reporting or ICEing, i.e., conventional vehicles blocking the access to a charging station \citep[][]{GuilletHiermannEtAl2021}. In such cases, drivers must take detours to reach another station, which may lead to increasing anxiety. 

	Moreover, simultaneous uncoordinated searches of multiple drivers may conflict if drivers head to the same charging station. 
	Navigation services platforms that offer services to find available charging stations through local navigation devices or an online API, may centralize driver requests or leverage active and passive driver community input, e.g., GPS trace data, to coordinate search recommendations.

	In the literature, stochastic charging station search methods offer a reliable alternative to existing search services, accounting for charging stations' availability uncertainty. Such methods consider charging stations as stochastic resources and aim to find a sequence of charging station visits – a search path – that minimizes the expected search cost to reach an available station. 
	These approaches are amenable for real-time applications, may significantly save drivers' time, and increase the search's reliability. However, such approaches so far focus always on a single-agent setting.
	 Accordingly, coordinating drivers in a multi-agent setting has not been studied so far, but may avoid possible visit conflicts and further improve the driver's charging experience by increasing the search's reliability and decreasing the search time.

	With this paper, we close this research gap by extending stochastic single-agent search algorithms to a stochastic multi-agent setting, accounting for information-sharing and possible requests centralization.
	We consider scenarios in which drivers may share their planned visits, their observations of a charging station's occupancy, or both.
	Our goal is to identify the coordination strategy that yields the best improvement on the drivers' search times and the search's reliability.

	\subsection{Related literature}


	In the following, we first review literature that relates to stochastic charging station search problems in a single-agent setting, including EV routing with uncertain charging station availabilities and more generic stochastic resource search problems. We then focus on multi-agent resource search problems with stochastic availability or locations.
	
	Only a few papers that deal with EV routing problems address charging station availability uncertainty but do not focus on open-search problems. 
	\cite{KullmanGoodsonEtlAl2021} solve the electric vehicle routing problem with a public-private recharging strategy, while \cite{SwedaDolinskayaEtAl2017} and \cite{JafariBoyles2017} focus on shortest paths with multiple charging stops for EVs. \cite{SwedaDolinskayaEtAl2017} propose adaptive charging and routing strategies when utilizing the public charging infrastructure, whereas \cite{JafariBoyles2017} additionally model both stochastic travel time and charging consumption.
	Alternatively, a few papers deal with stochastic resource search problems in general settings \citep{GuoWolfsonEtAl2018,SchmollSchubert2018} or more specific settings, e.g., on-street parking spots \citep{ArndtHafnerEtAl2016} or stochastic taxi customer demand \citep{TangKerberEtAl2013}. \cite{GuilletHiermannEtAl2021} are the first to cover multiple variants of the stochastic charging station search problem for EVs, considering charging or waiting times at stations.
	However, all aforementioned papers, including \cite{GuilletHiermannEtAl2021}, are limited to single-agent settings and ignore possible agents coordination.
			
	Focusing on multi-agent settings, most works on resource search problems under uncertainty focus on cooperative searches, i.e., settings in which agents share a unique common goal. In \cite{BourgaultFurukawaEtAl2003}, all agents aim at locating a resource with no a-priori information on its location in a decentralized decision-making setting. \cite{ChungBurdick2008} solve a similar setting  with centralized decision-making, while \cite{WongBourgaultEtAl2005}, and \cite{DaiSartoretti2020}, extend the single resource target search problem to a multi-target search problem. \cite{QinShaoEtAl2020} address the multi-agent travel officer problem in a centralized decision-making setting, in which agents must cooperatively collect resources with stochastic availability in given locations. In all of these papers, the cooperative search terminates after the (all) resource(s) have been found. This is not the case in our multi-agent charging station search setting: here, each agent terminates her search when she found at least one non-shareable available resource. 
	Existing work on multi-agent settings for EVs mostly focuses on autonomous EV fleet management, such as ride-sharing planning \citep{AlkanjiNascimentoEtAl2020} or online requests matching for ride-hailing \citep{KullmanCousineauEtAl2020} and do not cover stochastic resource search problems.
	
	In summary, most papers that deal with heterogeneous resource search problems with stochastic availability focus on a single-agent setting. Multi-agent search problems mostly focus on cooperative (multi) resource search problems with unknown resource locations.
	In contrast, the search problem studied in this work is rather collaborative than cooperative because there exists a trade-off between an individual agent's search experience and the system performance. Furthermore, while agents usually synchronously search for one or multiple resources, our setting requires the agents' search to start at different times and in different locations which are sequentially revealed.

\subsection{Aims and scope}

With this work, we close the research gap outlined above by providing coordinated stochastic search algorithms, that consider both resources and agents heterogeneity, tailored to various information-sharing and decision-making scenarios of high practical relevance.
Specifically, our contribution is three-fold. 
First, we formalize the underlying centralized multi-agent decision-making problem.
To this end, we define the planning problem as a single-decision-maker \gls{mdp} and show that with an additional policy constraint, this MDP can be alternatively represented as a set of single-agent MDPs, which enables decentralized and static planning. 
Second, we present several online algorithms that allow to solve a variety of real-world scenarios. 
Here, we consider both settings in which a navigation device addresses a charging request on-the-fly, by providing either a full search path (static planning) or only the next best charging station to visit (dynamic planning) to the driver. Moreover, we consider centralized and decentralized planning settings with different levels of information-sharing in which drivers share either their planned visits, or their charging station occupancy observations, or both. 
Third, we conduct extensive numerical studies based on real-world instances to analyze which coordination strategy yields the highest improvement potential from a system and a driver perspective.

Focusing on a short planning horizon, our results show that, from a system perspective, a centralized coordination strategy can decrease the system cost by 28\%, and that a static decentralized coordination strategy already achieves a 26\% cost decrease if visit intentions are shared.
In a decentralized setting with intention-sharing, observation-sharing does not increase the system's performance further, but enforcing agents' collaboration is required when drivers depart within a short time span.
While a decentralized setting with only observation-sharing performs worse than intention-sharing settings, it provides a computationally efficient implementation in practice. When implemented in a dynamic setting, it yields a 10\% cost decrease when drivers depart within a short time span, but achieves a 26\% cost decrease with larger departure time span.
From a driver perspective, coordination may save up to 23\% (intention-sharing setting) of a driver's search time, while increasing her search reliability.
Our results further show that a coordinated search outperforms uncoordinated searches, with respect to both best and worst solutions that an individual driver may obtain.
Finally, we show in additional analyses that coordination also positively impacts a driver's search for a longer planning horizons, yielding up to 46\% decrease in system cost.

\subsection{Organization}

The remainder of this paper is as follows. Section~\ref{sec:problem_definition} introduces our problem setting. In Section~\ref{sec:mdp_representation}, we formalize our multi-agent decision making problem as an MDP, before we characterize MDP policies that allow for decentralized policy execution and planning. Section~\ref{sec:online_heur} introduces our solution framework for online charging requests. In Section~\ref{sec:exp_design}, we describe our case study and the corresponding experimental design. We discuss numerical results in Section~\ref{sec:results}. Section~\ref{sec:conclusion} concludes this paper and provides an outlook on future research. 


\section{Problem setting}\label{sec:problem_definition}


We focus on a non-adversarial multi-agent search problem, with stochastic charging station availability, where multiple drivers seek to find an unoccupied charging station in their vicinity at the earliest possible time.
The drivers start their search asynchronously; each driver departs from a given location and may visit multiple occupied charging stations before reaching an available charging station. A driver visits the stations recommended by her navigation device, which synchronizes with a central navigation service platform.
In this setting, an EV driver's objective is to minimize her expected time to reach an available station and any related charging costs. The navigation service provider's objective is to satisfy as many drivers as possible by minimizing the sum of all drivers' individual search cost as well as the likelihood that a driver does not reach any available station within a given time budget.

We assume each individual search to be spatially and temporally bounded to account for a driver's limited time budget. Every unsuccessful search induces an individual penalty. 
Moreover, we assume that a driver cannot wait at an occupied station nor visit a station twice and that she stops at the first available station she visits. Stations are heterogeneous and using a station yields a cost. 
Drivers are heterogeneous with respect to the charging and penalty costs, their time budget, and the radius delimiting their (circular) search area.

Practically, the navigation device transmits a search solution to the driver: the driver may either receive a full sequence of stations to visit until an available station is reached, i.e., a search path, or she may receive the next station to visit dynamically. We refer to the former as static planning and to the latter as dynamic planning. Moreover, the solution planning can be (i) centralized  within the navigation service platform or (ii) decentralized, i.e., at agent-level. In the latter case, solution planning can happen directly within the local navigation device and devices only use the platform to share information with each other.
In both cases, agents may share their station occupancy observations intermittently or in real-time with the central platform. In the decentralized case, they may additionally share the planned charging station visits. 
To capture these varying characteristics which are of practical relevance, we introduce the following problem settings as summarized in Table~\ref{tab:settings}. 

\begin{table}[bp]
	\centering
	\caption{Problem settings overview}\label{tab:settings}
	{\small
		\setlength{\tabcolsep}{4pt} 
		\renewcommand{\arraystretch}{1} 
		\begin{tabular}{p{5.5em}p{4em}p{6em}p{4em}p{5.835em}p{5em}p{7em}}
			\toprule
			& \multicolumn{1}{l}{visit intentions} & availability observations   & path planning &  decision-making & type  & user-dependent solutions \\
			\midrule
			DEC   &       & \multicolumn{1}{r}{}  & static & decentralized & selfish & no \\
			DEC-I & \multicolumn{1}{p{5em}}{$\checkmark$} & \multicolumn{1}{r}{} & static & decentralized & collaborative & no \\
			DEC-O &       & $\checkmark$  & static & decentralized & informative & no \\
			DEC-IO & \multicolumn{1}{p{5em}}{$\checkmark$} & $\checkmark$  & static & decentralized & collaborative & no \\
			\midrule
			CEN & \multicolumn{1}{p{5em}}{$\checkmark$} & $\checkmark$  & dynamic & centralized & collaborative & yes \\
			DEC-O-d &       & $\checkmark$  & dynamic & decentralized & informative & yes \\
			\bottomrule
		\end{tabular}
	}
\end{table}%

\begin{description}
	\item[Static planning:] In such settings, a search path is planned upon request of the agent and cannot be dynamically updated once the agent started her search. Accordingly, an agent's actions only depend on the agent's own observations and the initial information available prior to her search. We assume that planned search paths and (if shared) collected observations, can be transmitted to the central platform to be available to subsequent agents.
	We introduce four decentralized decision-making settings according to the type of shared information. 
	In the \gls{1} setting, no information is shared between agents, while each agent is aware of prior availability observations of other agents when computing her search path in the \gls{3} setting.
	Both settings \gls{1} \& \gls{3} are purely informative, such that agents are unaware of other agents' planned search paths.
	The \gls{2} and \gls{4} extend the \gls{1} \& \gls{3} settings: each agent additionally knows previous agents' search paths, i.e., their intended station visits and visit times. Agents may use the information selfishly, e.g., by visiting other agents' target stations first, or they may use the information collaboratively. 
	
	\item[Dynamic planning:]
	In such settings, an agent does not receive the whole search path at once. Instead, it receives the next station to visit at each occupied visited station until it reaches an available station.
	Here, we consider a centralized-decision making setting~(\gls{5}) in which a central planner, i.e., the navigation platform, is fully aware of all agents' observations and actions, and assigns station visits on the fly to minimize the total expected search cost of all agents, while maximizing the likelihood that all agents find an available station. Here, each action assigned to an agent depends on the other agents' actions and all decisions are system-optimized as the platform does not prefer any agent. 
	Further, we consider a dynamic variant (\gls{9}) of the decentralized observation-sharing setting DEC-O. In this setting, the initial solution is dynamically re-planned upon each occupied station visit to account for the latest shared observations.
		
\end{description}

A few comments on our problem setting are in order.
	\begin{figure}[tbp]
		\centering
		\scalebox{0.65}{\input{./chapters/figures/scaled_traffic.tex}}
		\caption{Number of ending trips on Tuesday 12.01.22 in the east-area of Berlin, Germany}\label{fig:traffic}
		\fnote{The figure shows $\nicefrac{1}{3}$ of all collected anonymized OD pairs of all vehicles using TomTom navigation services, that end their trips in the east-area of Berlin.}
	\end{figure}
	First, we model our coordinated charging station search as a closed system.
	While this assumptions seems to be a limitation from a theoretical perspective, it is appropriate from a practitioners perspective as it reflects real-world planning scenarios, where charging requests usually accumulate during certain periods of the days (see Figure~\ref{fig:traffic}).
	Even for scenarios with charging requests homogeneously distributed during a day, we observe a sufficiently long period without charging requests during the night which imposes a natural system boundary.
	Second, we assume in this context constant availability probabilities and observations persistency over the planning horizon. 
This assumption is in line with practice for the following reason: in a single-agent scenario, modeling time-dependent recovering availability probabilities has no impact on the computed search path as an agent either succeeds or gives up the search after a limited time budget \cite[cf.][]{GuilletHiermannEtAl2021}. By restricting our experimental studies to short planning horizon scenarios, we can analogously ignore time-dependent recovering probability modeling. 
However, we discuss how to account for relaxed constant probability assumptions (see Section~\ref{subsec:observation_validity}), and detail the impact of coordination in longer planning horizon settings (see Section~\ref{subsubsec:large-horizon_exp}).
Third, we assume that all the shared information is available to the other drivers or the central decision-maker, but limited to a certain vicinity. This assumption allows to reduce computational effort and is reasonable in practice as far-distanced drivers do not interfere with each other.

\section{MDP representation}\label{sec:mdp_representation}

We refer to the problem setting discussed in Section~\ref{sec:problem_definition} as the multi-agent stochastic charging pole search (MSCPS) problem and model it as a sequential multi-agent decision making problem with a finite time-horizon. 
In the following, we first consider a centralized representation of the system states and represent the offline multi-agent search with an omniscient single decision-maker as a finite-horizon MDP in Section~\ref{subsec:mdp_central}, i.e., assuming that all future requests are known ahead.
We then reduce the solution space such that the solution policy can be decentrally executed in Section~\ref{sec:pol_category}, and show that in this case the MDP can be represented by a set of individual MDPs in Section~\ref{subsec:indiv_mdp}.

We formalize the MSCPS problem on a complete directed graph $\mathcal{G} = (\mathcal{V}, \mathcal{A})$ consisting of a set of vertices $\mathcal{V}$ and a set of arcs $(v, \acute{v}) \in \mathcal{A}$, where a vertex $v \in \mathcal{V}$ represents a charging station. 
We consider a set of $\mathcal{D} = \{1, ..., N\}$ agents. For each agent~$i\in\mathcal{D}$ let $t^i_0$  be her departure time and $v^i_0$ be her start vertex. 
Each agent~$i$ has a defined time budget $\bar{T}^{i}$ and we define the total planning horizon $T = [0, \max_{i\in\mathcal{D}}(t^i_0 + \bar{T}^i)]$ during which all searches occur.
Each agent can charge at any station located at a vertex $v \in \mathcal{V}$ within a limited search radius $\bar{S}^i$. We denote with $t_{v,\acute{v}}\geq 0$ the time to drive from $v$ to $\acute{v}$. 
An unsuccessful search yields a penalty cost $\bar{\beta}^i$ for the agent. We let $\gamma^i_{v}$ be the (time-equivalent) cost for using pole $v$ for agent~$i$, if $v$ is available upon the arrival of $i$.
We denote with $a_v\in\{0,1\}$ a binary random variable that models the availability of a station~$v\in\mathcal{V}$ at time $t\in T$, which the agent observes by visiting $v$, and let $p_v$ be the probability that station~$v$ is initially free. Finally, we assume that an occupied station remains occupied during the whole planning horizon~$T$.

\subsection{MDP notation}\label{subsec:mdp_central}

An agent triggers a new decision epoch either by requesting to charge her vehicle or by observing a new station.
We refer to the requesting or observing agent as the \textit{deciding agent} denoted with $\lambda$ although a central planner may take the actual decision. 
If the observed station is occupied and there is at least one reachable station within her remaining time budget, $\lambda$ selects her next station visit; otherwise $\lambda$ has terminated her search.
Note that each station visited by $\lambda$ can no longer be used by any succeeding agent, since the station is either (i) already occupied, or (ii) available and thus occupied by $\lambda$. 
Accordingly, the set of visited stations, i.e., stations that have been observed, corresponds to the set of occupied stations.

\paragraph{State space:}
We represent a system state~$x \in \mathcal{X}$ out of state space $\mathcal{X}$ as	
\begin{equation}
x = (\vec{x_{d}}, \mathcal{J}, \mathcal{T}, \mathcal{O})\;,
\end{equation}
with $\mathcal{J}\subseteq \mathcal{D}$ being the set of active agents,  $\mathcal{T}\subseteq \mathcal{D}$ being the set of successfully terminated agents, $\mathcal{O}\subseteq \mathcal{V}$ being the set of all visited stations, and $\vec{x_{d}} = (x^i)_{\forall i \in \mathcal{J} \bigcup \mathcal{T}}$ being the vector that describes the state of each agent. Here, we define an agent's state~$x^i$ as
\begin{equation}
x^i = (v^i, t^i, s^i)\;,
\end{equation}
with  $v^i \in \mathcal{V}$ being the station assigned to agent~$i$ in state~$x$; $t^i$ being the arrival time at $v^i$ and $s^i \in \{ \text{d}, \text{f}, \text{t}, \text{r} \}$ being the status of the agent: an agent can either (i) be en-route to the station ($s^i=\text{r}$), unaware of $v^i$'s realized availability, (ii) observe $v^i$ to be available, which successfully terminates her search ($s^i=\text{f}$), (iii) observe $v^i$ to be occupied, having sufficient time to reach a new station ($s^i=\text{d}$) or (iv) not ($s^i=\text{t}$), which unsuccessfully terminates her search. The observation of $v^i$ in (ii) and (iii) triggers a new decision epoch. 

\paragraph{Action space and immediate cost:}
We denote with $u$, the action taken in state~$x$ for agent~$\lambda$, and by $\mathcal{U}(x)$ the set of feasible actions, such that $u \in \mathcal{U}(x)$. We let $d(x, u)$ be the cost immediately induced by taking decision $u$ in state~$x$, which does not depend on any future uncertainty realization. 
For clarity, we refer to state~$x$ as $x^{\text{s}}$ if the station observed by $\lambda$ is available, or as $x^{\text{f}}$ if $\lambda$ observes an occupied station or begins her search.

	\noindent \textit{Available station} ($x^{\text{s}}$) : 
	Here, the status of $\lambda$ is $s^{\lambda} = \text{f}$, such that no further decision can be made; accordingly $u = \emptyset$ such that $\lambda$ belongs to the set of terminated agents ($\mathcal{T}$).
	The immediate cost is the cost for $\lambda$ to use $v^\lambda$, such that $d(x, u) = \gamma^\lambda_{v^\lambda}$.
	
	\noindent \textit{Occupied station} ($x^{\text{f}}$):
	Here, the status of $\lambda$ is $s^{\lambda} = \text{d}$ if $\lambda$ can select an unvisited station~$v$ to visit next. Accordingly we get $u \in \{v: v \in \mathcal{V}, v \notin \mathcal{O},t^\lambda - t^\lambda_0 + t_{v^\lambda, v}  \leq \bar{T}^{\lambda} \}$. 
	The immediate cost results to the driving time for $\lambda$ from her current station to the chosen station, such that $d(x, u) = t_{v^{\lambda}, u}$. At the next stage, its new assigned station is $\acute{v}^{\lambda} = u$, while its new arrival time  is $\acute{t}^{\lambda} = t^\lambda +  t_{v^{\lambda}, u}$. If $\lambda$ remains the deciding agent in the next stage, i.e., $\lambda=\acute{\lambda}$ then $\acute{s}^{\lambda} = \text{d}$, and otherwise $\acute{s}^{\lambda} = \text{r}$.
	If no station can be reached within $\lambda$'s remaining time budget, $\lambda$ has failed her search, which induces an immediate driver penalty $\bar{\beta}^\lambda$.
	 In this case, $s^{\lambda} = \text{t}$ and we set $d(x, u) = \bar{\beta}^\lambda$.
	 
	We define a policy $\pi \in \Pi$ as the state-action mapping function, such that $\pi(x)\in \mathcal{U}(x)$.

\paragraph{State transition:}
Upon a single-agent's action, the system transitions from state~$x$ to the next state~$\acute{x}$, with $\acute{\lambda}$ being the new deciding agent.
The new state~$\acute{x}$ can either be a successful state~${\acute{x}}^{\text{s}}$ for $\acute{\lambda}$ with probability~$p_{\acute{v}}$ or an unsuccessful state~${\acute{x}}^{\text{f}}$ with probability~$1 - p_{\acute{v}}$.

We introduce the value function $V^\pi(x)$, that corresponds to the expected sum of future costs obtained when executing $\pi$ from state $x$, and that can be recursively expressed as
\begin{equation}\label{eq:central_cost}
V^\pi(x) = d(x, \pi(x)) +  p_{\acute{v}}V^\pi({\acute{x}}^{\text{s}}) + (1 - p_{\acute{v}})V^\pi({\acute{x}}^{\text{f}}),
\end{equation}
with $x\in \{ x^{\text{s}}, x^{\text{f}}\}$. 
%
Then, our objective is to find a policy $\pi^*$ that minimizes the accumulated costs when executing $\pi^*$ from the initial state $x_0$, i.e, $V^{\pi^*}(x_0)\leq V^\pi(x_0) \;\;\forall \pi$. 

To increase the robustness of our solution, we introduce a global penalty cost $\beta^G$ that is induced in a termination state in case at least one agent unsuccessfully terminated her search. We note that a higher $\beta^G$ may decrease the quality disparities of the single-agent solutions by favoring a conservative system with equally bad solutions. On the contrary, a lower $\beta^G$ may favor single-agent high quality solutions at the detriment of other agents. 
We define $x_n$ as a termination state as soon as the set of active agents $\mathcal{J}$ becomes empty. In state~$x_n$, each agent~$i \in \mathcal{T}$ has either successfully (\mbox{$s^i=\text{f}$}) or unsuccessfully (\mbox{$s^i=\text{t}$}) terminated her search. We define $V(x_n)$, with $x_n$ being a termination state, as
\begin{equation}
V^{\pi}(x_n)  =  d(x_n, \pi(x_n)) + \beta^G \delta\;,
\end{equation} with $\delta$ being the binary variable that indicates whether at least one agent~$i \in \mathcal{T}$ has failed her search.
Table~\ref{tab:notation} summarizes the notation used to define the MSCPS problem and the related MDP.
\begin{table}[tbp]
	\caption{Notation}\label{tab:notation}
	\scalebox{0.9}{
			\begin{tabular}{rl}
				\toprule
				\multicolumn{2}{l}{Notation used to define the MSCPS problem}\\
				\midrule
				$\mathcal{G} = (\mathcal{V}, \mathcal{A})$ & Complete charging station graph \\
				$\mathcal{D}$ & Set of agents\\
				$t^i_0$ & Departure time of agent~$i$\\
				$v^i_0$& Start vertex of agent~$i$\\
				$\bar{S}^i$ & Maximal distance allowed between any vertex and the start vertex $v^i_0$ for agent~$i$ \\
				$\bar{T}^i$ & Time budget for agent~$i$ \\
				$T$ & Planning horizon \\
				$a_v$ &Binary random variable modeling the availability of $v$\\
				$p_v$& Initial probability that charging station~$v$ is available before any visit \\
				$t_{v,\acute{v}}$ & Driving time on arc $(v,\acute{v})$\\
				$\bar{\beta}^i$ & Termination penalty cost at an occupied station~$v$ for agent~$i$ \\
				$\gamma^i_{v}$  & Termination cost at an available station~$v$ for agent~$i$ \\  
				\midrule
				\multicolumn{2}{l}{Notation used to define the MDP}\\
				\midrule
				$\mathcal{X}$ & (Multi-agent) State Space\\
				$\mathcal{U}$ & (Multi-agent) Action Space\\
				$\mathcal{J}$ & Set of active drivers in state~$x\in\mathcal{X}$\\
				$\mathcal{T}$ & Set of terminated drivers in state~$x\in\mathcal{X}$\\
				$\mathcal{O}$ & Set of visited stations in state~$x\in\mathcal{X}$\\
				$v^i$ & Station assigned to driver $i$ in state~$x\in\mathcal{X}$\\
				$t^i$ & time when $i$ reaches $v^i$\\
				$s^i\in\{ \text{d}, \text{f}, \text{t}, \text{r} \} $ & status of $i$ in state~$x\in\mathcal{X}$\\
				$d(x,u)$& immediate cost induced by taking action $u\in\mathcal{U}(x)$ in state~$x\in\mathcal{X}$\\
				$\pi$ & System policy\\
				$V^\pi$ & Value function\\
				$\pi^i$ & Single-agent~$i$ policy\\
				$\delta$ & Binary variable indicating if at least one driver failed her search in termination state~$x_n \in \mathcal{X}$\\
				$\Pi^{\text{dep}}$ & Set of policy $\pi$ with \textit{user-dependent} single-agent policies\\
				$\Pi^{\text{ind}}$ & Set of policy $\pi$ with \textit{user-independent} single-agent policies\\
				\bottomrule
		\end{tabular}}
\end{table}%

\subsection{Policy representation}\label{sec:pol_category}

In Section~\ref{subsec:mdp_central}, we introduced the policy function $\pi$ which indicates for any encountered state~$x$ the action~$\pi(x)$ to take, assuming that a state-action pair~$(x, \pi(x))$ can apply to any agent.
Alternatively, we can represent $\pi$ as a set of single agent policies $\pi^i$ by defining $\pi^i$ only for states where agent~$i$ takes a decision.
With $\lambda$ being the deciding agent in state~$x$, we let $\pi^i(x) = \pi(x)$ if $i=\lambda$ in state~$x$ and $\pi^i(x) = \emptyset $ if $i \neq \lambda$. 
We refer to this joint set of single-agent policies $\{ \pi^i \}_{i \in \mathcal{D}}$ as the agent-based representation of $\pi$.

\paragraph{User-dependent single-agent policies:}
In general, the execution of a single-agent policy $\pi^i$ is user-dependent and must be centrally coordinated as each agent's action depends on other agents' observations. 
%
In state~$x$, the deciding agent~$\lambda$ is aware of the full state~$x$, i.e., aware of all other availability realizations observed by all other agents. From a single-agent perspective, the station selection for $\lambda$ depends on whether and where other agents terminated. For all states $x$ corresponding to $\lambda$ located at $v$, selected actions $\pi^i(x)$ might not be identical. 
We let $\Pi^{\text{dep}}$ be the set that contains user-dependent policies $\pi = \{ \pi^i \}_{i \in \mathcal{D}}$, with $\pi^i$ being user-dependent.

\paragraph{User-independent single-agent policies:}

While coordinated search in general requires central coordination, it is possible to ensure a-priori coordination between agents without central coordination, i.e., preserving user-independent search policies. In this special case, coordination accounts for other agent’s visit intentions when deriving an agent’s search policy but excludes observation sharing during the search. In such a setting, an agent $i$, executing a search policy $\pi$, being located at station $v$ with $v$ being occupied, will always visit the same next station. We can formally describe this condition as
\begin{equation}\label{eq:cond_dec}
	\pi^i(x) = \pi^i(\bar{x}), \; \forall (x,\bar{x}) \in \bar{\mathcal{X}} \times \bar{\mathcal{X}}\text{ st. }  v^i = {\bar{v}}^i = v,
\end{equation}
with $v^i$ being the location of agent~$i$ in state~$x$, ${\bar{v}}^i$ being the location of $i$ in state~$\bar{x}$, and  $\bar{\mathcal{X}}$ being the subset of states possibly reachable from $x_0$ through $\pi$.
We let $\Pi^{\text{ind}}$ be the set that contains user-independent policies $\pi = \{ \pi^i \}_{i \in \mathcal{D}}$, with $\pi^i$ being user-independent.

\subsection{MDP representation with a user-independent policy constraint}\label{subsec:indiv_mdp}

The MDP defined in Section~\ref{subsec:mdp_central} considers a user-dependent global policy that can be represented as a set of single-agent user-dependent policies (see Section~\ref{sec:pol_category}). We show that by constraining single-agent policies to be user-independent, this global MDP representation simplifies to a set of single-agent MDPs, which allows to easily plan each agent's solution decentrally.
We introduce the representation of a single-agent MDP in Section~\ref{subsubsec:indiv_mdp} accordingly. Here, the objective function equals the sum of single-agent MDP objective functions with an extra penalty cost.
%
In this case, a single-agent policy $\pi^i$ translates to an ordered sequence of station visits $C^i=(v^i_0,\ldots,v^i_n)$ with $i$ starting at $v^i_0$, following the vertices in sequence, and terminating either the search at any first available charging station~$v\in C^i$, or unsuccessfully at $v^i_n$. 
Section~\ref{subsubsec:mdp_eq} then focuses on representation equivalence.

\subsubsection{Single-agent MDP notation:}\label{subsubsec:indiv_mdp}

Analogously to \cite{GuilletHiermannEtAl2021}, we model each driver's individual search process as a single-agent finite-horizon MDP. Let $\mathcal{S}$ be the (single-agent) state space and $x\in \mathcal{S}$ be a state defined as $x=(C^i,a)$, with $C^i=(v^i_0,...,v^i_k)$ being the sequence of visited stations, and $a$ being the realized availability at the last visited station~$v^i_k$ of $C$. If $a=0$, then the agent takes an action $u$ that consists of the single station selection decision $u=(v)$, with $v\in \tilde{\mathcal{V}}, v\notin C^i$, and $\tilde{\mathcal{V}}$ being the set of reachable stations from $v^i_k$. 
In this case, the transition function $p^i_t(\acute{x}|x,u)$ describes the probability for the agent $i$ to be in state~$\acute{x}$ after having taken action $u$ in state~$x$. The immediate cost induced for taking action $u=(v)$ in state~$x$ results to the travel time $t_{v^i_k,v}$ between $v^i_k$ and $v$, or to the penalty cost $\bar{\beta}^i$, if the agent already spent her time budget. 
If $a=1$, then the agent successfully terminated her search at $v^i_k$ and the agent-specific station usage cost $\gamma^i_{v^i_k}$ results. 

Policy $\pi^i$ is the function that maps the planned action $u$ for agent~$i$ in each encountered state~$x$, and defines the related search path $C^i(\pi^i) = (v^i_0,...,v^i_n)$, with $\pi^i((v^i_0,...,v^i_k),0) = v^i_{k+1} \; \; \forall \; k \in [0,...,n-1]$.


Agent~$i$ aims to find a policy $\pi^i$ that minimizes her search cost $F^{\pi^i}((v^i_0),0)$, with $F$ being the single-agent value function, which is defined as
\begin{equation}\label{eq:single_cost}	
\begin{split}
F^{\pi^i}(x) =& t_{v_k,v} +  \sum_{\acute{x}\in \mathcal{S}} p^i_t(\acute{x}|x ,\pi^i(x)) F^{\pi^i}(\acute{x}) \\
\Leftrightarrow F^{\pi^i}(C,0) = t_{v_k,v}  &+  p^i_t((C',1)|(C,0) ,\pi^i(C,0)).\gamma^i_{v} \\
&+  p^i_t((C',0)|(C,0) ,\pi^i(C,0))F^{\pi^i}(C',0) \;,\\
\end{split}
\end{equation}
with $x=(C,0)$ being a non-termination state, $\acute{x}=(C',a')$ and $C'=(v_0,...,v_k,v)$.

\paragraph{Transition functions:}

To define our transition functions, we recall that station occupancy observations are persistent for the whole planning horizon and a-priori availability probabilities are identical for all agents by assumption, which leads us to the following observation:
if the first station $v$ visited by $j$ is available, $j$ stops and charges there, otherwise $v$ is occupied and remains occupied during the search horizon. Accordingly, $v$ will never be available to another agent $i$ who intends to visit station~$v$ after~$j$.
More generally, the probability that $v$ is available to agent~$i$ equals the probability that all agents~$j$ intending to visit~$v$ before~$i$ ($t^i_{v} \geq t^j_{v}$) will have found a free station before their expected visit to~$v$, and that station~$v$ is available during the search. 

Formally, we denote the probability that station~$v$ is available for agent~$i$ at time~$t$ with~$p^i_v(t)$. 
We let $\rho^i(t)$ be the probability that agent~$i$ found at least one station available among all stations $v$, from sequence $C^i = (v^i_1,...,v^i_n)$, for which the visit time $t^i_v$ is lower than~$t$, and define $\rho^i(t)$ as
\begin{equation}\label{eq:successRate}
\rho^i(t) =1 - \prod_{\mathclap{v \in C^i, t^i_v \leq t }} (1 -  p^i_{v}(t) ).
\end{equation}

\noindent Then, $p^i_v(t) $ reads
\begin{equation}\label{eq:probaVisible}
p^i_v(t) = p_v \prod_{\mathclap{j \in \mathcal{D},  j \neq i, t^{j}_v \leq t}} \rho^{j}(t^{j}_v).
\end{equation}
Note that both definitions are finitely nested as visits are ranked by agent's  arrival time.

We now define user-dependent transition function as 
\begin{equation}\label{eq:trans_functions_ma}
\begin{split}
p^i_t((C',1)|(C,0),(v)) &= p^i_v(t'),\\
p^i_t((C',0)|(C,0),(v)) &= 1 - p^i_v(t') ,
\end{split}
\end{equation}
with $C:=C^i$. We let $C'$ be sequence $C$ extended by station~$v$, and $t'$ be the accumulated driving time for sequence $C'$.

In a single-agent setting, transition functions are independent of other agents planned actions and simplify to
\begin{equation}\label{eq:trans_functions_sa}
\begin{split}
p_t((C',1)|(C,0),(v)) &= p_v,\\
p_t((C',0)|(C,0),(v)) &= 1 - p_v.
\end{split}
\end{equation}

\subsubsection{System evaluation cost:}\label{subsubsec:mdp_eq}
With $F^{\pi^i}$ being the single-agent value function for agent~$i$, we now introduce a cost that jointly evaluates all single-agent policies.
Let $\alpha^i = F^{\pi^i}(x^i_0)$ be the cost that explicitly evaluates the expected value of the policy cost assigned to $i$ in its initial state. 
Cost $\alpha^i$ can be decomposed \citep[cf.][]{GuilletHiermannEtAl2021} as 
 \begin{equation}\label{eq:cost_decompos}
 \alpha^i =  (1 - {\rho}^i)  \beta^i  + A^i(\pi^i) \;.
 \end{equation}
 We let $\bar{\rho}^i(\pi,k)$ be the probability that agent~$i$ fails in finding at least one free station in $C^i_{[0:k]}$, while $\rho^i$ denotes the probability to find at least one free station in the whole sequence~$C^i$. Here we define $A^i(\pi^i)$ as 
 \begin{equation}\label{eq:def_A}
 A^i(\pi^i) = \sum_{k = 0}^{n-1} [ t_{v^i_k,v^i_{k+1}} \bar{\rho}^i(\pi^i, k)] + \sum_{k = 1}^{n} \gamma^i_{v^i_k} p^i_{v^i_k}(C_{[0:k-1]}) \bar{\rho}^i(\pi^i, k-1) \;.
 \end{equation}

We define the cost $\alpha^\pi$ that jointly evaluates the system policy $\pi =  \{ \pi^i \}_{i \in \mathcal{D}}$, i.e., the set of single-agent user-independent policies, as
\begin{equation}\label{eq:global_cost}
\alpha^\pi  = \sum_{i \in \mathcal{D}}\alpha^i + (1 - \prod_{i \in \mathcal{D}} \rho^i) \beta^G,
\end{equation}
with $\alpha^\pi$ being the sum of all expected single-agent MDP costs and a cost $(1 - \prod_{i \in \mathcal{D}} \rho^i) \beta^G$ that penalizes the system with respect to the number of agents that could not successfully finish their search.
Quantity $(1 - \prod_{i \in \mathcal{D}} \rho^i)$ represents the likelihood that at least one agent did not successfully terminate her search. 

We now show that for such a  policy $\pi$, the joint cost $\alpha^\pi$ equals the value function $V^{\pi}$ evaluated in the initial global state~$x_0$. 

\begin{prop}\label{prop:equal_MDP}
Let policy $\pi$ be a set of single-agent user-independent policies, such that $\pi \in \Pi^{\text{ind}}$.
Then
 $$\alpha^{\pi} = V^{\pi}(x_0),$$
with $V^{\pi}$ being defined in Equation~\ref{eq:central_cost} and $\alpha^\pi$ being defined in Equation~\ref{eq:global_cost}.
\end{prop}

Proposition~\ref{prop:equal_MDP} simplifies the representation of the centralized MDP to a set of single-agent MDPs, which enables us to devise decentralized online algorithms in Section~\ref{sec:online_heur}.


\section{Online solution planning}\label{sec:online_heur}

In the following, we present our online heuristics to process sequentially revealed charging requests. In this setting, the set of agents $\mathcal{D}$ is initially empty and we update $\mathcal{D}$ each time a new charging request enters the system.

\subsection{Static policy planning}\label{subsec:decentralized}

For static policy planning, we focus on decentralized decision-making settings (see Section~\ref{sec:problem_definition}). 
We plan each agent's search path with a modified version of the stochastic search algorithm developed in \cite{GuilletHiermannEtAl2021}, by taking into account the latest available information, i.e., the latest shared visit intentions or the latest availability observations.
In the following, we first briefly outline our algorithm (Section~\ref{subsubsec:dec}) in its basic variant without any information-sharing, before we detail the required changes to account for observation-sharing (Section~\ref{subsubsec:dec-o}), visit intention-sharing (Section~\ref{subsubsec:label_coop}), or both (Section~\ref{subsubsec:dec-io}).

\subsubsection{No information-sharing (\gls{1}):}\label{subsubsec:dec}

This setting corresponds to a (fully-decentralized) single-agent setting, in which each agent is unaware of any prior requested search paths and availability observations. In practice, this setting equals planning routes on individual non-communicating navigation devices or with a stateless navigation service platform, which does not retain information about past requests.
Here, each agent aims to minimize her individual cost~$\alpha^i$, with $\alpha^i$ being defined based on a single-agent MDP with \textit{user-independent} transition functions (see Equation~\ref{eq:trans_functions_sa}).
As shown in \cite{GuilletHiermannEtAl2021}, a multi-label setting algorithm with a heuristic dominance criterion, which we refer to as LH, can efficiently solve this problem setting. 

In general, a multi-label setting algorithm propagates partial policies to find a cost-optimal policy, maintaining a list of all explored and non-dominated partial policies in cost-increasing order.
%
%
A partial policy~$\pi^i_v$  describes a given search path starting from the initial vertex~$v_0$ and ending at $v$. 
We associate each partial policy~$\pi^i_v$ with a label $L^i_v$, associated to vertex $v$ and agent~$i$, defined as  $L^i_v = (t^i_v, A^i_v, \rho^i_v, \alpha^i_v)$. A label~$L^i_v$ consists of the accumulated driving time~$t^i_v$, the partial cost~$A^i_v$, the overall probability of success~$\rho^i_v$, and the total cost for agent~$i$ (see Equation~\ref{eq:cost_decompos}). 
We recall that $A^i_v$ and $\rho^i_v$ result from the decomposition of cost~$\alpha^i_v$ (see Section~\ref{subsubsec:indiv_mdp}).
At each exploration step, the algorithm retrieves the minimum-cost partial policy $\pi^i_v$ and propagates her related label $L^i_v$ to all unvisited vertices $\acute{v}$, reachable from $v$. For each vertex $\acute{v}$, the algorithm discards the propagated label $L^i_{\acute{v}}$ if it is dominated by any other label at vertex $\acute{v}$, and otherwise discards any labels that $L^i_{\acute{v}}$ may dominate.
Specifically, considering two partial policies $\pi^i_1$ and $\pi^i_2$ for agent~$i$ that end with the same vertex visit $v$ and their associated labels~$L^i_v(\pi^i_1)$ and $L^i_v(\pi^i_2)$, we say that $L^i_v(\pi^i_1)$ dominates $L^i_v(\pi^i_2)$ ($L^i_v(\pi^i_1) \succ L^i_v(\pi^i_2)$), if
\begin{equation} \label{eq:dom_1}
1- \rho^i_v(\pi_1) \leq 1- \rho^i_v(\pi_2)\; ,
\end{equation}
\begin{equation} \label{eq:dom_2}
A^i_{v}(\pi_1) \leq A^i_{v}(\pi_2)
\end{equation}
 are true.

Finally, the labeling procedure returns the -- non-dominated -- minimum-cost label, that describes the search policy~$\pi^i$ with minimum cost~$\alpha^i$. 
The used dominance relation does not guarantee the optimality of $\pi^i$  in a single agent setting but provides close to optimal solutions with significant runtime savings compared to an exact dominance relation \citep[cf.][]{GuilletHiermannEtAl2021}. For a detailed pseudo-code of this algorithm, we refer to Appendix~\ref{app:lh_algo}. 
We solve each incoming request with this algorithm and refer to its sequential application as hierarchical label-setting, denoted with \gls{HLH}.

\subsubsection{Occupancy observation-sharing (\gls{3}):}\label{subsubsec:dec-o}

In this setting, each agent~$i$ knows about occupied stations observed by other agents prior to her search.
To account for this knowledge, we remove observed occupied stations from an agent's action space because occupied stations cannot be freed up during the remaining planning horizon. 
%
%
Accordingly, an action $u$ for $i$ consists of the single station selection decision $u=(v)$ with $v\in \tilde{\mathcal{V}}, v\notin C^i, v \notin \mathcal{O}$, $\tilde{\mathcal{V}}$ being the set of reachable stations from~$v^i_k$, and $\mathcal{O}$ being the set of observed occupied stations. 
To account for this modified action space, we reduce the charging station network graph to unvisited stations whenever we compute a new search path.
Accordingly, this allows us to use the HLH algorithm for the no-information-sharing setting to compute each agent's search path.

\subsubsection{Visit-intentions sharing (\gls{2}):}\label{subsubsec:label_coop}

In this setting, an agent~$i$ knows all station visit intentions of agents who started their search prior to $i$, i.e., all $\pi^{j} \text{ } \forall j \in \tilde{\mathcal{D}}$, with $\tilde{\mathcal{D}}$ being the set of preceding agents.
Here, we define cost $\alpha^i$ based on a single-agent MDP with \textit{user-dependent} transition functions (see Equation~\ref{eq:trans_functions_ma}).
Given $i$ is the $i^{th}$ requesting agent, we let $\pi_{i-1} = \{\pi^{j}\}_{\forall j \in \tilde{\mathcal{D}}} $ be the joint policy of the (sub)system prior to $i$'s request, i.e., the joint set of all single-agent policies for agents in $\{1,...,i-1\}$.
Then, the probability of station~$v$ to be available to~$i$ results from Equation~\ref{eq:probaVisible}, by replacing~$\mathcal{D}$ by~$\tilde{\mathcal{D}}$
\begin{equation}
p_v(t) = p_v \prod_{\mathclap{j \in \tilde{\mathcal{D}}, j \neq i, t^{j}_v \leq t^i_v}} \rho^{j}(t)\;,
\end{equation}
with $\rho^{j}(t)$ being defined in Equation~\ref{eq:successRate}.

Each agent~$i$ uses the information about other agent's visit intentions to individually optimize her own search path, which may occur at the detriment of the agents that started their search earlier. In some cases, agent~$i$ may bypass another agent~$j$ by visiting the intended stations of~$j$ before~$j$'s expected visit times.

\paragraph{Collaborative intention-sharing:}
To avoid the selfish use of intention-sharing, we design a hierarchical solution method such that agents minimize their search times without compromising other agents' success. Here, agent $i$ does not optimize her search path with respect to her individual cost; instead she optimizes her search path with respect to the cost of the subsytem that includes herself and already planned policies $\pi_i$ of other agents.
We refer to this collaborative implementation as \gls{HLHC}.
To optimize the agent's policy with respect to the subsystem cost, we compute $n$ candidate policies for agent $i$ using our LH algorithm, which queues the evaluated policies by cost-increasing order.
Out of these candidate policies, we then select the policy that yields the lowest subsystem cost. 

Formally, let $\Gamma^i = \{\pi^i_k: \forall k \in [0,n]\} $ be the set that contains these $n$ candidate policies for agent~$i$.
Then, we (i) compute $\Gamma^i$ using LH, evaluate the joint cost of the subsystem policy $\pi_{i-1}$ and the newly planned policy, i.e., $\pi_i = \pi_{i-1} \circ \{\pi^i\}$ for all $\pi^i \in\ \Gamma^i $, and (ii) select ${\pi^i}^*$ among $\Gamma^i $ that minimizes $\alpha^{{\pi_i}}$ as
\begin{equation}
{\pi^i}^* = \argmin_{\pi^i  \in\ \Gamma^i } \alpha^{\pi_{i-1} \circ \{\pi^i\}}\; .
\end{equation}
If agents are heterogeneous, an agent~$i$ needs to know other agents' parameters $\bar{\beta}^i$ and $\gamma^i_v$ to compute the exact value $\alpha^{\pi_{i-1} \circ \{\pi\}}$. In practice, these parameters can be either shared or approximated.

\subsubsection{Intentions \& Occupancy observations sharing (\gls{4}):}\label{subsubsec:dec-io}

In this setting, an agent~$i$ combines both knowledge about past agents' observations and visit intentions when planning her search path at time $t^i$.
Similar to \gls{3}, we remove occupied stations from an agent's action space, and similar to \gls{2}, transition functions are user-dependent.
In this setting, we do not need to account for an agent $j$'s remaining visit intentions, if $j$ started earlier than $i$ and successfully finished her search already. If this agent $j$ is not terminated yet at planning time $t^i$, we know that all stations visited by $j$ before $t^i$ are occupied, such that we only account for $j$'s visit intentions that would occur later than $t^i$.
Accordingly, we truncate $j$'s search path $C^{j}$ to the stations not visited yet at $t^i$ and refer to the truncated sequence as $\bar{C}^j$. We obtain availability probabilities by replacing $C^j$ by $\bar{C}^j$ in Equations~\ref{eq:successRate}\&\ref{eq:probaVisible}.

\subsection{Dynamic policy planning}

In the following, we describe the methodology for dynamic policy planning. We first focus on a centralized decision-making setting in which all observations are shared and all intentions are known to a centralized decision-maker (Section~\ref{subsubsec:centralized}) before we briefly discuss a decentralized planning setting with observation-sharing (Section~\ref{subsubsec:dec_online}).

\subsubsection{Centralized decision-making:}\label{subsubsec:centralized}

To dynamically solve the large-scale MDP introduced in Section~\ref{subsec:mdp_central}, we utilize two different algorithms. The first algorithm is a rollout algorithm (RO) with a one-step decision rule as described in \citet{GoodsonThomasEtAl2017}, which is known to provide high-quality solutions in similar settings.
This rollout algorithm explores the MDP solution tree partially, using a base-policy to approximate the value function. In each state, the algorithm selects the action that yields the lowest approximated cost. 
The second algorithm bases on a dynamic implementation of our HLH-c algorithm (see Section~\ref{subsubsec:label_coop}). Instead of selecting the next best station visit based on a partial MDP solution tree exploration, this algorithm (re)computes an agent's individual search path using the latest observations and visit intentions available at each decision step. We then use the first station visit of the recomputed search path as the next station visit. 
We refer to this second algorithm as LH-RO and note that it combines dynamic and offline planning similar to the work of \cite{UlmerGoodsonEtAl2019}.

\paragraph{Rollout algorithm (\gls{RO}):\label{par:RO}}
Figure~\ref{alg:rollout} details the pseudo-code of our algorithm which dynamically solves the MDP defined in Section~\ref{subsec:mdp_central}. 
We initialize the set of active and terminated agents, the set of observed stations, and the vector that describes the state of each agent~(l.1). 
A new request or a new station visit triggers a new decision epoch~(l.3). In case this is not the first decision epoch, we update the status of the last deciding agent from $s^\lambda = \text{d}$ to $s^\lambda = \text{r}$ and her assigned station to $v^*$. We assume that the system state~$x$ gets implicitly updated upon each new decision epoch.
The current agent~$\lambda$ can observe $v^\lambda$ as available and successfully terminate her search~(l.7\&8), as occupied~(l.9), or may start her search, such that we add $\lambda$ to $\mathcal{J}$~(l.10\&11).
In case $v^\lambda$ is not available, the deciding agent~$\lambda$ must select the next best station~$v^*$ to visit. 
For each feasible station that $\lambda$ may visit~(l.13) and for both possible states ${\acute{x}}^{\text{f}}$ and ${\acute{x}}^{\text{s}}$, the function $\texttt{greedyCost}(\acute{x})$ approximates the related cost-to-go $V^{\tilde{\pi}}(\acute{x})$ by following the greedy base-policy $\tilde{\pi}$ until a termination state or until $K$ decision epochs are reached. 
%
Given the explicit values of $V^{\tilde{\pi}}({\acute{x}}^{\text{s}})$ and $V^{\tilde{\pi}}({\acute{x}}^{\text{f}})$, the best station $\pi(x)=v$ minimizes the cost-to-go~(l.16,17\&18) as
\begin{equation}\label{eq:rollout}
\pi(x) = \argmin_{u=(v) \in \mathcal{U}(x)} d(x, u) +  p_{v}V^{\tilde{\pi}}({\acute{x}}^{\text{s}}) + (1 - p_{v})V^{\tilde{\pi}}({\acute{x}}^{\text{f}}) \;.
\end{equation}
If there is no reachable station within $\lambda$'s remaining time budget (l.19), $\lambda$ has failed her search. The search continues over the overall planning horizon as long as new decision epochs are triggered~(l.2\&3).

\begin{figure}[tbp]
	\footnotesize
	\caption{Online rollout algorithm}
	\label{alg:rollout}
	\begin{algorithmic}[1]
	\State$\mathcal{J} \gets \emptyset$,  $\mathcal{T} \gets \emptyset$,  $\mathcal{O} \gets \emptyset$, $\vec{x} \gets \vec{0}$, $x\gets(\mathcal{J}, \mathcal{T}, \mathcal{O}, \vec{x})$
	\While{True}
		\If{$\texttt{newEpochTriggered}()$}		
			\If{$\mathcal{J} \neq \emptyset \land s^\lambda == \text{d}$}
				\State $s^\lambda \gets \text{r}$, $v^\lambda \gets v^*$, $t^\lambda \gets t^\lambda + t_{v_k,v^*}$
			\EndIf			
			\State $\lambda \gets \texttt{decidingAgent}()$, $v^\lambda \gets \texttt{observedStation}()$		
			\If{$a_{v^\lambda} == 1$}
				\State$\mathcal{O}$.add($v^\lambda$), $\mathcal{T}$.add($\lambda$), $\mathcal{J}$.pop($\lambda$), $s^\lambda \gets \text{f}$		
			\Else
				\If{$\lambda \notin \mathcal{J}$}
					\State$\mathcal{J}$.add($\lambda$), $s^\lambda \gets \text{d}$
				\EndIf						
					\State $v^* \gets 0$, $Q^*	 \gets \infty$
					\For{$v \in \mathcal{U}(x)$} 
						\For{${\acute{x}}^{\text{s}}, {\acute{x}}^{\text{f}} \in T(x,v)$}
							\State $V^s \gets \texttt{greedyCost}({\acute{x}}^{\text{s}}) $, $V^f \gets \texttt{greedyCost}({\acute{x}}^{\text{f}}) $
							\State $Q \gets t_{v_k,v} + (1 - {p}_{v}) V^f + {p}_{v}V^s$
							\If{$Q < Q^* $}
								\State  $Q^* \gets Q $, $v^* \gets v$
							\EndIf
						\EndFor
					\EndFor		
					\If{$v^* == 0$}
						\State $s^\lambda \gets \text{t}$
				\EndIf
				\EndIf
			\EndIf
	\EndWhile		
	\end{algorithmic}
\end{figure}

\paragraph{Dynamic \gls{HLHC} algorithm (\gls{LHRO}):}
We now use the dynamic variant of the \gls{HLHC} algorithm to solve the CEN setting and detail the pseudo code in Figure~\ref{alg:lhro_algo}.
\begin{figure}[tbp]
	\footnotesize
	\caption{Label-based heuristic (\gls{LHRO})}
	\label{alg:lhro_algo}
	\begin{algorithmic}[1]
		\State $\mathcal{J} \gets \emptyset$,  $\mathcal{T} \gets \emptyset$,  $\mathcal{O} \gets \emptyset$, $\vec{x} \gets \vec{0}$, $x\gets(\mathcal{J}, \mathcal{T}, \mathcal{O}, \vec{x})$, $\Pi \gets \emptyset$
		\While{True}
		\If{$\texttt{newEpochTriggered}$}		
		\If{$\mathcal{J} \neq \emptyset \land s^\lambda == \text{d}$}
		\State $s^\lambda \gets \text{r}$, $t^\lambda \gets t^\lambda + t_{v^\lambda,v^*}$, $v^\lambda \gets v^*$, 
		\EndIf			
		\State $\lambda \gets \texttt{decidingAgent}()$, $v^\lambda \gets \texttt{observedStation}()$, $\Pi$.pop($\pi^\lambda$)
		\If{$a_{v^\lambda} == 1$}
		\State $\mathcal{O}$.add($v^\lambda$), $\mathcal{T}$.add($\lambda$), $\mathcal{J}$.pop($\lambda$), $s^\lambda \gets \text{f}$,$\Pi$.pop($\pi^\lambda$)
		\Else
		\If{$\lambda \notin \mathcal{J}$}
		\State $\mathcal{J}$.add($\lambda$), $s^\lambda \gets \text{d}$
		\EndIf			
		\State $\Pi$.pop($\pi^\lambda$)		
		\State $\pi^* \gets 0$, $\alpha^*	\gets \infty$
		\State $T \gets \bar{T}^\lambda - t^\lambda$
		\For{$\pi \in \texttt{getBestPaths}(v^\lambda, T, \Pi, \mathcal{O})$}
			\State $\alpha \gets \texttt{getCost}(\Pi\circ \pi)$
				\If{$\alpha < \alpha^* $}
					\State $\alpha^* \gets \alpha $, $v^* \gets \pi(x)$, $\pi^* \gets \pi$
				\EndIf 
		\EndFor		
		\If{$v^* == 0$}
		\State $s^\lambda \gets \text{t}$
		\Else	
		\State $\Pi$.add($\pi^*$)
		\EndIf
		\EndIf
		\EndIf
		\EndWhile		
	\end{algorithmic}
\end{figure}
Upon each agent's request, the algorithm plans a user-independent policy (see Section~\ref{subsubsec:dec-o}) and dynamically refines it at each agent's decision epoch using the latest updated information, i.e., station occupancy observations and visit intention updates. 

Similar to the online rollout algorithm (see Section~\ref{par:RO}), we initialize our variables and compute a visit recommendation each time an agent starts her search or visits an occupied station. We add the agent to the set of terminated agents if it finds an available station~(l.7\&8). In addition, we initialize the set $\Pi$ that contains for each active agent $i \in \mathcal{J}$ her last computed user-independent policy.
The procedure $\texttt{getBestPaths}$~(l.15) executes the single-agent label-setting algorithm LH (see Section~\ref{subsubsec:dec-o}) that returns the $n$ best candidate single-agent policies for $\lambda$.
We then select policy $\pi^{*}$ that minimizes the subsystem cost $\alpha^{\Pi\circ{\pi^{*}}}$~(l.16,17\&18) as
\begin{equation}\label{eq:Pi_refinement}
\pi^{*} = \argmin_{\pi \in \Gamma^\lambda } \alpha^{\Pi\circ{\pi}},
\end{equation}
with $\Gamma^{\lambda} $ being the set that contains the $n$ best policies for $\lambda$, and with $\Pi$ being the set of individual policies for all agents except $\lambda$.
%
We let the procedure $\texttt{getCost}$~(l.16) compute the value of $\alpha^{\Pi \circ{\pi}}$.
Finally, as long as unvisited stations are left, the minimum cost policy $\pi^{*}$ provides the next station visit $v^*$ for $\lambda$ from the current state~$x$~(l.19) with $\lambda$ being located at $v^\lambda_0 := v^\lambda$. Since $\pi^{*}$ maps to a search path $(v^\lambda_0, v^\lambda_1, ..., v^\lambda_n)$, $v^*$ corresponds to the first unvisited station, i.e., $v^* := v^\lambda_1$.

\subsubsection{Decentralized decision-making:}\label{subsubsec:dec_online}

In the decentralized decision-making settings with observation-sharing (DEC-O-d), we compute an agent's search path using the algorithm developed for the static planning setting DEC-O (see Section~\ref{subsubsec:dec-o}). However, we recompute the initially planned search path each time the agent visits an occupied station, using the latest observations shared by all agents
\subsection{Observation validity}\label{subsec:observation_validity}

For the sake of completeness, we discuss how to relax station occupancy persistency in the following. To do so, we assume that availability probabilities of already visited stations can recover over time following a time-dependent exponential function \citep[cf.][]{GuilletHiermannEtAl2021} defined as
\begin{equation}\label{eq:time_dependent_occ}
p^r_v(\Delta^i_v)  = p_v(1 - e^{-(\frac{\mu_v}{p_v})(\Delta^i_v)})\;,
\end{equation}
with $\Delta^i_v$ being the elapsed time since $i$ visited $v$ for the last time.
Here, $\frac{1}{\mu_v}$, denotes the average time station~$v$ remains occupied, and $p_v$ the probability that $v$ is available prior to any visit. Both values remain constant over the total planning horizon.
To account for time-dependent recovering functions, we consider observed stations as candidate stations with an availability probability recovered according to Equation~\ref{eq:time_dependent_occ}. We note that to reduce the computation overhead, one could chose to exclude latest observed stations from candidate stations, formally if $\Delta^i_v \leq T_{\text{thres}}$. 

In this paper, we assume that agents can only passively communicate about occupied charging stations. These observations could be deduced from sharing GPS trace data, if a vehicle drives by a station without stopping, then the station is assumed to be occupied. However, future work could assume agents being able to communicate actively about station availability when they leave the station after having (re)charged their car. In this case, the station is available upon the agent's departure and the probability availability decreases from $1$ to her initial value $p_v$, analogously to Equation~\ref{eq:time_dependent_occ} as
\begin{equation}\label{eq:time_dependent_avail}
p^r_v(\Delta^i_v)  = p_v  +  (1-p_v)(e^{-(\frac{\mu_v}{p_v})(\Delta^i_v)}).
\end{equation}
Moreover, Equation~\ref{eq:time_dependent_avail} could be used in a setting that allows an agent not to select the first (or next) available station, while reporting about the availability status of the non-selected stations.
\section{Experimental design}\label{sec:exp_design}
\begin{figure}[hbp]		
	\centering
	\includegraphics[width=.35\linewidth]{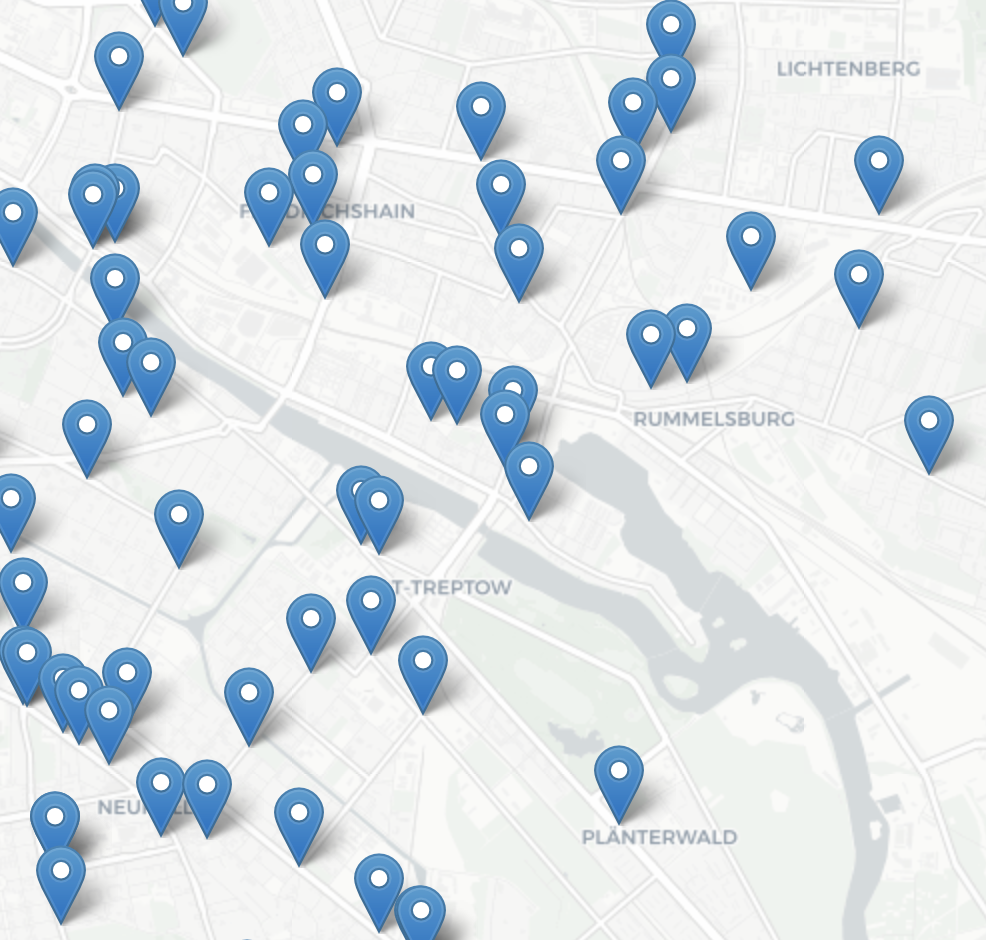}
	\caption{Charging stations distribution in a part of Berlin}\label{fig:berlin}
	\fnote{\footnotesize Charging station network in the city center of Berlin, Germany used to build the respective instance graph. The Image was created using Folium, which is data licensed under MIT License.}
\end{figure}

To analyze the effectiveness of the different coordination strategies, we conduct extensive numerical simulation experiments on real-world test instances for the city of Berlin (see Figure~\ref{fig:berlin}). In the following, we first detail our instance generation, before we describe additional benchmark algorithms, and elaborate on the metrics used to evaluate our algorithms' performance.

\subsection{Instance generation}

Besides the charging station availability distribution, the ratio of candidate stations per number of drivers and the departure time horizon are the main factors that impact our results. Accordingly, we vary the number of drivers, the global and individual search area dimensions, and the planning horizon to create a diverse set of scenarios as follows.

To account for varying spatial overlaps in between multiple drivers' search areas as well as for a varying number of drivers, we randomly draw departure locations within a radius of $r^s \in \{100,300,700\}$ meters for a total number of $N \in \{2,...,10\}$ drivers. Additionally, we consider two different driver search radii of $\bar{S} = 1$ km and $\bar{S} = 2$ km.
Table~\ref{tab:stations} specifies the available number of stations that result from varying those parameters.
Moreover, we account for varying temporal overlaps between multiple drivers by equally distributing the drivers' search start time within a varying time horizon $t^s \in \{0,1,5,15\}$~min. 
Independent of those characteristics, drivers have a search time budget of $\bar{T} = 5$~min. 
Utilizing a full-factorial design, we thus obtain $216$ different test instances for our studies.

\begin{table}[htp]
	\caption{Number of stations depending on the search area dimension\label{tab:stations}}
	\centering
	\begin{tabular}{crrrrrrrrrr}
		\toprule
		\multirow{2}[4]{*}{$\bar{S}$} & \multicolumn{1}{c}{\multirow{2}[4]{*}{$r^s$}} & \multicolumn{9}{c}{$N$} \\
		\cmidrule{3-11}          &       & 2     & 3     & 4     & 5     & 6 & 7 & 8 & 9 & 10 \\
		\midrule
		\multirow{3}[1]{*}{1000} & 100   & 11    & 13    & 13    & 13  &  13    & 13    & 13   & 13 & 13 \\
		& 300   & 11    & 13    & 14    & 15 & 16 & 17& 18 & 18  & 18 \\
		& 700   & 10    & 13    & 15    & 14 & 16 &18 & 22 & 22 & 21 \\
		\multirow{3}[1]{*}{2000} & 100   & 22    & 23    & 23  & 23 & 23 &  23    & 23    & 23    & 23 \\
		& 300   & 23    & 25    & 27    & 28   & 29 & 30 & 31 & 31 & 31 \\
		& 700   & 26    & 31    & 36    & 40   & 40 &  35  & 52 & 53  & 40 \\
		\bottomrule
	\end{tabular}%
\end{table}

To analyze the impact of the varying charging station availability, we consider a low and a high charging station availability scenario.
We generate those scenarios based on probability distributions centered on an expected mean $d_a$  with $d_a=0.25$ for the low-availability scenario (\gls{low_avail}) and $d_a=0.60$ for the high-availability scenario (\gls{high_avail}). 
For each instance and availability scenario, we perform $100$ simulation runs and use the same realized availability values to compute simulated estimates over all test instances.

We analyzed the sensitivity of our algorithms with respect to the penalty term $\beta^G$ in a preliminary study (see Appendix~\ref{app:beta_sensitivity}).
In this study, we observed only little sensitivity of the results to $\beta^G$, such that there exists only a minor trade-off between system robustness and quality performances, that is best addressed by setting $\beta^G=700$~min.
%


We further consider agents' heterogeneity to account for both heterogeneous and homogeneous use cases. 
We focus the main results discussion on the homogeneous agents use-case, because a service provider does not want to bias the search towards single agents by considering heterogeneous parameters in practice.
%
Accordingly, we set the station's utilization cost $\gamma^i_{v}=0$ for all drivers $i\in \mathcal{D}$ and all stations $v\in \mathcal{V}$, and consider an identical time budget $\bar{T}$ and search radius $\bar{S}$ for all agents.
We additionally briefly analyze the impact of heterogeneous agents' time budget and search radius to complete our analyses.

\subsection{Benchmark algorithms}
We evaluate the quality performances of the RO and LH-RO in the centralized setting~(CEN). Since the MDP is too large to be solved optimally, we benchmark both policies against a myopic greedy algorithm~G and a deterministic offline solution OFF, as suggested in \cite{Powell2009}. 

In G, we greedily decide on the next station visit $v^*$ for the deciding agent~$\lambda$ in each decision epoch based on a cost combining the driving time from its current station $v^\lambda$ to available stations and the individual driver's time-based penalty weighted by stations' availability probability. Formally, 
\begin{equation}
v^* = \argmin_{v \in \bar{\mathcal{V}}} t_{v^\lambda, v} + (1 - p_v) \bar{\beta}\; ,
\end{equation}
with $\bar{\mathcal{V}}$ being the set of candidate stations that agent~$\lambda$ can visit.

In OFF, we assume the charging demand, i.e., all drivers' departure time and location, to be known for the overall planning horizon.
We then compute for each realization \mbox{$k \in [0,...,100]$} of simulated stations availability, the minimum-cost assignment of drivers to stations, using a weighted bipartite graph $G'=(\mathcal{V}',\mathcal{A}')$, with $(v,i) \in \mathcal{A}'$ such that $i\in \mathcal{D}$ and $v \in \mathcal{V}_{\text{avail}}(k)$ is the set of available stations for realization $k$. 
Let the weight for arc $(v,i)$ be the driving time from the agent $i$'s start location to station $v$, such that $w_{v,i} = t_{v^i_0,v}$. We add dummy station vertices $\nu$ if $| \mathcal{V}_{\text{avail}}| \leq |\mathcal{D}| $, such that $w_{\nu,i} = \beta \;  \forall i \in \mathcal{D} $ and we add dummy driver vertices $\iota$ if $| \mathcal{V}_{\text{avail}}| \geq |\mathcal{D}| $, such that $w_{v,\iota} = 0 \; \forall v \in \mathcal{V}_{\text{avail}}$. We then solve the resulting assignment problem with the Karp algorithm \citep{Karp1980}.

\subsection{Performance evaluation}

For each test instance, we successively compute the search path (static planning) or next station visit (dynamic planning) for all drivers, according to their departure order and the selected setting.
We then evaluate the performance from both a driver and a system perspective, based on 100 simulation runs for both \gls{low_avail} and \gls{high_avail} scenarios.
From a driver perspective, we compute the realized search time $\hat{t}^k$ for each simulation run~$k$, which allows us to compute the simulated estimate of a driver $i$'s individual cost $\hat{\alpha}^i$ as
\begin{equation}
\hat{\alpha}^i = \nicefrac{\sum_{k=0}^{100}  \hat{t}^k + \hat{\delta}^k  \bar{\beta}}{100}\;,
\end{equation}
with $\hat{\delta}^k$ being the binary variable that indicates whether the $k^{\text{th}}$ search was successful. We obtain the driver's success rate $\hat{\rho}^i$ as 
\begin{equation}
\hat{\rho}^i = \nicefrac{\sum_{k=0}^{100} \hat{\delta}^k}{100}\;.
\end{equation}
From a system perspective, we compute the simulated estimate of the expected system cost $\hat{\alpha}$ as
\begin{equation}
\hat{\alpha}  = \sum_{i \in \mathcal{D}} \hat{\alpha}^i +  (1 - \prod_{i \in \mathcal{D}} \hat{\rho}^i) \beta^G\;.
\end{equation}
The quantity $\hat{\rho} =\prod_{i \in \mathcal{D}} \hat{\rho}^i$ describes the realized system success rate, i.e., the simulated estimate of the likelihood that all drivers successfully finished their search.

\section{Results}\label{sec:results}

We first discuss our results from a system perspective (Section~\ref{subsec:system_eval}) before we focus on a driver perspective (Section~\ref{subsec:user_eval}).

\subsection{System perspective}\label{subsec:system_eval}

In the following, we first analyze the impact of collaboration in intention-sharing settings, before we focus on the impact of dynamic planning in observation-sharing settings.
We then benchmark our algorithms for the centralized setting, and finally draw general conclusions on the performance of all possible algorithmic settings.

\subsubsection{Collaboration in intention-sharing settings:}

To analyze the benefit of collaboration, we compare the performances of \gls{HLH} and \gls{HLHC} (see Section~\ref{subsubsec:label_coop}), in the \gls{2} setting, with respect to the system cost $\hat{\alpha}$ and the solution's fairness. Here, we consider a solution to be fair if each driver obtains a similar cost $\alpha^i$ independent of her departure position. 

Figures~\ref{fig:dec_quality}a and \ref{fig:dec_quality}b 
\begin{figure}[tbp]
	\begin{tabular}{c @{\qquad} c @{\qquad}}
		\scalebox{0.85}{
\begin{tikzpicture}

\definecolor{color0}{rgb}{0.194607843137255,0.453431372549019,0.632843137254902}
\definecolor{color1}{rgb}{0.881862745098039,0.505392156862745,0.173039215686275}

\begin{axis}[
legend cell align={left},
legend style={fill opacity=0.8, draw opacity=1, text opacity=1, draw=white!80!black, at={(0.97,0.75)}, anchor=south east, legend columns=1, /tikz/every even column/.append style={column sep=0.5cm}, /tikz/every odd column/.append style={column sep=0.15cm}},
tick align=outside,
tick pos=left,
x grid style={white!69.0196078431373!black},
xlabel={Driver departure position},
xmin=-0.459, xmax=4.359,
xtick style={color=black},
xticklabels={0,1,2,3,4,5},
y grid style={white!69.0196078431373!black},
ylabel={Individual cost [min]},
ymin=-0.634561445199996, ymax=124.5943547892,
ytick style={color=black}
]

\draw[draw=white!23.921568627451!black,fill=color1,line width=0.3pt] (axis cs:0,0) rectangle (axis cs:0,0);
\addlegendimage{ybar,ybar legend,draw=white!23.921568627451!black,fill=color1,line width=0.3pt};
\addlegendentry{\gls{HLHC}}

\draw[draw=white!23.921568627451!black,fill=color0,line width=0.3pt] (axis cs:0,0) rectangle (axis cs:0,0);
\addlegendimage{ybar,ybar legend,draw=white!23.921568627451!black,fill=color0,line width=0.3pt};
\addlegendentry{\gls{HLH}}

\addplot [black, forget plot]
table {%
-0.15 30.907808388
-0.15 22.928076444
};
\addplot [black, forget plot]
table {%
-0.15 81.410503503
-0.15 97.26932672
};
\addplot [black, forget plot]
table {%
-0.195 22.928076444
-0.105 22.928076444
};
\addplot [black, forget plot]
table {%
-0.195 97.26932672
-0.105 97.26932672
};
\addplot [black, forget plot]
table {%
0.85 33.007725552
0.85 30.666449604
};
\addplot [black, forget plot]
table {%
0.85 40.339262091
0.85 41.796556548
};
\addplot [black, forget plot]
table {%
0.805 30.666449604
0.895 30.666449604
};
\addplot [black, forget plot]
table {%
0.805 41.796556548
0.895 41.796556548
};
\addplot [black, mark=o, mark size=3, mark options={solid,fill opacity=0}, only marks, forget plot]
table {%
0.85 54.536631192
};
\addplot [black, forget plot]
table {%
1.85 43.942671725
1.85 34.676722168
};
\addplot [black, forget plot]
table {%
1.85 62.407317804
1.85 74.58120252
};
\addplot [black, forget plot]
table {%
1.805 34.676722168
1.895 34.676722168
};
\addplot [black, forget plot]
table {%
1.805 74.58120252
1.895 74.58120252
};
\addplot [black, forget plot]
table {%
2.85 39.830412882
2.85 39.765755092
};
\addplot [black, forget plot]
table {%
2.85 50.259569916
2.85 51.140978756
};
\addplot [black, forget plot]
table {%
2.805 39.765755092
2.895 39.765755092
};
\addplot [black, forget plot]
table {%
2.805 51.140978756
2.895 51.140978756
};
\addplot [black, mark=o, mark size=3, mark options={solid,fill opacity=0}, only marks, forget plot]
table {%
2.85 15.333383388
2.85 73.5333795
};
\addplot [black, forget plot]
table {%
3.85 21.614147505
3.85 11.90507908
};
\addplot [black, forget plot]
table {%
3.85 28.216128588
3.85 33.370008048
};
\addplot [black, forget plot]
table {%
3.805 11.90507908
3.895 11.90507908
};
\addplot [black, forget plot]
table {%
3.805 33.370008048
3.895 33.370008048
};
\addplot [black, forget plot]
table {%
0.05 39.045318675
0.05 33.069422212
};
\addplot [black, forget plot]
table {%
0.05 102.119703462
0.05 118.902131324
};
\addplot [black, forget plot]
table {%
0.00500000000000002 33.069422212
0.095 33.069422212
};
\addplot [black, forget plot]
table {%
0.00500000000000002 118.902131324
0.095 118.902131324
};
\addplot [black, forget plot]
table {%
1.05 58.590360531
1.05 28.749248304
};
\addplot [black, forget plot]
table {%
1.05 93.345396096
1.05 108.003231132
};
\addplot [black, forget plot]
table {%
1.005 28.749248304
1.095 28.749248304
};
\addplot [black, forget plot]
table {%
1.005 108.003231132
1.095 108.003231132
};
\addplot [black, forget plot]
table {%
2.05 23.424078705
2.05 10.20151788
};
\addplot [black, forget plot]
table {%
2.05 68.929148503
2.05 96.394309044
};
\addplot [black, forget plot]
table {%
2.005 10.20151788
2.095 10.20151788
};
\addplot [black, forget plot]
table {%
2.005 96.394309044
2.095 96.394309044
};
\addplot [black, forget plot]
table {%
3.05 30.117286546
3.05 11.08635204
};
\addplot [black, forget plot]
table {%
3.05 43.948446543
3.05 62.529175872
};
\addplot [black, forget plot]
table {%
3.005 11.08635204
3.095 11.08635204
};
\addplot [black, forget plot]
table {%
3.005 62.529175872
3.095 62.529175872
};
\addplot [black, forget plot]
table {%
4.05 8.34565935300001
4.05 5.05766202
};
\addplot [black, forget plot]
table {%
4.05 24.125741811
4.05 26.542085196
};
\addplot [black, forget plot]
table {%
4.005 5.05766202
4.095 5.05766202
};
\addplot [black, forget plot]
table {%
4.005 26.542085196
4.095 26.542085196
};
\path [draw=black, fill=color1]
(axis cs:-0.24,30.907808388)
--(axis cs:-0.06,30.907808388)
--(axis cs:-0.06,81.410503503)
--(axis cs:-0.24,81.410503503)
--(axis cs:-0.24,30.907808388)
--cycle;
\path [draw=black, fill=color1]
(axis cs:0.76,33.007725552)
--(axis cs:0.94,33.007725552)
--(axis cs:0.94,40.339262091)
--(axis cs:0.76,40.339262091)
--(axis cs:0.76,33.007725552)
--cycle;
\path [draw=black, fill=color1]
(axis cs:1.76,43.942671725)
--(axis cs:1.94,43.942671725)
--(axis cs:1.94,62.407317804)
--(axis cs:1.76,62.407317804)
--(axis cs:1.76,43.942671725)
--cycle;
\path [draw=black, fill=color1]
(axis cs:2.76,39.830412882)
--(axis cs:2.94,39.830412882)
--(axis cs:2.94,50.259569916)
--(axis cs:2.76,50.259569916)
--(axis cs:2.76,39.830412882)
--cycle;
\path [draw=black, fill=color1]
(axis cs:3.76,21.614147505)
--(axis cs:3.94,21.614147505)
--(axis cs:3.94,28.216128588)
--(axis cs:3.76,28.216128588)
--(axis cs:3.76,21.614147505)
--cycle;
\path [draw=black, fill=color0]
(axis cs:-0.04,39.045318675)
--(axis cs:0.14,39.045318675)
--(axis cs:0.14,102.119703462)
--(axis cs:-0.04,102.119703462)
--(axis cs:-0.04,39.045318675)
--cycle;
\path [draw=black, fill=color0]
(axis cs:0.96,58.590360531)
--(axis cs:1.14,58.590360531)
--(axis cs:1.14,93.345396096)
--(axis cs:0.96,93.345396096)
--(axis cs:0.96,58.590360531)
--cycle;
\path [draw=black, fill=color0]
(axis cs:1.96,23.424078705)
--(axis cs:2.14,23.424078705)
--(axis cs:2.14,68.929148503)
--(axis cs:1.96,68.929148503)
--(axis cs:1.96,23.424078705)
--cycle;
\path [draw=black, fill=color0]
(axis cs:2.96,30.117286546)
--(axis cs:3.14,30.117286546)
--(axis cs:3.14,43.948446543)
--(axis cs:2.96,43.948446543)
--(axis cs:2.96,30.117286546)
--cycle;
\path [draw=black, fill=color0]
(axis cs:3.96,8.34565935300001)
--(axis cs:4.14,8.34565935300001)
--(axis cs:4.14,24.125741811)
--(axis cs:3.96,24.125741811)
--(axis cs:3.96,8.34565935300001)
--cycle;
\addplot [black, forget plot]
table {%
-0.24 44.736351678
-0.06 44.736351678
};
\addplot [black, forget plot]
table {%
0.76 35.349050226
0.94 35.349050226
};
\addplot [black, forget plot]
table {%
1.76 49.676837902
1.94 49.676837902
};
\addplot [black, forget plot]
table {%
2.76 43.819864824
2.94 43.819864824
};
\addplot [black, forget plot]
table {%
3.76 22.945692426
3.94 22.945692426
};
\addplot [black, forget plot]
table {%
-0.04 69.382532946
0.14 69.382532946
};
\addplot [black, forget plot]
table {%
0.96 85.4892474
1.14 85.4892474
};
\addplot [black, forget plot]
table {%
1.96 48.07338782
2.14 48.07338782
};
\addplot [black, forget plot]
table {%
2.96 40.839781982
3.14 40.839781982
};
\addplot [black, forget plot]
table {%
3.96 13.18750239
4.14 13.18750239
};
\end{axis}

\end{tikzpicture}} &
		\scalebox{0.85}{
\begin{tikzpicture}

\definecolor{color0}{rgb}{0.194607843137255,0.453431372549019,0.632843137254902}
\definecolor{color1}{rgb}{0.881862745098039,0.505392156862745,0.173039215686275}

\begin{axis}[
legend cell align={left},
legend style={fill opacity=0.8, draw opacity=1, text opacity=1, draw=white!80!black, at={(0.97,0.75)}, anchor=south east, legend columns=1, /tikz/every even column/.append style={column sep=0.5cm}, /tikz/every odd column/.append style={column sep=0.15cm}},
tick align=outside,
tick pos=left,
x grid style={white!69.0196078431373!black},
xlabel={Driver departure position},
xmin=-0.459, xmax=4.359,
xtick style={color=black},
xticklabels={0,1,2,3,4,5},
y grid style={white!69.0196078431373!black},
ylabel={Individual cost [min]},
ymin=-3.268346466, ymax=75.39499737,
ytick style={color=black}
]

\draw[draw=white!23.921568627451!black,fill=color1,line width=0.3pt] (axis cs:0,0) rectangle (axis cs:0,0);
\addlegendimage{ybar,ybar legend,draw=white!23.921568627451!black,fill=color1,line width=0.3pt};
\addlegendentry{\gls{HLHC}}

\draw[draw=white!23.921568627451!black,fill=color0,line width=0.3pt] (axis cs:0,0) rectangle (axis cs:0,0);
\addlegendimage{ybar,ybar legend,draw=white!23.921568627451!black,fill=color0,line width=0.3pt};
\addlegendentry{\gls{HLH}}

\addplot [black, forget plot]
table {%
-0.15 3.023877812
-0.15 0.678967379999999
};
\addplot [black, forget plot]
table {%
-0.15 6.749801761
-0.15 6.93051353200001
};
\addplot [black, forget plot]
table {%
-0.195 0.678967379999999
-0.105 0.678967379999999
};
\addplot [black, forget plot]
table {%
-0.195 6.93051353200001
-0.105 6.93051353200001
};
\addplot [black, mark=o, mark size=3, mark options={solid,fill opacity=0}, only marks, forget plot]
table {%
-0.15 18.244960064
};
\addplot [black, forget plot]
table {%
0.85 2.185477149
0.85 0.945826472000001
};
\addplot [black, forget plot]
table {%
0.85 17.397731461
0.85 37.505820144
};
\addplot [black, forget plot]
table {%
0.805 0.945826472000001
0.895 0.945826472000001
};
\addplot [black, forget plot]
table {%
0.805 37.505820144
0.895 37.505820144
};
\addplot [black, forget plot]
table {%
1.85 1.2554549
1.85 0.515378579999999
};
\addplot [black, forget plot]
table {%
1.85 11.59619522
1.85 14.71333908
};
\addplot [black, forget plot]
table {%
1.805 0.515378579999999
1.895 0.515378579999999
};
\addplot [black, forget plot]
table {%
1.805 14.71333908
1.895 14.71333908
};
\addplot [black, mark=o, mark size=3, mark options={solid,fill opacity=0}, only marks, forget plot]
table {%
1.85 36.18188976
};
\addplot [black, forget plot]
table {%
2.85 2.973032832
2.85 2.745624036
};
\addplot [black, forget plot]
table {%
2.85 12.979853304
2.85 13.847210244
};
\addplot [black, forget plot]
table {%
2.805 2.745624036
2.895 2.745624036
};
\addplot [black, forget plot]
table {%
2.805 13.847210244
2.895 13.847210244
};
\addplot [black, mark=o, mark size=3, mark options={solid,fill opacity=0}, only marks, forget plot]
table {%
2.85 53.593745828
};
\addplot [black, forget plot]
table {%
3.85 0.950436029999999
3.85 0.307260072
};
\addplot [black, forget plot]
table {%
3.85 4.556207272
3.85 6.512847156
};
\addplot [black, forget plot]
table {%
3.805 0.307260072
3.895 0.307260072
};
\addplot [black, forget plot]
table {%
3.805 6.512847156
3.895 6.512847156
};
\addplot [black, forget plot]
table {%
0.05 6.986599391
0.05 3.324451028
};
\addplot [black, forget plot]
table {%
0.05 16.631001081
0.05 18.244960064
};
\addplot [black, forget plot]
table {%
0.00500000000000002 3.324451028
0.095 3.324451028
};
\addplot [black, forget plot]
table {%
0.00500000000000002 18.244960064
0.095 18.244960064
};
\addplot [black, mark=o, mark size=3, mark options={solid,fill opacity=0}, only marks, forget plot]
table {%
0.05 37.270077124
};
\addplot [black, forget plot]
table {%
1.05 8.390488401
1.05 0.945826472000001
};
\addplot [black, forget plot]
table {%
1.05 39.483466212
1.05 71.819390832
};
\addplot [black, forget plot]
table {%
1.005 0.945826472000001
1.095 0.945826472000001
};
\addplot [black, forget plot]
table {%
1.005 71.819390832
1.095 71.819390832
};
\addplot [black, forget plot]
table {%
2.05 0.698777459999999
2.05 0.636955219999999
};
\addplot [black, forget plot]
table {%
2.05 1.164209535
2.05 1.25753132
};
\addplot [black, forget plot]
table {%
2.005 0.636955219999999
2.095 0.636955219999999
};
\addplot [black, forget plot]
table {%
2.005 1.25753132
2.095 1.25753132
};
\addplot [black, mark=o, mark size=3, mark options={solid,fill opacity=0}, only marks, forget plot]
table {%
2.05 6.61256750000001
};
\addplot [black, forget plot]
table {%
3.05 11.897538448
3.05 1.148717172
};
\addplot [black, forget plot]
table {%
3.05 27.686055951
3.05 50.316618
};
\addplot [black, forget plot]
table {%
3.005 1.148717172
3.095 1.148717172
};
\addplot [black, forget plot]
table {%
3.005 50.316618
3.095 50.316618
};
\addplot [black, forget plot]
table {%
4.05 0.509896389000001
4.05 0.307260072
};
\addplot [black, forget plot]
table {%
4.05 3.236880844
4.05 3.792752304
};
\addplot [black, forget plot]
table {%
4.005 0.307260072
4.095 0.307260072
};
\addplot [black, forget plot]
table {%
4.005 3.792752304
4.095 3.792752304
};
\addplot [black, mark=o, mark size=3, mark options={solid,fill opacity=0}, only marks, forget plot]
table {%
4.05 10.039271652
};
\path [draw=black, fill=color1]
(axis cs:-0.24,3.023877812)
--(axis cs:-0.06,3.023877812)
--(axis cs:-0.06,6.749801761)
--(axis cs:-0.24,6.749801761)
--(axis cs:-0.24,3.023877812)
--cycle;
\path [draw=black, fill=color1]
(axis cs:0.76,2.185477149)
--(axis cs:0.94,2.185477149)
--(axis cs:0.94,17.397731461)
--(axis cs:0.76,17.397731461)
--(axis cs:0.76,2.185477149)
--cycle;
\path [draw=black, fill=color1]
(axis cs:1.76,1.2554549)
--(axis cs:1.94,1.2554549)
--(axis cs:1.94,11.59619522)
--(axis cs:1.76,11.59619522)
--(axis cs:1.76,1.2554549)
--cycle;
\path [draw=black, fill=color1]
(axis cs:2.76,2.973032832)
--(axis cs:2.94,2.973032832)
--(axis cs:2.94,12.979853304)
--(axis cs:2.76,12.979853304)
--(axis cs:2.76,2.973032832)
--cycle;
\path [draw=black, fill=color1]
(axis cs:3.76,0.950436029999999)
--(axis cs:3.94,0.950436029999999)
--(axis cs:3.94,4.556207272)
--(axis cs:3.76,4.556207272)
--(axis cs:3.76,0.950436029999999)
--cycle;
\path [draw=black, fill=color0]
(axis cs:-0.04,6.986599391)
--(axis cs:0.14,6.986599391)
--(axis cs:0.14,16.631001081)
--(axis cs:-0.04,16.631001081)
--(axis cs:-0.04,6.986599391)
--cycle;
\path [draw=black, fill=color0]
(axis cs:0.96,8.390488401)
--(axis cs:1.14,8.390488401)
--(axis cs:1.14,39.483466212)
--(axis cs:0.96,39.483466212)
--(axis cs:0.96,8.390488401)
--cycle;
\path [draw=black, fill=color0]
(axis cs:1.96,0.698777459999999)
--(axis cs:2.14,0.698777459999999)
--(axis cs:2.14,1.164209535)
--(axis cs:1.96,1.164209535)
--(axis cs:1.96,0.698777459999999)
--cycle;
\path [draw=black, fill=color0]
(axis cs:2.96,11.897538448)
--(axis cs:3.14,11.897538448)
--(axis cs:3.14,27.686055951)
--(axis cs:2.96,27.686055951)
--(axis cs:2.96,11.897538448)
--cycle;
\path [draw=black, fill=color0]
(axis cs:3.96,0.509896389000001)
--(axis cs:4.14,0.509896389000001)
--(axis cs:4.14,3.236880844)
--(axis cs:3.96,3.236880844)
--(axis cs:3.96,0.509896389000001)
--cycle;
\addplot [black, forget plot]
table {%
-0.24 4.766058738
-0.06 4.766058738
};
\addplot [black, forget plot]
table {%
0.76 6.02028281
0.94 6.02028281
};
\addplot [black, forget plot]
table {%
1.76 1.75114748
1.94 1.75114748
};
\addplot [black, forget plot]
table {%
2.76 6.705680958
2.94 6.705680958
};
\addplot [black, forget plot]
table {%
3.76 3.31726889
3.94 3.31726889
};
\addplot [black, forget plot]
table {%
-0.04 11.45896789
0.14 11.45896789
};
\addplot [black, forget plot]
table {%
0.96 33.97807155
1.14 33.97807155
};
\addplot [black, forget plot]
table {%
1.96 0.884244179999998
2.14 0.884244179999998
};
\addplot [black, forget plot]
table {%
2.96 18.953301116
3.14 18.953301116
};
\addplot [black, forget plot]
table {%
3.96 1.343535902
4.14 1.343535902
};
\end{axis}

\end{tikzpicture}} \\
		\small (a) \gls{low_avail} scenario & \small(b) \gls{high_avail} scenario \\
	\end{tabular}	
	\vskip 0.2 in
	\begin{center}
		\stackunder[5pt]{
			\footnotesize
			\begin{tabular}{ccrrrrrr}
				\toprule
				& \multicolumn{1}{l}{} & \multicolumn{1}{l}{$\Delta(\text{mean})$}& \multicolumn{1}{l}{$\Delta(\text{median})$} & \multicolumn{1}{l}{$\Delta(\text{max})$} & \multicolumn{1}{l}{$\Delta(\text{min})$} & \multicolumn{1}{l}{$\Delta(\text{q1})$} & \multicolumn{1}{l}{$\Delta(\text{q3})$} \\
				\midrule
				\multirow{3}[2]{*}{\gls{low_avail}} & \multicolumn{1}{l}{HLH-c} &  26.7 & 31.2 & 63.9  & 22.7 & 22.3 & 53.2 \\
				&\multicolumn{1}{l}{HLH}      & 60.1 & 72.3  & 92.4  & 28.0 & 50.2 & 78.0 \\
				&$\Delta_{\text{ref}}$ [\%]     & -56 & -57 & -31 & -19 & -56 & -32 \\
				\midrule
				\multirow{3}[2]{*}{\gls{high_avail}} &\multicolumn{1}{l}{HLH-c} & 11.3 & 4.95  & 47.1  & 2.44  & 2.07  & 12.8 \\
				&\multicolumn{1}{l}{HLH}  & 28.5  & 33.1 & 65.2  & 3.02  & 11.4 & 38.3 \\
				&$\Delta_{\text{ref}}$ [\%]  & -60  & -85 & -28 & -19 & -82 & -66 \\
				\bottomrule
		\end{tabular}}
		{\small (c) Comparison of the distributions of driver's costs per departure position}
	\end{center}
	\vskip 0.2 in
	\caption{Comparison of \gls{HLH} and \gls{HLHC} in the \gls{2} setting for $t^s=0$~min} \label{fig:dec_quality}
	\footnotesize \textit{Note.} 
	Each subplot shows for each driver $i$ the distribution of the realized individual cost $\hat{\alpha}^i$ depending on her departure position, over all test instances that correspond to $t^s=0$~min, $r^s\in \{100,300,700\}$ m, $\bar{S} \in \{1000,2000\}$ meters, for $N=5$ drivers.
	In the table, we compute $\Delta$ as follows for each metric $m \in \{ \text{mean}, \text{median}, \text{q1}, \text{q3}, \text{max}, \text{min}\}$: $\Delta(m) = \max_{i} (m(i)) - \min_{i} (m(i))$, with $i$ being the departure position and $m(i)$ being the statistic of the costs distribution corresponding to the $i^{\text{th}}$ departure position. We compute $\Delta_{\text{ref}}$ as follows: $\nicefrac{\Delta_{\text{HLH-c}}(m) - \Delta_{\text{HLH}}(m)}{\Delta_{\text{HLH}}(m)}$.
\end{figure}	
show the distribution of individual drivers' cost depending on their departure order, obtained with the collaborative algorithm (\gls{HLHC}), respectively the non-collaborative algorithm (\gls{HLH}) for $t^s=0$~min, in low- and high-availability scenarios for all instances with $N=5$ drivers.
We note that five simultaneously active drivers represent a very likely real-world scenario that we observed in practice.
For each setting, Table~\ref{fig:dec_quality}c compares the difference between the highest and the lowest value (median, mean, 1\textsuperscript{st} and 3\textsuperscript{rd} quartiles, max, min) obtained for these individual drivers' cost distributions.
Figure~\ref{fig:dec_collab} shows the cost distribution depending on the starting time horizon $t^s$ for both \gls{HLH} and \gls{HLHC}.

As can be seen in Figure~\ref{fig:dec_quality}, solutions obtained with \gls{HLHC} outperform the solutions obtained with \gls{HLH} on average with respect to both the system cost $\hat{\alpha}$ and the solution's fairness.
Table~\ref{fig:dec_quality}c shows that the departure position has a smaller impact on the results with than without collaboration. The difference between highest and lowest mean, median, maximum, minimum, 1\textsuperscript{st} quartile, 3\textsuperscript{rd} quartile values decreases with collaboration.
In particular, collaboration decreases the difference for mean individual costs by 56\% in the \gls{low_avail} scenario and by 60\% in the \gls{high_avail} scenario. 
In the analyzed case, drivers start their search closely one after another, such that the last navigated drivers have a high chance to start while the preceding drivers are still searching. Without collaboration, the latest drivers may use the visit intention information to their advantage by earlier visiting the stations targeted by the preceding drivers.
Here, the collaborative procedure \gls{HLHC} yields lower realized system cost $\hat{\alpha}$ than \gls{HLH} as it ensures that the use of visit intention information benefits all drivers by avoiding conflicts at the stations already included in their search paths.
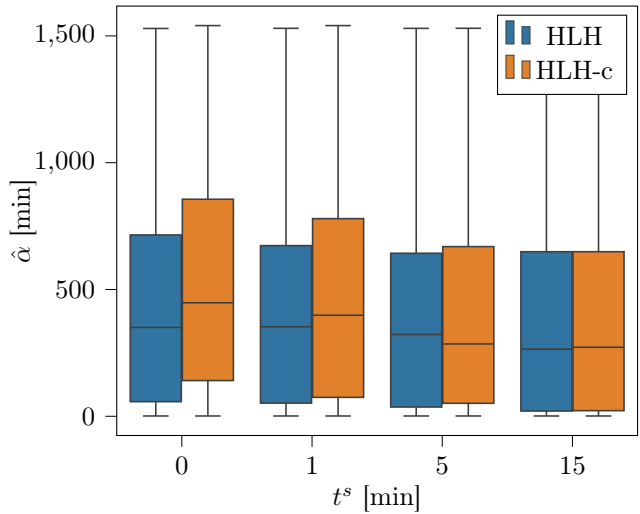
\begin{figure}[tbp]
	\centering
	\scalebox{1}{
\begin{tikzpicture}[font=\small]

\definecolor{color0}{rgb}{0.194607843137255,0.453431372549019,0.632843137254902}
\definecolor{color1}{rgb}{0.881862745098039,0.505392156862745,0.173039215686275}

\begin{axis}[
tick align=outside,
tick pos=left,
x grid style={white!69.0196078431373!black},
xlabel={$t^s$ [min]},
xmin=-0.5, xmax=3.5,
xtick style={color=black},
xtick={0,1,2,3},
xticklabels={0,1,5,15},
y grid style={white!69.0196078431373!black},
ylabel={$\hat{\alpha}$ [min]},
ymin=-75.821873, ymax=1617.710913,
ytick style={color=black}
]
\path [draw=white!24.7058823529412!black, fill=color0, semithick]
(axis cs:-0.396,57.5958875)
--(axis cs:-0.004,57.5958875)
--(axis cs:-0.004,715.204535)
--(axis cs:-0.396,715.204535)
--(axis cs:-0.396,57.5958875)
--cycle;
\path [draw=white!24.7058823529412!black, fill=color1, semithick]
(axis cs:0.004,140.8572825)
--(axis cs:0.396,140.8572825)
--(axis cs:0.396,855.855715)
--(axis cs:0.004,855.855715)
--(axis cs:0.004,140.8572825)
--cycle;
\path [draw=white!24.7058823529412!black, fill=color0, semithick]
(axis cs:0.604,52.00347)
--(axis cs:0.996,52.00347)
--(axis cs:0.996,673.0963675)
--(axis cs:0.604,673.0963675)
--(axis cs:0.604,52.00347)
--cycle;
\path [draw=white!24.7058823529412!black, fill=color1, semithick]
(axis cs:1.004,74.6166825)
--(axis cs:1.396,74.6166825)
--(axis cs:1.396,779.3935675)
--(axis cs:1.004,779.3935675)
--(axis cs:1.004,74.6166825)
--cycle;
\path [draw=white!24.7058823529412!black, fill=color0, semithick]
(axis cs:1.604,36.48036)
--(axis cs:1.996,36.48036)
--(axis cs:1.996,642.86837)
--(axis cs:1.604,642.86837)
--(axis cs:1.604,36.48036)
--cycle;
\path [draw=white!24.7058823529412!black, fill=color1, semithick]
(axis cs:2.004,51.0981575)
--(axis cs:2.396,51.0981575)
--(axis cs:2.396,669.159165)
--(axis cs:2.004,669.159165)
--(axis cs:2.004,51.0981575)
--cycle;
\path [draw=white!24.7058823529412!black, fill=color0, semithick]
(axis cs:2.604,20.64179)
--(axis cs:2.996,20.64179)
--(axis cs:2.996,649.018505)
--(axis cs:2.604,649.018505)
--(axis cs:2.604,20.64179)
--cycle;
\path [draw=white!24.7058823529412!black, fill=color1, semithick]
(axis cs:3.004,22.145915)
--(axis cs:3.396,22.145915)
--(axis cs:3.396,649.018505)
--(axis cs:3.004,649.018505)
--(axis cs:3.004,22.145915)
--cycle;
\draw[draw=white!24.7058823529412!black,fill=color0,line width=0.3pt] (axis cs:0,0) rectangle (axis cs:0,0);
\addlegendimage{ybar,ybar legend,draw=white!24.7058823529412!black,fill=color0,line width=0.3pt};
\addlegendentry{\gls{HLH}}

\draw[draw=white!24.7058823529412!black,fill=color1,line width=0.3pt] (axis cs:0,0) rectangle (axis cs:0,0);
\addlegendimage{ybar,ybar legend,draw=white!24.7058823529412!black,fill=color1,line width=0.3pt};
\addlegendentry{\gls{HLHC}}

\addplot [semithick, white!24.7058823529412!black]
table {%
-0.2 57.5958875
-0.2 1.15689
};
\addplot [semithick, white!24.7058823529412!black]
table {%
-0.2 715.204535
-0.2 1529.36676
};
\addplot [semithick, white!24.7058823529412!black]
table {%
-0.298 1.15689
-0.102 1.15689
};
\addplot [semithick, white!24.7058823529412!black]
table {%
-0.298 1529.36676
-0.102 1529.36676
};
\addplot [semithick, white!24.7058823529412!black]
table {%
0.2 140.8572825
0.2 1.15689
};
\addplot [semithick, white!24.7058823529412!black]
table {%
0.2 855.855715
0.2 1540.72614
};
\addplot [semithick, white!24.7058823529412!black]
table {%
0.102 1.15689
0.298 1.15689
};
\addplot [semithick, white!24.7058823529412!black]
table {%
0.102 1540.72614
0.298 1540.72614
};
\addplot [semithick, white!24.7058823529412!black]
table {%
0.8 52.00347
0.8 1.15689
};
\addplot [semithick, white!24.7058823529412!black]
table {%
0.8 673.0963675
0.8 1529.97354
};
\addplot [semithick, white!24.7058823529412!black]
table {%
0.702 1.15689
0.898 1.15689
};
\addplot [semithick, white!24.7058823529412!black]
table {%
0.702 1529.97354
0.898 1529.97354
};
\addplot [semithick, white!24.7058823529412!black]
table {%
1.2 74.6166825
1.2 1.15689
};
\addplot [semithick, white!24.7058823529412!black]
table {%
1.2 779.3935675
1.2 1540.73215
};
\addplot [semithick, white!24.7058823529412!black]
table {%
1.102 1.15689
1.298 1.15689
};
\addplot [semithick, white!24.7058823529412!black]
table {%
1.102 1540.73215
1.298 1540.73215
};
\addplot [semithick, white!24.7058823529412!black]
table {%
1.8 36.48036
1.8 1.15689
};
\addplot [semithick, white!24.7058823529412!black]
table {%
1.8 642.86837
1.8 1529.86002
};
\addplot [semithick, white!24.7058823529412!black]
table {%
1.702 1.15689
1.898 1.15689
};
\addplot [semithick, white!24.7058823529412!black]
table {%
1.702 1529.86002
1.898 1529.86002
};
\addplot [semithick, white!24.7058823529412!black]
table {%
2.2 51.0981575
2.2 1.15689
};
\addplot [semithick, white!24.7058823529412!black]
table {%
2.2 669.159165
2.2 1530.11141
};
\addplot [semithick, white!24.7058823529412!black]
table {%
2.102 1.15689
2.298 1.15689
};
\addplot [semithick, white!24.7058823529412!black]
table {%
2.102 1530.11141
2.298 1530.11141
};
\addplot [semithick, white!24.7058823529412!black]
table {%
2.8 20.64179
2.8 1.15689
};
\addplot [semithick, white!24.7058823529412!black]
table {%
2.8 649.018505
2.8 1528.28104
};
\addplot [semithick, white!24.7058823529412!black]
table {%
2.702 1.15689
2.898 1.15689
};
\addplot [semithick, white!24.7058823529412!black]
table {%
2.702 1528.28104
2.898 1528.28104
};
\addplot [semithick, white!24.7058823529412!black]
table {%
3.2 22.145915
3.2 1.15689
};
\addplot [semithick, white!24.7058823529412!black]
table {%
3.2 649.018505
3.2 1528.54787
};
\addplot [semithick, white!24.7058823529412!black]
table {%
3.102 1.15689
3.298 1.15689
};
\addplot [semithick, white!24.7058823529412!black]
table {%
3.102 1528.54787
3.298 1528.54787
};
\addplot [semithick, white!24.7058823529412!black]
table {%
-0.396 350.119965
-0.004 350.119965
};
\addplot [semithick, white!24.7058823529412!black]
table {%
0.004 447.66341
0.396 447.66341
};
\addplot [semithick, white!24.7058823529412!black]
table {%
0.604 353.06969
0.996 353.06969
};
\addplot [semithick, white!24.7058823529412!black]
table {%
1.004 398.45106
1.396 398.45106
};
\addplot [semithick, white!24.7058823529412!black]
table {%
1.604 322.827305
1.996 322.827305
};
\addplot [semithick, white!24.7058823529412!black]
table {%
2.004 285.19604
2.396 285.19604
};
\addplot [semithick, white!24.7058823529412!black]
table {%
2.604 264.932105
2.996 264.932105
};
\addplot [semithick, white!24.7058823529412!black]
table {%
3.004 272.2813
3.396 272.2813
};
\end{axis}

\end{tikzpicture}}
	\caption{Comparison of the $\hat{\alpha}$-distribution obtained with \gls{HLH} and \gls{HLHC} in the \gls{2} setting}\label{fig:dec_collab}
	\fnote{\footnotesize The figure shows the $\hat{\alpha}$-distribution for each considered setting, depending on the starting time horizon $t^s$.}
\end{figure}

Figure~\ref{fig:dec_collab}, as well as an additional analysis in Appendix~\ref{app:num_res}, show a similar trend for $t^s=1$~min, but decreasing effects for larger departure time horizon, $t^s=5$ and  $t^s=15$~min. In the latter case, we observe no significant benefit in a collaborative setting.

\begin{res}
	Visit intentions sharing can – and should – be used collaboratively. Without collaboration, visit intention may favor selfish solutions that negatively affect early departing drivers' search paths. 
\end{res}

\subsubsection{Dynamic planning for decentralized observation-sharing:}\label{subsec:dynamic_impact}

We analyze the impact of dynamically re-planning solutions in the observation-sharing setting.
Table~\ref{tab:dec-on-off} compares the system cost $\hat{\alpha}$ for \gls{3} and \gls{9}, in low- and high-availability scenarios. 
As can be seen, the effectiveness of dynamic and static planning depends on the length of the drivers' departure time horizon: if $t^s$ is small, drivers start their search almost simultaneously and cannot benefit from prior observations of preceding drivers. Here, dynamically sharing observations during the search significantly increases the available information for all drivers and thus leads to significant improvements, by decreasing cost $\hat{\alpha}$ on average by 8\% and up to 53\%.
If $t^s$ is large, subsequently searching drivers benefit from observations shared by prior drivers already in a static planning approach. Accordingly, the benefit of dynamic observation-sharing decreases with an average cost decrease limited to 2\%.
Here, DEC-O even outperforms \gls{9} in some cases (e.g., $N=4$ \& $t^s\in\{5,15\}$).

\begin{res}
	Without intention-sharing, dynamic observation-sharing in addition to static observation-sharing improves the system performances for short departure time horizons by decreasing cost $\hat{\alpha}$ by 8\% on average. 
\end{res}

\begin{table}[hbp]
	\centering
	\caption{System cost comparison between static and dynamic decentralized policy planning strategies}\label{tab:dec-on-off}
	{
		\centering
		\begin{tabular}{rrrrrrrrrr}
			\toprule
			& \multicolumn{4}{c}{\gls{low_avail}}        &       & \multicolumn{4}{c}{\gls{high_avail}} \\
			\cmidrule{2-5}\cmidrule{7-10}    \multicolumn{1}{l}{n} & \multicolumn{1}{l}{$t^s$=0} & \multicolumn{1}{l}{$t^s$=1} & \multicolumn{1}{l}{$t^s$=5} & \multicolumn{1}{l}{$t^s$=15} &       & \multicolumn{1}{l}{$t^s$=0} & \multicolumn{1}{l}{$t^s$=1} & \multicolumn{1}{l}{$t^s$=5} & \multicolumn{1}{l}{$t^s$=15} \\
			\midrule  
			2     &    -13.4 & -6.94 & 0.09  & 0.09  &       & -7.14 & -1.54 & 0.00    & 0.00\\
			3     &    -27.7 & -21.9 & -7.67 & -1.16 &       & -53.1 & -37.2 & 0.65  & 0.20 \\
			4     &    -21.7 & -9.37 & -4.42 & 5.30   &       & -15.5 & -34.6 & 33.1 & 27.7 \\
			5     &    -14.0   & -11.3 & -2.81 & -0.75 &       & 10.4 & -25.2 & -12.7 & 7.62 \\
			6     &    -10.2 & -4.52 & -1.90  & -0.58 &       & -22.4 & -3.84 & -4.56 & 0.75 \\
			7     &    -6.14 & -2.12 & -1.47 & 0.77  &       & 0.86  & 3.87  & -5.12 & 1.45 \\
			8     &    -7.96 & -3.94 & -2.07 & -0.13 &       & -6.32 & -12.9 & -5.01 & -5.31 \\
			9     &    -6.80  & -5.75 & 0.18  & -0.10  &       & -11.2 & -11.1 & -2.75 & -4.72 \\
			10     &    -6.38 & -3.78 & -1.51 & 0.27  &       & -5.21 & -3.58 & -3.56 & -5.50 \\
			\bottomrule
		\end{tabular}%
	}
	\fnote{\footnotesize The table shows $\hat{\alpha}$'s relative improvement $\Delta [\%] = \nicefrac{(\hat{\alpha}_{\text{dyn}}-\hat{\alpha}_{\text{stat} } )}{\hat{\alpha}_{\text{stat}}}$ of the dynamic setting to the static setting for \gls{3} in percentages. A negative $\Delta$ shows that the dynamic setting outperforms the static counterpart.}
\end{table}%

\subsubsection{Centralized planning:}

In the following, we compare the performance of RO's and \gls{LHRO}'s policies to the greedy (G) and offline (OFF) benchmark described in Section~\ref{sec:exp_design}.
%
Here, G provides an upper bound to \gls{RO} and \gls{LHRO}, allowing to analyze the performance gain by looking ahead rather than myopically deciding. Contrary, OFF provides a lower bound for each availability realization, as all uncertain information -- station availability and future charging demand -- is known, allowing to study an artificial perfect information setting. 
\begin{figure}[tbp]
		\centering
		\scalebox{1}{
			\small
			\begin{tabular}{c @{\qquad} }
				\input{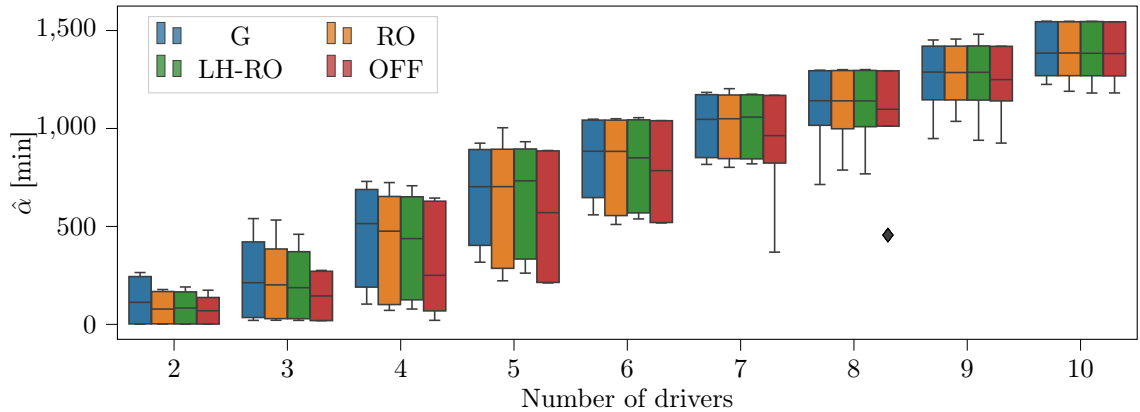} \\
				(a) $\bar{S}=1000$ meters \\
				\\
				\input{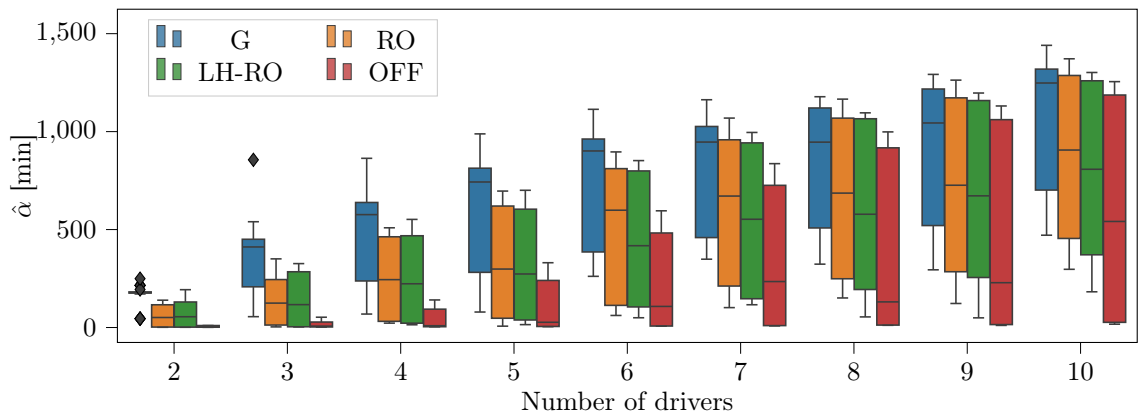} \\
				(b) $\bar{S}=2000$ meters \\
		\end{tabular}}
	\caption{Distribution of the realized system cost $\hat{\alpha}$ in \gls{5} for \gls{RO}, \gls{LHRO}, G, and OFF per number of drivers, separated for small and large search areas}\label{fig:boxplot_online}
\end{figure}
Figure~\ref{fig:boxplot_online} shows the detailed distribution of the realized cost $\hat{\alpha}$ for \gls{RO}, \gls{LHRO}, G, and OFF, for a varying number of drivers and search radii $\bar{S}$. Here, we note that we applied two-tailed Wilcoxon signed-rank tests to verify the significance of the average performance comparison.

As can be seen, both \gls{RO} and \gls{LHRO} outperform the myopic policies (G) by decreasing cost~$\hat{\alpha}$ by 14\% (RO) and 16\% (LH-RO) on average.
The cost reduction depends however on the search radius~$\bar{S}$. 
For small search areas ($\bar{S} = 1000$ m), both RO and LH-RO decrease the cost obtained with G by 2\% on average, while the perfect-information setting OFF yields a 9\% cost decrease on average. We observe that the benefits of using \gls{RO}, \gls{LHRO}, or even OFF, decrease with the number of drivers as in this case, there is only little room for improvement over the myopic policy, due to the very limited number of candidate stations available for all drivers.
For larger search areas ($\bar{S} = 2000$ m), all non-myopic settings achieve a significantly higher cost reduction of 29\% (RO), 34\% (LH-RO) and 63\% (OFF) compared to G. 
According to LH-RO and RO's performances, we decide to use \gls{LHRO} in the \gls{5} setting.

\begin{res}
	On average, the \gls{LHRO} algorithm decreases the cost $\hat{\alpha}$ obtained by RO by 3\% and the cost obtained by G by 16\%  in a centralized setting.
	Specifically, for larger search areas ($\bar{S} = 2000$ m), \gls{LHRO} achieves a cost reduction of 7\% (compared to RO) and 34\% (compared to G).	
\end{res}

\subsubsection{General performance evaluation:}

We compare the performances of all decentralized settings, \gls{1}, \gls{2}, \gls{3}, \gls{4} and \gls{9}, the centralized setting CEN, and a decentralized myopic benchmark, \gls{1}(N), that greedily computes search paths in a decentralized way, to reflect drivers' behavior without any search assistance. 

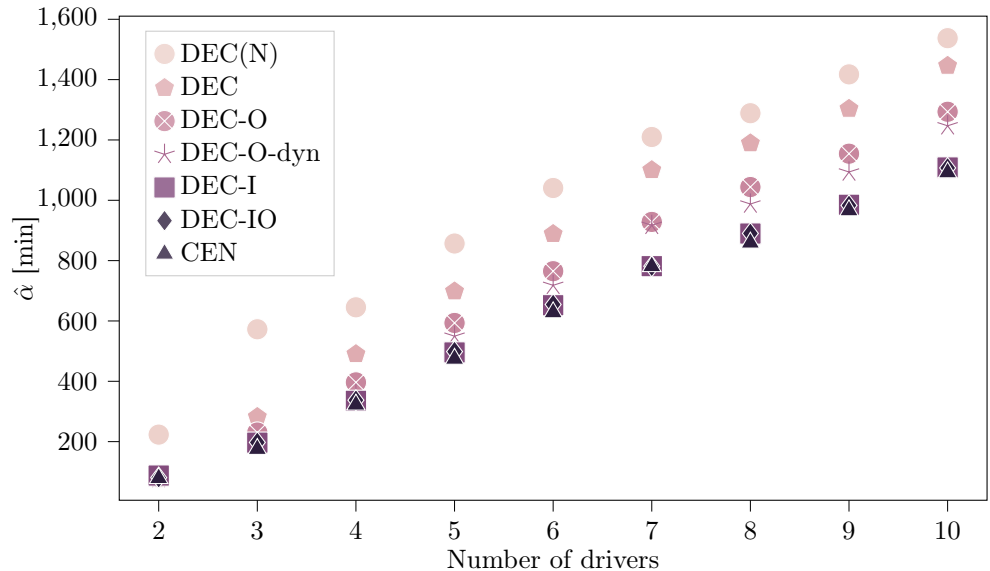
\begin{figure}[tbp]
	\centering
\begin{tikzpicture}[font=\small]

\definecolor{color0}{rgb}{0.931269222332537,0.820192179608212,0.797148097466359}
\definecolor{color1}{rgb}{0.87313570987373,0.676097952777522,0.694695586642386}
\definecolor{color2}{rgb}{0.784044088059945,0.529266054426589,0.620056892694176}
\definecolor{color3}{rgb}{0.662652747046639,0.402798937278128,0.559929395373557}
\definecolor{color4}{rgb}{0.515106903685576,0.298010475350561,0.490506191393007}
\definecolor{color5}{rgb}{0.175086564895221,0.118400233069168,0.242159891378365}
\definecolor{color6}{rgb}{0.175086564895221,0.118400233069168,0.242159891378365}

\begin{axis}[
width=13cm,
height=8cm,
legend cell align={left},
legend style={fill opacity=0.8, draw opacity=1, text opacity=1, at={(0.03,0.97)}, anchor=north west, draw=white!80!black},
tick align=outside,
tick pos=left,
x grid style={white!69.0196078431373!black},
xlabel={Number of drivers},
xmin=1.6, xmax=10.4,
xtick style={color=black},
y grid style={white!69.0196078431373!black},
ylabel={$\hat{\alpha}$ [min]},
ymin=5.68248258333328, ymax=1610.34342783333,
ytick style={color=black}
]
\addplot [only marks, color0, mark=*, mark size=4, mark options={solid,draw=white}]
table {%
2 223.248253333333
3 572.475963333333
4 645.256748541667
5 856.733000833333
6 1040.72635125
7 1209.801589375
8 1288.63279708333
9 1417.40049958333
10 1537.40429395833
};
\addlegendentry{DEC(N)}
\addplot [only marks, color1, mark=pentagon*, mark size=4, mark options={solid,draw=white}]
table {%
2 90.6440864583333
3 282.036779583333
4 490.312206458333
5 698.267130416667
6 888.528719375
7 1100.300895
8 1189.153985625
9 1303.47381916667
10 1446.00160270833
};
\addlegendentry{DEC}
\addplot [only marks, color2, mark=otimes*, mark size=4, mark options={solid,draw=white}]
table {%
2 82.1270325
3 230.173131875
4 396.413104166667
5 593.36435
6 764.970520833333
7 928.262400208333
8 1043.90776708333
9 1154.443388125
10 1293.42485041667
};
\addlegendentry{DEC-O}
\addplot [only marks, color3, dash pattern=on 4pt off 1pt on 4pt off 1pt on 1pt off 1pt, mark=star, mark size=4, mark options={solid}]
table {%
	2 77.634184375
	3 188.609589166667
	4 357.165916458333
	5 549.549111458333
	6 718.101064583333
	7 917.036355625
	8 987.632244583333
	9 1093.77881833333
	10 1246.919229375
};
\addlegendentry{\gls{9}}
\addplot [only marks, color4, dash pattern=on 1pt off 1pt, mark=square*, mark size=4, mark options={solid,draw=white}]
table {%
	2 87.5876572916667
	3 196.666367916667
	4 335.2926375
	5 495.629834583333
	6 652.146175208333
	7 781.609086666667
	8 889.808953333334
	9 985.183819375
	10 1108.766715
};
\addlegendentry{DEC-I}
\addplot [only marks, color5, dash pattern=on 5pt off 1pt on 1pt off 1pt, mark=diamond*, mark size=4, mark options={solid,draw=white}]
table {%
2 80.6598316666666
3 197.412985416667
4 337.313776875
5 498.378998125
6 654.563127083333
7 781.451397083333
8 890.19100875
9 982.890975625
10 1107.21106875
};
\addlegendentry{DEC-IO}
\addplot [only marks, color6, dash pattern=on 5pt off 1pt on 2pt off 1pt on 2pt off 1pt, mark=triangle*, mark size=4, mark options={solid,draw=white}]
table {%
2 78.6216164583333
3 175.283775833333
4 322.184230625
5 474.18149625
6 628.157073333333
7 781.18007375
8 859.196300208333
9 966.208985416667
10 1092.14277375
};
\addlegendentry{CEN}
\end{axis}
\end{tikzpicture}
	\caption{Average realized system cost $\hat{\alpha}$ for all information-sharing settings and both selfish settings, aggregated over all instances, per number of drivers}\label{fig:boxplot_all}
\end{figure}

\begin{figure}[tbp]
		\centering
		\begin{tabular}{c @{\qquad} c @{\qquad}}
				\scalebox{0.115}{\includegraphics{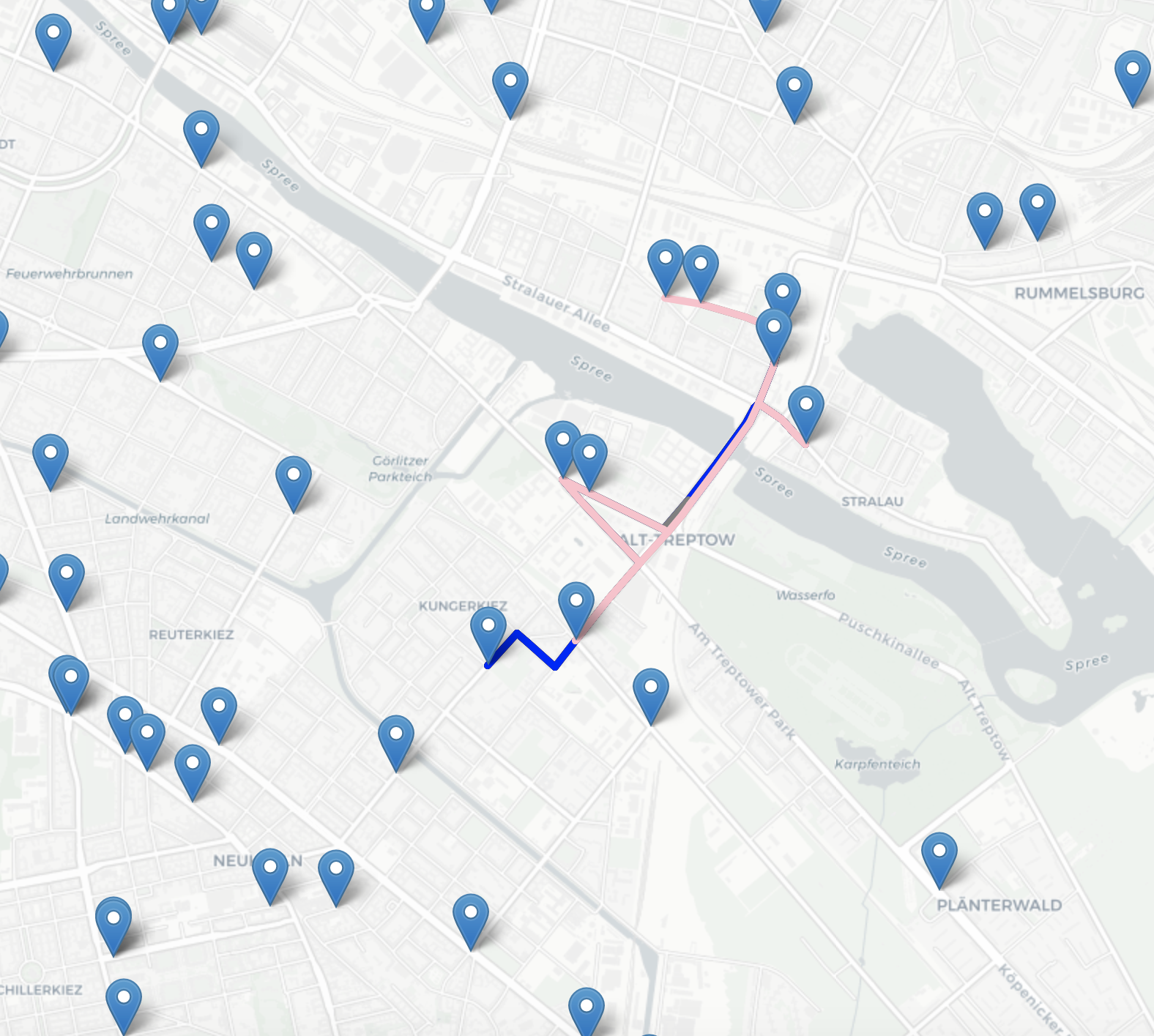}} &
				\hspace{-0.35cm} \scalebox{0.115}{\includegraphics{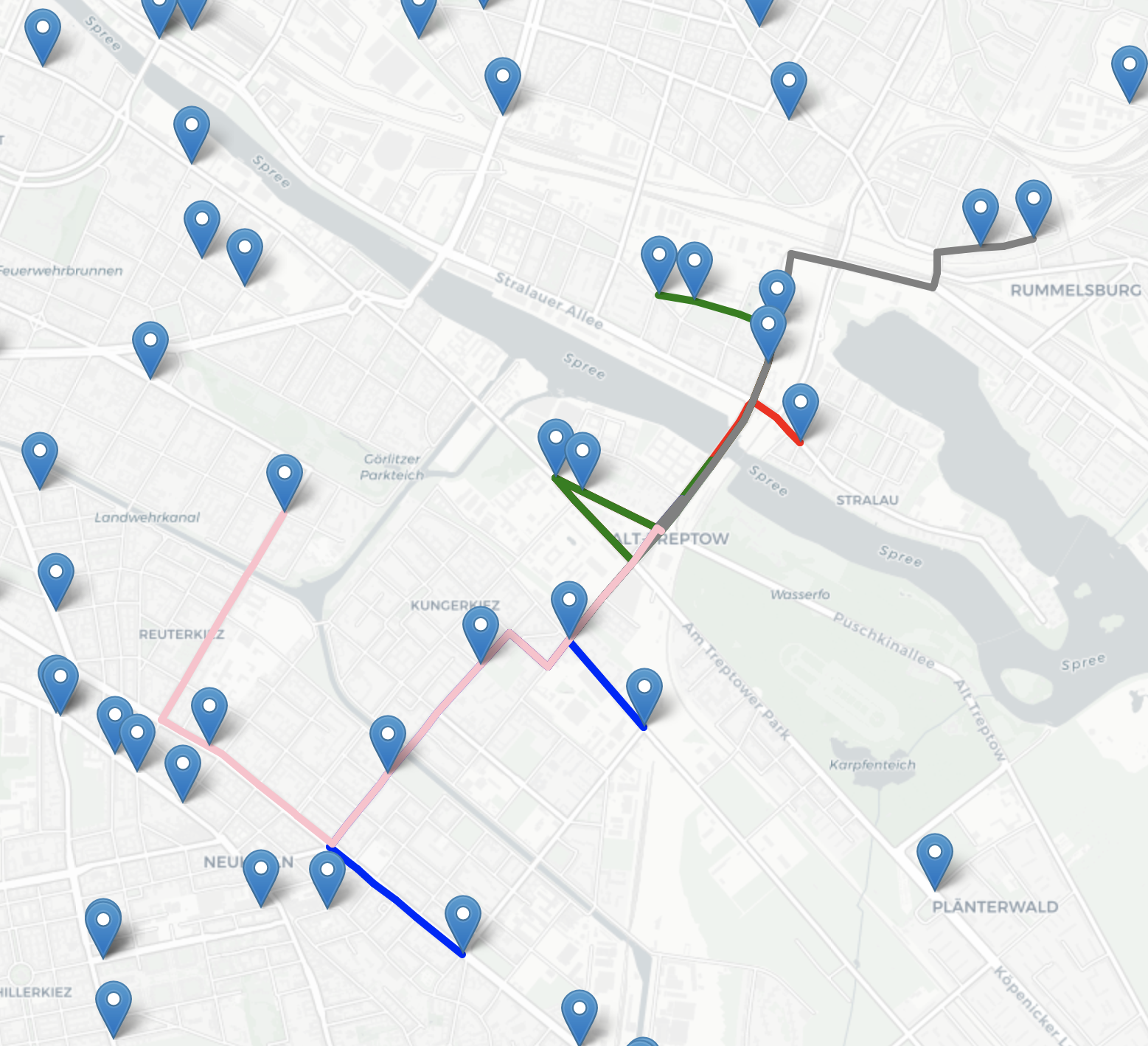}} \\
				\small (a) DEC & \small (b) DEC-I \\
		\end{tabular}
	\caption{Visual representation of uncoordinated (DEC) and coordinated (DEC-I) search routes}\label{fig:folium}
\end{figure}

Figure~\ref{fig:boxplot_all} provides a detailed comparison of all settings with respect to the system cost~$\hat{\alpha}$ averaged over all test instances. 
As can bee seen, an advanced search strategy without coordination (\gls{1}) already decreases the system cost obtained with a naive search strategy~(\gls{1}(N)) by 15\% on average. 
With coordination, the DEC-O setting decreases the cost compared to \gls{1}(N) by 26\% in a static setting and by 30\% in a dynamic setting (\gls{9}), while DEC-I and DEC-IO lead to a 37\% decrease, and CEN yields a 39\% cost decrease.
Comparing uncoordinated and coordinated settings, we note that DEC-O-d decreases the cost obtained in \gls{1} by 18\% while both DEC-I and DEC-IO decrease it by 26\%, and CEN leads to a 28\% decrease.
The static observation-sharing setting DEC-O performs on average worse than both static intention-sharing settings due to the algorithm's sensitivity to the departure time horizon $t^s$ (see Section~\ref{subsec:dynamic_impact}) and yields only a 13\% cost decrease.
Both DEC-I and DEC-IO show performances very close to CEN, which indicates that a decentralized and static setting performs nearly as well as a dynamic and centralized setting, as long as drivers share intentions. 
%
Figure~\ref{fig:folium} visualizes two static search paths computed in the DEC and DEC-I settings, for $N=5$ drivers, $\bar{S}=2000$~m, \mbox{$r^s=300$~m} and $t^s=0$~min: the benefit of coordination translates into routes with fewer overlapping stations.

\begin{res}
	Compared to an uncoordinated setting (DEC), decentralized static intention-sharing (DEC-I, DEC-IO) reduces $\hat{\alpha}$ on average by 26\%. 
	A centralized dynamic setting (CEN) leads to a slightly higher cost reduction of 28\%.
\end{res} 

\begin{figure}[tbp]
	{
		\begin{tabular}{x{0.527\textwidth}x{.527\textwidth}}
			\input{chapters/figures/st_0_1_25_1000.tex} &
			\hspace{-0.5cm} \input{chapters/figures/st_5_15_25_1000.tex}\\
			\small (a) $t^s\in \{0,1\}$~min \& $\bar{S}=1000$ meters & \small (b) $t^s \in \{5,15\} $~min \& $\bar{S}=1000$ meters \\
			&\\
			\input{chapters/figures/st_0_1_25_2000.tex} &
			\hspace{-0.5cm} \input{chapters/figures/st_5_15_25_2000.tex}\\
			\small (c) $t^s\in \{0,1\}$~min \& $\bar{S}=2000$ meters & \small (d) $t^s \in \{5,15\} $~min \& $\bar{S}=2000$ meters \\
		\end{tabular}
	}
	\caption{Comparison of decentralized and centralized decision-making in low-availability scenarios}\label{fig:dec_comp_low}
	\fnote{\footnotesize Each subplot shows the distribution of the realized system cost $\hat{\alpha}$  for each considered setting, per number of drivers $N$ and search radius $\bar{S}$.}
\end{figure}

Figure~\ref{fig:dec_comp_low} compares \gls{1}, \gls{2}, \gls{3}, \gls{4}, \gls{9} and \gls{5} with respect to cost $\hat{\alpha}$ in low-availability scenarios.  
We divide plots between short departure time horizons ($t^s\in\{0,1\} $~min) and large departure time horizons ($t^s\in\{5,15\} $~min), as well as between small search areas ($\bar{S}=1000$ m) and large search areas ($\bar{S}=2000$ m).

As can be seen, a larger search area ($\bar{S}=2000$~m) decreases the system cost~$\hat{\alpha}$ for coordinated settings by at least 30\% in \gls{low_avail} and by at least 90\% in \gls{high_avail} scenarios. In contrast, $\hat{\alpha}$ remains constant, or in some cases increases without coordination.

For short departure time horizons (see Figures~\ref{fig:dec_comp_low}a\&\ref{fig:dec_comp_low}c), we observe that pure intention-sharing (DEC-I) slightly outperforms combined observation- and intention-sharing (DEC-IO) for static policies, mostly when $\bar{S}=2000$~m.
%
In this case, \gls{4} tends to provide lower individual cost solutions for the early departing drivers than \gls{2} due to the additional observation information.
Here, improving the solutions of early drivers worsens the solution quality of subsequent drivers due to the limited number of possible station visits.
In contrast, \gls{2} achieves better performances from a system perspective by negatively affecting the early drivers' search, such that subsequent drivers can obtain higher-quality solutions.
In observation-sharing settings, results confirm that dynamic policies (DEC-O-d) outperform static policies (DEC-O).

For large departure time horizons (see Figures~\ref{fig:dec_comp_low}b\&\ref{fig:dec_comp_low}d), \gls{4} improves over \gls{2} and even outperforms \gls{5} for smaller search areas. Notably, \gls{3} performs very similar to the other coordinated settings in this case.
As the departure time horizon increases, the likelihood that searches temporally overlap decreases, which in turn decreases the benefits of intention-sharing in addition to observation-sharing. We further observe that \gls{3} slightly outperforms \gls{9} in this case for small search areas ($\bar{S}=1000$ m).
We note that in this case, DEC-O is well suited for a practical implementation as it has lower computational requirements than \gls{4} or \gls{5}.

Figure~\ref{fig:dec_comp_high} in Appendix~\ref{app:num_res} contains similar analyses for high-availability scenarios.
While these results show similar trends in general, we note that the benefit of dynamic observation-sharing in a decentralized setting (\gls{9}) is less consistent in this scenario for smaller departure time horizons.

\begin{res}
	Sharing occupancy observations in addition to intentions is not beneficial for almost simultaneous searches, i.e., $t^s \in \{0,1\}$~min. In this case, observation-sharing improves the early departing drivers’ solution to the detriment of succeeding drivers, which worsens the total system performance.
\end{res}
\begin{res}
	For larger departure time horizons, i.e., $t^s \in \{5,15\}$~min, the decentralized observation-sharing setting (DEC-O) performs similar as the decentralized information-sharing settings (DEC-I, DEC-IO) and as the centralized setting (CEN).
\end{res}

\subsection{Single-driver perspective}\label{subsec:user_eval}

In the following, we evaluate how coordination impacts an individual driver's solution, before we analyze the impact of driver heterogeneity.

\subsubsection{Drivers' benefits of coordination:}

We refer to the uncoordinated solutions (DEC) as selfish solutions, in which a driver obtains her solution independently, without additional information and in her own best interest.
First, we analyze the worst and best realized solutions obtained among all drivers for each test instance. For all decentralized settings (\gls{2}, \gls{3}, \gls{4}, \gls{9}) as well as for the centralized dynamic setting~(\gls{5}), we compare the results to the uncoordinated setting~(DEC). We then analyze the impact of a driver's departure position on her individual solution, before we analyze each driver's success rate and search time deviation between the selfish and any coordinated solution.

\begin{figure}[tbp]
		\centering
		\begin{tabular}{c}
			\input{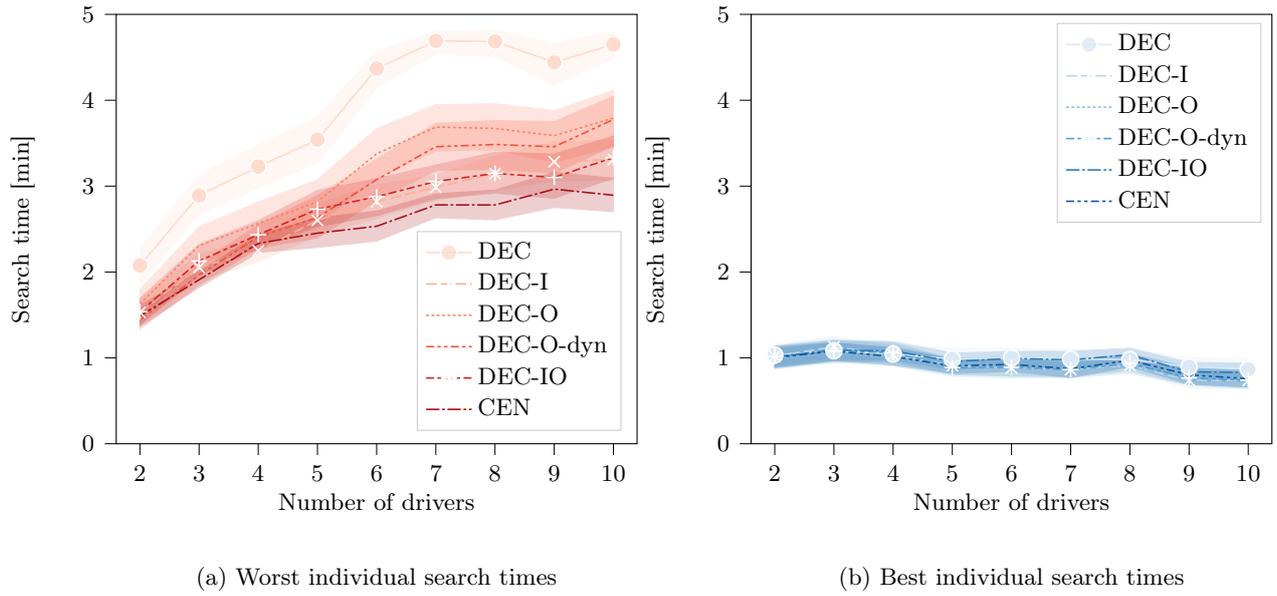} \\		
		\end{tabular}
	\caption{Distribution of the worst and best search times in the low-availability scenario \label{fig:Anarchy_time}}
	\fnote{\footnotesize In Subfigure (a), we show the mean value of all highest search times and a corridor corresponding to a 95\% confidence interval, over all test instances of the low-availability scenario. In Subfigure (b), we analogously show the mean values of all lowest search times.}
\end{figure}

\begin{figure}[tbp]
		\centering
		\begin{tabular}{c}
			\input{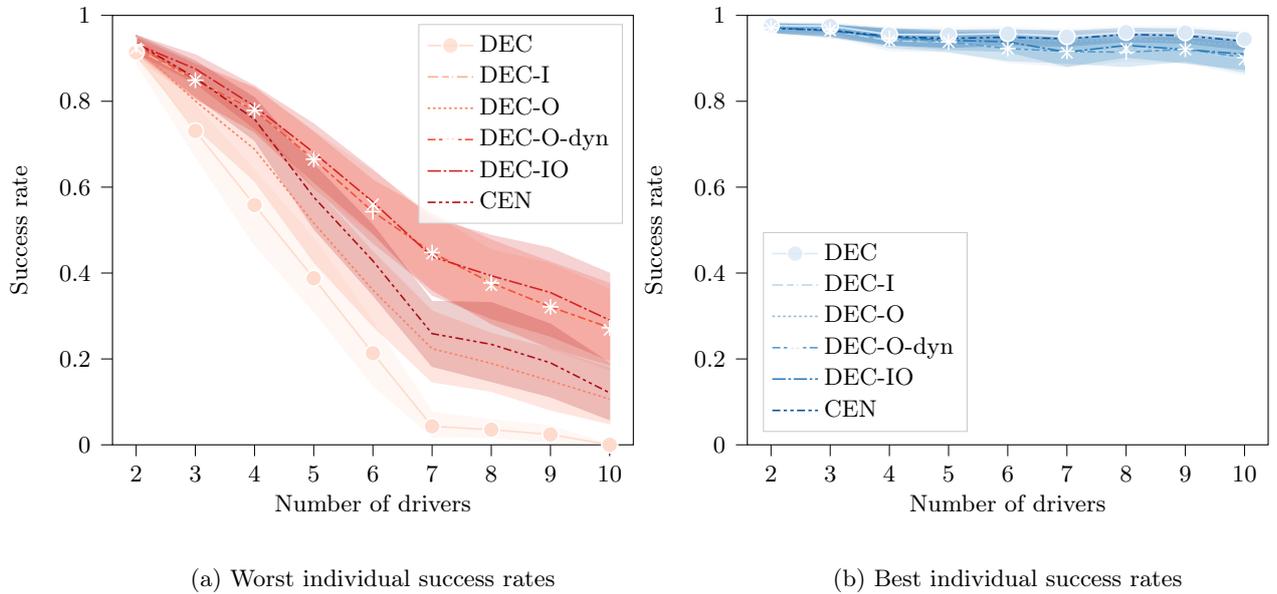} \\
		\end{tabular}
	\caption{Distribution of the worst and best success rate in the low-availability scenario \label{fig:Anarchy_success}}
	\fnote{\footnotesize In Subfigure (a), we show the mean value of all lowest success rates and a corridor corresponding to a 95\% confidence interval, over all test instances of the low-availability scenario. In Subfigure (b), we analogously show the mean values of all highest success rates.}
\end{figure}

Figure~\ref{fig:Anarchy_time} shows the distribution of the lowest and highest realized individual search times over all test instances corresponding to a low-availability scenario for each analyzed setting.
Figure~\ref{fig:Anarchy_success} analogously shows the distribution of the lowest and highest realized individual success rates.
As can be seen, the lowest search times are comparable with and without coordination, such that a selfish solution does not improve the best case scenario with respect to individual search times.
However, all coordinated settings decrease the maximal individual search times, in particular by 30 \% in DEC-I and DEC-IO, and by 35\% in CEN.
Analyzing performances with respect to individual success rates (see Figure~\ref{fig:Anarchy_success}), there exists a minor trade-off between the best and the lowest individual success rates. While the highest individual success rate appears to be slightly higher in a selfish environment, the lowest success rate increases significantly in all coordinated settings, by up to 30 \% (0.02 to 0.32).
We note that \gls{5} yields the most robust solutions, by providing lowest worst individual search times and highest worst individual success rates.

\begin{res}
	Coordinated searches outperform selfish searches, because they significantly improve the worst-possible solution that a driver may obtain while preserving her best-possible solution.
\end{res}

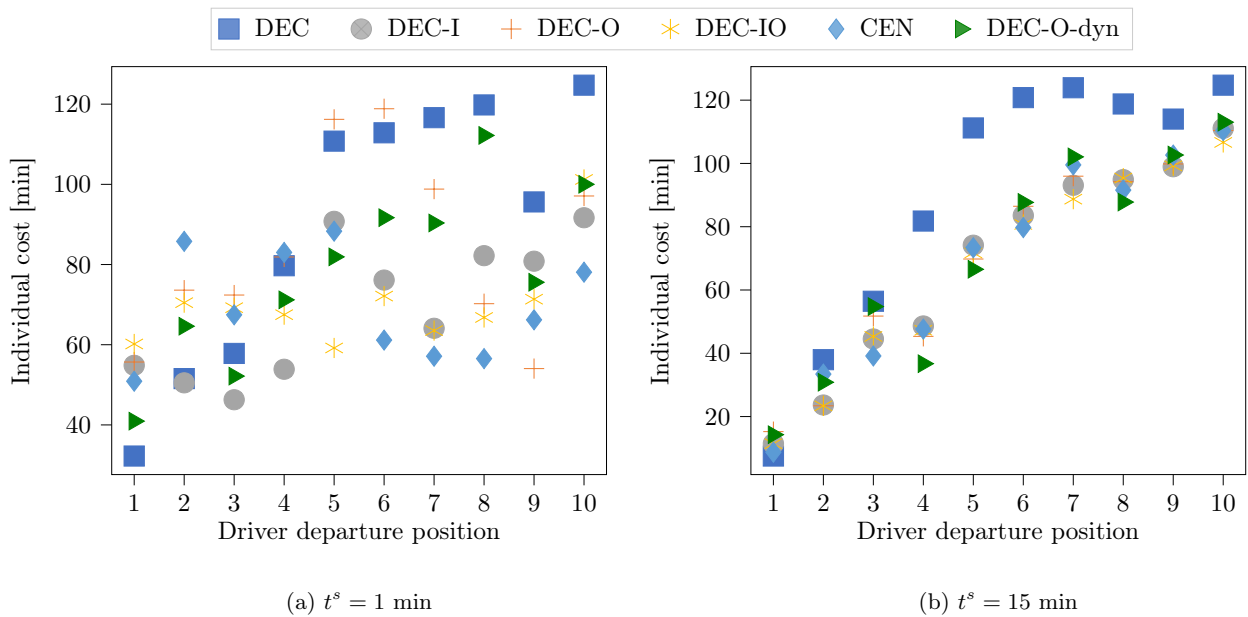
\begin{figure}[bp]	
	\scalebox{0.95}{
\begin{tikzpicture}[font=\small]

\definecolor{color0}{rgb}{0.266666666666667,0.447058823529412,0.768627450980392}
\definecolor{color1}{rgb}{0.929411764705882,0.490196078431373,0.192156862745098}
\definecolor{color2}{rgb}{1,0.752941176470588,0}
\definecolor{color3}{rgb}{0.356862745098039,0.607843137254902,0.835294117647059}

\begin{groupplot}[group style={group size=2 by 1, horizontal sep=2cm}]
\nextgroupplot[
legend cell align={left},
legend style={fill opacity=0.8, draw opacity=1, text opacity=1, at={(0.2,1.04)}, anchor=south west, draw=white!80!black, legend columns=6, /tikz/every even column/.append style={column sep=0.5cm}, /tikz/every odd column/.append style={column sep=0.15cm}},
tick align=outside,
tick pos=left,
title style={at={(0.5,-0.3)},anchor=north,yshift=-0.1},
title={\footnotesize (a) $t^s=1$ min},
x grid style={white!69.0196078431373!black},
xlabel={Driver departure position},
xmin=0.55, xmax=10.45,
xtick distance=1,
xtick style={color=black},
y grid style={white!69.0196078431373!black},
ylabel={Individual cost [min]},
ymin=27.6152326224, ymax=129.358793932267,
ytick style={color=black}
]
\addplot [only marks, color0, mark=square*, mark size=4, mark options={solid}]
table {%
1 32.2399399546667
2 51.5531833426667
3 57.8291044333333
4 79.6980933173333
5 110.776353200667
6 112.853096476
7 116.628966514667
8 119.8031792
9 95.6126770786667
10 124.7340866
};
\addlegendentry{DEC}
\addplot [only marks, white!64.7058823529412!black, mark=otimes*, mark size=4, mark options={solid}]
table {%
1 54.8604269006667
2 50.495853854
3 46.2802222186667
4 53.88381408
5 90.774047866
6 76.1257574173333
7 64.1013579226667
8 82.2126455953333
9 80.8321645866667
10 91.670436388
};
\addlegendentry{DEC-I}
\addplot [only marks, color1, mark=+, mark size=4, mark options={solid}]
table {%
1 55.6872297366667
2 73.6139829866667
3 72.3925726093333
4 81.8870645753333
5 116.211233442667
6 118.853168464
7 98.814417004
8 70.2336760753333
9 54.0474358293333
10 97.0992278006667
};
\addlegendentry{DEC-O}
\addplot [only marks, color2, mark=asterisk, mark size=4, mark options={solid}]
table {%
1 60.1883759946667
2 70.4950568013333
3 69.24208798
4 67.4987499633333
5 59.191823638
6 72.1321982173333
7 63.5321937713333
8 66.797648106
9 71.3577778273333
10 101.219491662667
};
\addlegendentry{DEC-IO}
\addplot [only marks, color3, mark=diamond*, mark size=4, mark options={solid}]
table {%
1 50.8950254093333
2 85.7793865233333
3 67.460542244
4 83.0138043946667
5 88.3163576046667
6 61.15861897
7 57.1263205293333
8 56.535538048
9 66.210491088
10 78.0898851
};
\addlegendentry{CEN}
\addplot [only marks, green!50!black, mark=triangle*, mark size=4, mark options={solid,rotate=270}]
table {%
1 40.9377214286667
2 64.6120603426667
3 52.1667493253333
4 71.1947914066667
5 81.91214134
6 91.687959056
7 90.3643944366667
8 112.199356062
9 75.5737992033333
10 100.004032853333
};
\addlegendentry{\gls{9}}

\nextgroupplot[
legend cell align={left},
tick align=outside,
tick pos=left,
x grid style={white!69.0196078431373!black},
xlabel={Driver departure position},
title style={at={(0.5,-0.3)},anchor=north,yshift=-0.1},
title={\footnotesize (b) $t^s=15$ min},
xmin=0.55, xmax=10.45,
xtick distance=1,
xtick style={color=black},
y grid style={white!69.0196078431373!black},
ylabel={Individual cost [min]},
ymin=1.6637155008, ymax=130.594580461867,
ytick style={color=black}
]
\addplot [only marks, color0, mark=square*, mark size=4, mark options={solid}]
table {%
	1 7.52420936266667
	2 37.9759792386667
	3 56.4429056933333
	4 81.7918364353333
	5 111.214831770667
	6 120.798677876
	7 123.919160837333
	8 118.790593746
	9 114.020272224
	10 124.7340866
};
\addplot [only marks, white!64.7058823529412!black, mark=otimes*, mark size=4, mark options={solid}]
table {%
	1 11.4343926146667
	2 23.653329566
	3 44.5231125173333
	4 48.633168694
	5 74.157579348
	6 83.5387066433333
	7 93.0905641313333
	8 94.8882583013333
	9 98.9491824333333
	10 111.092799445333
};
\addplot [only marks, color1, mark=+, mark size=4, mark options={solid}]
table {%
	1 15.2412838946667
	2 23.396531238
	3 51.7582027313333
	4 45.3483626966667
	5 69.792020996
	6 86.4610184646667
	7 95.9684020526667
	8 94.23001406
	9 99.8320420773333
	10 110.414664888667
};
\addplot [only marks, color2, mark=asterisk, mark size=4, mark options={solid}]
table {%
	1 11.4343926146667
	2 23.3843414466667
	3 45.319203772
	4 47.2707610146667
	5 71.718254406
	6 80.7161610673333
	7 88.7060758926667
	8 95.3638750373333
	9 99.230333098
	10 106.635439048
};
\addplot [only marks, color3, mark=diamond*, mark size=4, mark options={solid}]
table {%
	1 8.73833839866667
	2 33.4225527886667
	3 39.1452722206667
	4 47.6383419326667
	5 73.353922954
	6 79.720637996
	7 99.528554276
	8 91.5774230353333
	9 102.633900231333
	10 110.386826764667
};
\addplot [only marks, green!50!black, mark=triangle*, mark size=4, mark options={solid,rotate=270}]
table {%
	1 14.2689944846667
	2 30.8418973553333
	3 54.7629862006667
	4 36.7092435893333
	5 66.5418761406667
	6 87.6860440473333
	7 102.073053145333
	8 87.783208426
	9 102.655101653333
	10 113.029878988
};
\end{groupplot}

\end{tikzpicture}}
	\caption{Single-agent cost ordered by departure times in a low-availability scenario (\gls{low_avail}) with larger search radius ($\bar{S}=2000 $ m, $N=10$ drivers)}\label{fig:indiv_cost}
\end{figure}

Figure~\ref{fig:indiv_cost} shows the individual costs $\hat{\alpha}^i$ obtained for all drivers $i$ depending on their departure position, averaged over all values of $r^s \in \{100,300,700\}$ for test instances with ten drivers. Figure~\ref{fig:indiv_cost}a shows results for a short departure time horizon ($t^s=1$~min), while Figure~\ref{fig:indiv_cost}b details results for a larger departure time horizon ($t^s=15$~min).
As can be seen, coordination reduces a driver's search cost (nearly) independent of her departure position.
In some specific cases (e.g., for the first driver with $t^s=15$~min), a driver may obtain a higher individual cost with coordination than without. However, these larger costs occur at the benefit of a smaller spread between the worst and the best solution that any driver may obtain.
Moreover, our results show that individual solutions are more homogeneous for shorter departure time horizons, as searches take place almost simultaneously. With a larger departure time horizon, early drivers are privileged against succeeding drivers as they have more chances to find a free station before parallel competing searches start.

\begin{res}
	At the exception of early departing drivers, the individual solution obtained by a driver with coordination outperforms the one obtained without coordination, independent of the driver's departure position. 
\end{res}

Table~\ref{tab:single_agent} shows the relative individual search time deviation ($\hat{\Delta} t_{rel}$) and the absolute individual success rate deviation ($\hat{\Delta} \rho$) per number of drivers $N$ and search radius $\bar{S}$ for all instances, aggregated over all, $r^s$, $t^s$, and availability values.

\begin{table}[tp]
	\centering
	\caption{Individual search time and reliability improvements with coordination\label{tab:single_agent}}%
	{
		\centering
		\scalebox{0.78}{\begin{tabular}{ccccccccccccc}
			\toprule
			&       & \multicolumn{5}{c}{$\hat{\Delta} t_{rel}$} &       & \multicolumn{5}{c}{$\hat{\Delta} \rho$} \\
			\cmidrule{3-7}\cmidrule{9-13}          & \multicolumn{1}{l}{$N$} & \multicolumn{1}{l}{DEC-I} & \multicolumn{1}{l}{DEC-O} & \multicolumn{1}{l}{DEC-O-d} & \multicolumn{1}{l}{DEC-IO} & \multicolumn{1}{l}{CEN} &       & \multicolumn{1}{l}{DEC-I} & \multicolumn{1}{l}{DEC-O} & \multicolumn{1}{l}{DEC-O-d} & \multicolumn{1}{l}{DEC-IO} & \multicolumn{1}{l}{CEN} \\
			\midrule
			\multicolumn{1}{c}{\multirow{9}[2]{*}{$\bar{S}=1000$ m}} & 2     & 2.21  & 2.06  & 2.20   & 2.12  & 1.71  &       & 0.00     & 0.00     & 0.00     & 0.00     & 0.00 \\
			& 3     & 1.40  & 2.39  & -2.33 & 2.21  & -2.33 &       & 0.01  & 0.01  & 0.01  & 0.01  & 0.01 \\
			& 4     & 3.72  & 4.52  & 1.21  & 2.98  & 0.08  &       & 0.03  & 0.02  & 0.02  & 0.03  & 0.03 \\
			& 5     & 5.86  & 7.27  & 4.52  & 4.49  & 2.23  &       & 0.05  & 0.03  & 0.04  & 0.05  & 0.05 \\
			& 6     & 16.8 & 10.5 & 9.13  & 15.5 & 12.0 &       & 0.07  & 0.04  & 0.06  & 0.07  & 0.08 \\
			& 7     & 18.0 & 13.6 & 7.26  & 15.7 & 9.88  &       & 0.09  & 0.05  & 0.06  & 0.09  & 0.08 \\
			& 8     & 21.3 & 14.5 & 9.85  & 21.3 & 15.4 &       & 0.08  & 0.04  & 0.05  & 0.08  & 0.08 \\
			& 9     & 23.4 & 18.0 & 11.8 & 23.4 & 14.9 &       & 0.07  & 0.04  & 0.06  & 0.07  & 0.08 \\
			& 10    & 21.1 & 17.7 & 7.32  & 21.4 & 20.5 &       & 0.09  & 0.05  & 0.08  & 0.09  & 0.09 \\
			\midrule
			\multicolumn{1}{c}{\multirow{9}[2]{*}{$\bar{S}=2000$ m}} & 2     & 1.99  & 2.11  & 2.39  & 2.11  & 2.50   &       & 0.00     & 0.00     & 0.00     & 0.00     & 0.00 \\
			& 3     & 3.79  & 3.02  & -1.83 & 4.00   & -1.73 &       & 0.01  & 0.01  & 0.02  & 0.02  & 0.02 \\
			& 4     & 3.50   & 4.07  & 0.35  & 2.82  & -0.34 &       & 0.03  & 0.02  & 0.03  & 0.03  & 0.04 \\
			& 5     & 1.97  & 4.87  & -3.50  & -0.40  & -3.37 &       & 0.06  & 0.03  & 0.05  & 0.06  & 0.07 \\
			& 6     & 7.68  & 7.92  & 0.15  & 6.77  & 1.01  &       & 0.10   & 0.05  & 0.08  & 0.10   & 0.11 \\
			& 7     & 14.9 & 11.0 & 4.17  & 13.3 & 8.70   &       & 0.16  & 0.09  & 0.10   & 0.16  & 0.17 \\
			& 8     & 1.36  & 4.61  & -9.01 & 0.03  & -9.16 &       & 0.19  & 0.09  & 0.13  & 0.19  & 0.21 \\
			& 9     & 2.28  & 6.54  & -6.56 & 0.51  & -13.3 &       & 0.23  & 0.12  & 0.16  & 0.23  & 0.25 \\
			& 10    & 9.38  & 7.43  & -2.38 & 7.84  & -3.01 &       & 0.29  & 0.13  & 0.18  & 0.29  & 0.30 \\
			\bottomrule
		\end{tabular}}%
	}%
	\fnote{
		\footnotesize The table compares the average search time deviation $\Delta t_{\text{rel}}$ and success rate deviation $\Delta \rho$. We denote with $\hat{\Delta} t_{\text{rel}}$~[\%] and $\hat{\Delta} \rho$ their average over all test instances corresponding to $r^s \in \{100,300,700\}$, $\bar{S} \in \{1000,2000\}$ and both \gls{low_avail} and \gls{high_avail}. Values are computed as follows: $\Delta t_{\text{rel}} =  -\nicefrac{1}{n} (\sum_{i=0}^{n} \nicefrac{(\hat{t}^i_{\text{setting}} -   \hat{t}^i_{\text{DEC}})}{\hat{t}^i_{\text{DEC}}} )$  and $\Delta \rho = - \nicefrac{1}{n} ( \sum_{i=0}^{n}  \hat{\rho}^i_{\text{DEC}} -   \hat{\rho}^i_{\text{setting}})$, with $n$ being the number of drivers considered in the repsective instance. Values are positive when the evaluated setting outperforms \gls{1} . 
	}
\end{table}%

As can be seen, all coordinated settings allow a driver to reduce her search time and increase her search reliability on average. Specifically, a driver may save 2\% (DEC-O-d), 3\% (CEN), 8\% (DEC-O, DEC-IO) and 9\% (DEC-I) of her search time, while she can increase her success rate by 0.05 (DEC-O), 0.06 (DEC-O-d) and 0.09 (DEC-I, DEC-IO, CEN).
In line with the system-perspective evaluation, DEC-I, DEC-IO, and CEN yield the best performances from a driver-perspective too, such that we further detail results only for these settings.

There exists a trade-off between search time savings and the reliability improvement. For small search areas ($\bar{S}$=1000 m), drivers save 8\% (CEN) or 12\% (DEC-I, DEC-IO) of their search times on average, and up to 23\% (DEC-I, DEC-IO) with a high number of drivers ($N=9$). The success rate increase is limited to 0.05 on average (DEC-I, DEC-IO, CEN).
For larger search areas ($\bar{S}=2000$ m), the time gain decreases. Drivers may only save 5\% of their search times on average (DEC-I, DEC-IO) or even slightly increase their search time by 2\% (CEN). In this case, however, the reliability gain significantly increases, with drivers increasing their success rates on average by 0.12 (DEC-I, DEC-IO) or 0.13 (CEN), and up to 0.30 (CEN with $N=10$).

\subsubsection{Impact of driver heterogeneity:}\label{subsec:heterogeneity}

\begin{figure}[bp]
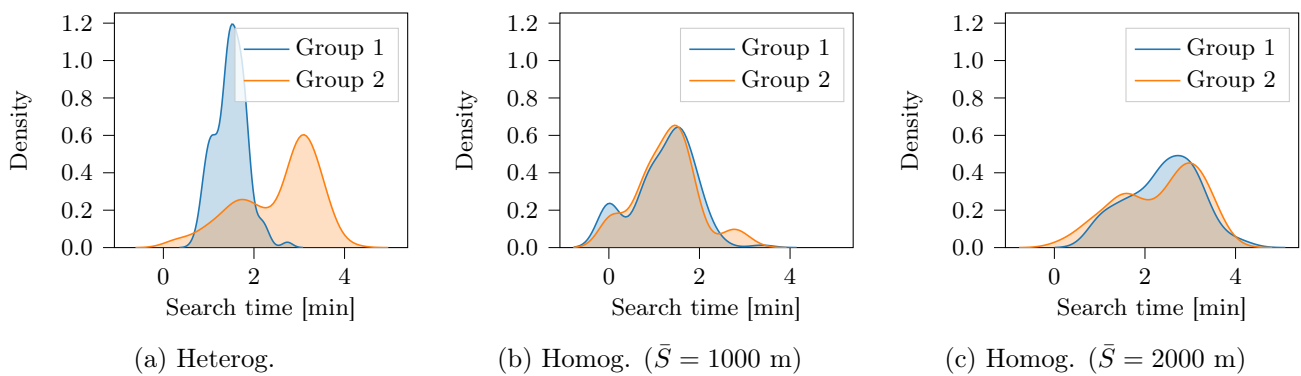

	\centering
	\hskip -2em
		\begin{tabular}{x{0.32\textwidth}x{.32\textwidth}x{.32\textwidth}}
			\input{./chapters/figures/hetero_rad_1.tex}	&
			\input{./chapters/figures/hetero_rad_2.tex}	 &
			\input{./chapters/figures/hetero_rad_3.tex}	 \\
			\small (a) Heterog. & \small(b) Homog. ($\bar{S}=1000$ m)& \small(c) Homog. ($\bar{S}=2000$ m) 
		\end{tabular}
	\caption{Impact of heterogeneous search radii on drivers' search times}\label{fig:heterog_rad}
	\fnote{\footnotesize All figures show the results obtained in the CEN-setting with $N=10$ drivers for $\bar{T}=5$~min.}
\end{figure}

To analyze the impact of driver heterogeneity, we split drivers into two distinct groups: the first group contains all drivers with an odd departure position, while the second group contains all drivers with an even departure position.
Figure~\ref{fig:heterog_rad} shows the impact of drivers with heterogeneous search radii in the CEN setting by analyzing three cases:
in the first case (\ref{fig:heterog_rad}a), drivers of the first group have a smaller search radius ($\bar{S}=1000$ m), while drivers of the second group have a larger search radius ($\bar{S}=2000$ m).
\begin{figure}[tp]
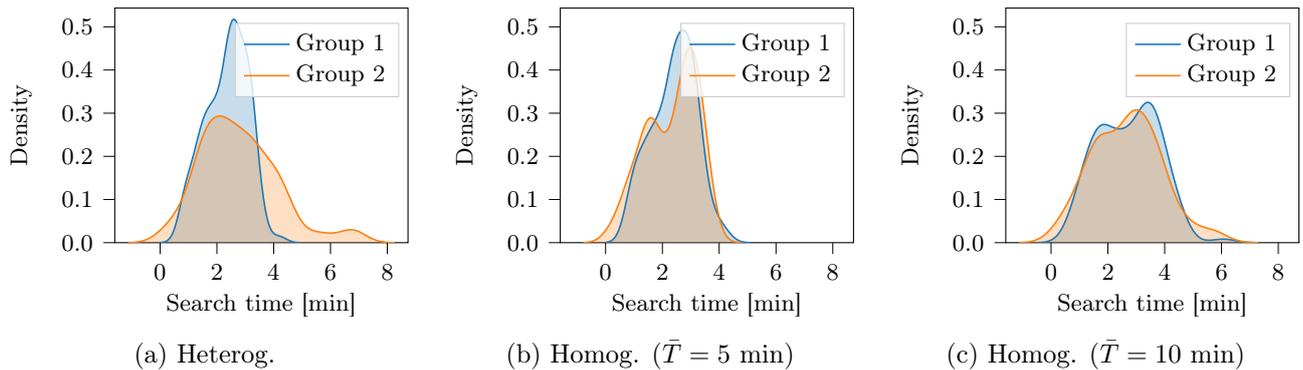

	\centering
		\begin{tabular}{x{0.32\textwidth}x{.32\textwidth}x{.32\textwidth}}
			\input{./chapters/figures/hetero_time_1.tex}	&
			\input{./chapters/figures/hetero_time_2.tex}	 &
			\input{./chapters/figures/hetero_time_3.tex}	 \\
			\small (a) Heterog. & \small(b) Homog. ($\bar{T}=5$ min) & \small(c) Homog. ($\bar{T}=10$ min)
		\end{tabular}
	\caption{Impact of heterogeneous time budgets on drivers' search times}\label{fig:heterog_time}
	\fnote{\footnotesize All figures show the results obtained in the CEN-setting with $N=10$ drivers for $\bar{S}=1000$~m.}
\end{figure}
In the second case (\ref{fig:heterog_rad}b), drivers of both groups have a smaller search radius ($\bar{S}=1000$ m) while in the third case (\ref{fig:heterog_rad}c), drivers of both groups have a larger search radius ($\bar{S}=2000$ m). Accordingly, drivers are heterogeneous in Figure~\ref{fig:heterog_rad}a and homogeneous in Figures~\ref{fig:heterog_rad}b\&\ref{fig:heterog_rad}c. 
Analogously, Figure~\ref{fig:heterog_time} shows the impact of smaller and larger time budgets. In the first case (\ref{fig:heterog_time}a), drivers from the first group have a smaller time budget (\mbox{$\bar{T}=5$~min}) while drivers from the second group have a larger time budget ($\bar{T}=10$~min), in the second case (\ref{fig:heterog_time}b) all drivers have a smaller time budget whereas in the third case (\ref{fig:heterog_time}c), all drivers have a larger time budget.
We average results for test instances with $N=10$ drivers.

As can be seen, the distribution of individual search times in homogeneous settings is (nearly) independent of the respective group. 
This is not the case in heterogeneous settings. Here, drivers that perform a more constrained search, i.e., have a smaller search radius (see Figure~\ref{fig:heterog_rad}a) or a smaller time budget (see Figure \ref{fig:heterog_time}a), obtain lower search times compared to drivers that perform a less constrained search, especially in the first case.
To reduce potential visit overlap, drivers with a larger search radius increase their chances of finding unoccupied stations by visiting far-distanced stations that are less affected by potential overlaps, which then contributes to increase their search times.
Finally, we note that performing searches with homogeneous parameters appears to be preferable from a practitioner's perspective, in order to provide fair and consistent service to all customers.

\begin{res}
	Time budget or search radius heterogeneity favors drivers with lower time budget and lower search radius.
\end{res}

\subsubsection{Coordination in longer planning horizons:}\label{subsubsec:large-horizon_exp}

	In the following, we analyze the impact of coordination in a longer planning horizon, by comparing three charging demand distribution scenarii, see Figure~\ref{fig:scnearios}. In \gls{sc1}, drivers are steadily entering the system, with $5$ requests per $15$ minutes time interval. 
	In \gls{sc2}, each fourth request is skipped, whereas in \gls{sc3}, every fourth and fifth requests are skipped.
	The first scenario reflects a homogeneous demand distribution, while the other scenarii reflect heterogeneous demand distributions, similar to peak demand that arises in practice.

	Table~\ref{tab:longsc} compares the relative cost improvement obtained on average between coordinated (CEN) and uncoordinated decentralized (DEC) planning in both short ($t^s = 15$ minutes) and long ($t^s=180$ minutes) planning horizon settings, depending on the radius of the area in which all agents start their search ($r^s \in \{100,300,700\}$ meters).
	Figure~\ref{fig:indiv_cost_60} further compares the individual drivers cost obtained in the long planning horizon setting with coordination (CEN) and without (DEC) depending on the driver's departure times, the search radius ($\bar{S} \in \{1000,2000\}$ meters) and the distribution scenario (\gls{sc1} or \gls{sc3}).

	\begin{table}[tbp]
		\begin{minipage}[t]{0.45\textwidth}
			\hspace{-0.5cm}
			\captionof{figure}{\scriptsize Requests distribution scenario}\label{fig:scnearios}
			\setlength\tabcolsep{4pt}
			\begin{tabular}
				{x{0.2\textwidth}x{.4\textwidth}}
				\gls{sc1} & 
\begin{tikzpicture}[font=\scriptsize]

\definecolor{color0}{rgb}{0.12156862745098,0.466666666666667,0.705882352941177}

\begin{axis}[
tick align=outside,
height=2cm,
width=6cm,
ytick distance=1,
tick pos=left,
x grid style={white!69.0196078431373!black},
xmin=-0.67, xmax=9.67,
xtick style={color=black},
y grid style={white!69.0196078431373!black},
ymin=0, ymax=1.2,
ytick style={color=black}
]
\draw[draw=none,fill=color0] (axis cs:-0.2,0) rectangle (axis cs:0.2,1);
\draw[draw=none,fill=color0] (axis cs:0.8,0) rectangle (axis cs:1.2,1);
\draw[draw=none,fill=color0] (axis cs:1.8,0) rectangle (axis cs:2.2,1);
\draw[draw=none,fill=color0] (axis cs:2.8,0) rectangle (axis cs:3.2,1);
\draw[draw=none,fill=color0] (axis cs:3.8,0) rectangle (axis cs:4.2,1);
\draw[draw=none,fill=color0] (axis cs:4.8,0) rectangle (axis cs:5.2,1);
\draw[draw=none,fill=color0] (axis cs:5.8,0) rectangle (axis cs:6.2,1);
\draw[draw=none,fill=color0] (axis cs:6.8,0) rectangle (axis cs:7.2,1);
\draw[draw=none,fill=color0] (axis cs:7.8,0) rectangle (axis cs:8.2,1);
\draw[draw=none,fill=color0] (axis cs:8.8,0) rectangle (axis cs:9.2,1);
\end{axis}

\end{tikzpicture} \\
				\gls{sc2} & 
\begin{tikzpicture}[font=\scriptsize]

\definecolor{color0}{rgb}{0.12156862745098,0.466666666666667,0.705882352941177}

\begin{axis}[
tick align=outside,
height=2cm,
width=6cm,
tick pos=left,
ytick distance=1,
x grid style={white!69.0196078431373!black},
xmin=-0.67, xmax=9.67,
xtick style={color=black},
y grid style={white!69.0196078431373!black},
ymin=0, ymax=1.05,
ytick style={color=black}
]
\draw[draw=none,fill=color0] (axis cs:-0.2,0) rectangle (axis cs:0.2,1);
\draw[draw=none,fill=color0] (axis cs:0.8,0) rectangle (axis cs:1.2,1);
\draw[draw=none,fill=color0] (axis cs:1.8,0) rectangle (axis cs:2.2,1);
\draw[draw=none,fill=color0] (axis cs:2.8,0) rectangle (axis cs:3.2,0);
\draw[draw=none,fill=color0] (axis cs:3.8,0) rectangle (axis cs:4.2,1);
\draw[draw=none,fill=color0] (axis cs:4.8,0) rectangle (axis cs:5.2,1);
\draw[draw=none,fill=color0] (axis cs:5.8,0) rectangle (axis cs:6.2,1);
\draw[draw=none,fill=color0] (axis cs:6.8,0) rectangle (axis cs:7.2,0);
\draw[draw=none,fill=color0] (axis cs:7.8,0) rectangle (axis cs:8.2,1);
\draw[draw=none,fill=color0] (axis cs:8.8,0) rectangle (axis cs:9.2,1);
\end{axis}

\end{tikzpicture} \\
				\gls{sc2} & 
\begin{tikzpicture}[font=\scriptsize]

\definecolor{color0}{rgb}{0.12156862745098,0.466666666666667,0.705882352941177}

\begin{axis}[
tick align=outside,
height=2cm,
width=6cm,
tick pos=left,
ytick distance=1,
x grid style={white!69.0196078431373!black},
xmin=-0.67, xmax=9.67,
xtick style={color=black},
y grid style={white!69.0196078431373!black},
ymin=0, ymax=1.05,
ytick style={color=black}
]
\draw[draw=none,fill=color0] (axis cs:-0.2,0) rectangle (axis cs:0.2,1);
\draw[draw=none,fill=color0] (axis cs:0.8,0) rectangle (axis cs:1.2,1);
\draw[draw=none,fill=color0] (axis cs:1.8,0) rectangle (axis cs:2.2,1);
\draw[draw=none,fill=color0] (axis cs:2.8,0) rectangle (axis cs:3.2,0);
\draw[draw=none,fill=color0] (axis cs:3.8,0) rectangle (axis cs:4.2,0);
\draw[draw=none,fill=color0] (axis cs:4.8,0) rectangle (axis cs:5.2,1);
\draw[draw=none,fill=color0] (axis cs:5.8,0) rectangle (axis cs:6.2,1);
\draw[draw=none,fill=color0] (axis cs:6.8,0) rectangle (axis cs:7.2,1);
\draw[draw=none,fill=color0] (axis cs:7.8,0) rectangle (axis cs:8.2,0);
\draw[draw=none,fill=color0] (axis cs:8.8,0) rectangle (axis cs:9.2,0);
\end{axis}

\end{tikzpicture} \\
			\end{tabular}
			\fnote{\footnotesize The figure shows for each scenario the requests distribution pattern, with each (missing) bar corresponding to a (missing) request.}
		\end{minipage}\qquad
		\begin{minipage}[t]{0.45\textwidth}
			\hspace{0.8cm}
			\caption{Relative cost improvement for \gls{5} over \gls{1} \label{tab:longsc}}
			\setlength\tabcolsep{3pt}
			\begin{tabular}{crrrr}
				&&&&\\
				\toprule
				\multicolumn{1}{l}{Horizon} & \multicolumn{1}{l}{$r^s$} & \multicolumn{1}{l}{\gls{sc1}} & \multicolumn{1}{l}{\gls{sc2}} & \multicolumn{1}{l}{\gls{sc3}} \\
				\midrule
				\multirow{3}[2]{*}{$t^s=15$ min} & 100 m   & -38.3\% & -41.3\% & -40.3\% \\
				& 300 m   & -38.3\% & -42.6\% & -49.0\% \\
				& 700 m   & -17.2\% & -11.8\% & -25.7\% \\
				\midrule
				\multirow{3}[2]{*}{$t^s=180$ min} & 100 m   & -21.7\% & -26.5\% & -29.6\% \\
				& 300 m   & -22.6\% & -26.2\% & -29.7\% \\
				& 700 m   & -23.6\% & -25.8\% & -27.7\% \\
				\bottomrule
			\end{tabular}
			\fnote{
				\footnotesize The table shows the relative improvement of the realized system cost $\hat{\alpha}$, computed as follows: $\Delta [\%] = \nicefrac{(\hat{\alpha}_{\text{CEN}}-\hat{\alpha}_{\text{DEC} } )}{\hat{\alpha}_{\text{DEC}}}$ of the CEN setting to the DEC setting in percentages, averaged over all instances corresponding to $\bar{S}\in\{1000,2000\} $ m, and both \gls{low_avail} and \gls{high_avail} scenarii, for each value of $r^s$. 
			} 
		\end{minipage} 
	\end{table}

	As can be seen, coordination significantly improves the overall search performances in longer horizon settings (cf. Table~\ref{tab:longsc}).
	Comparing the results of the short and long planning horizon settings, our results show that cost savings in the long planning horizon setting decrease for a limited departure area ($r^s \in \{100,300\} $ meters), but increase when drivers are initially better distributed over the search area (i.e., $r^s=700$ meters). 
	%
	When accounting for a long planning horizon, the overall station availability in the system decreases slightly due to an increasing number of drivers entering the system and possibly blocking charging stations for a longer period of time. 
	However, additional information related to stations getting freed can be shared, such that the results illustrate the trade-off that exists between the performance loss due to the availability decrease and the performance gain due to the information increase.
	
	Our results further show that a decreased charging demand (i.e., \gls{sc3}) increases the cost reduction obtained with coordination in long planning horizons, with an up to 46\% $\hat{\alpha}$ cost decrease in the \gls{high_avail} scenario.
	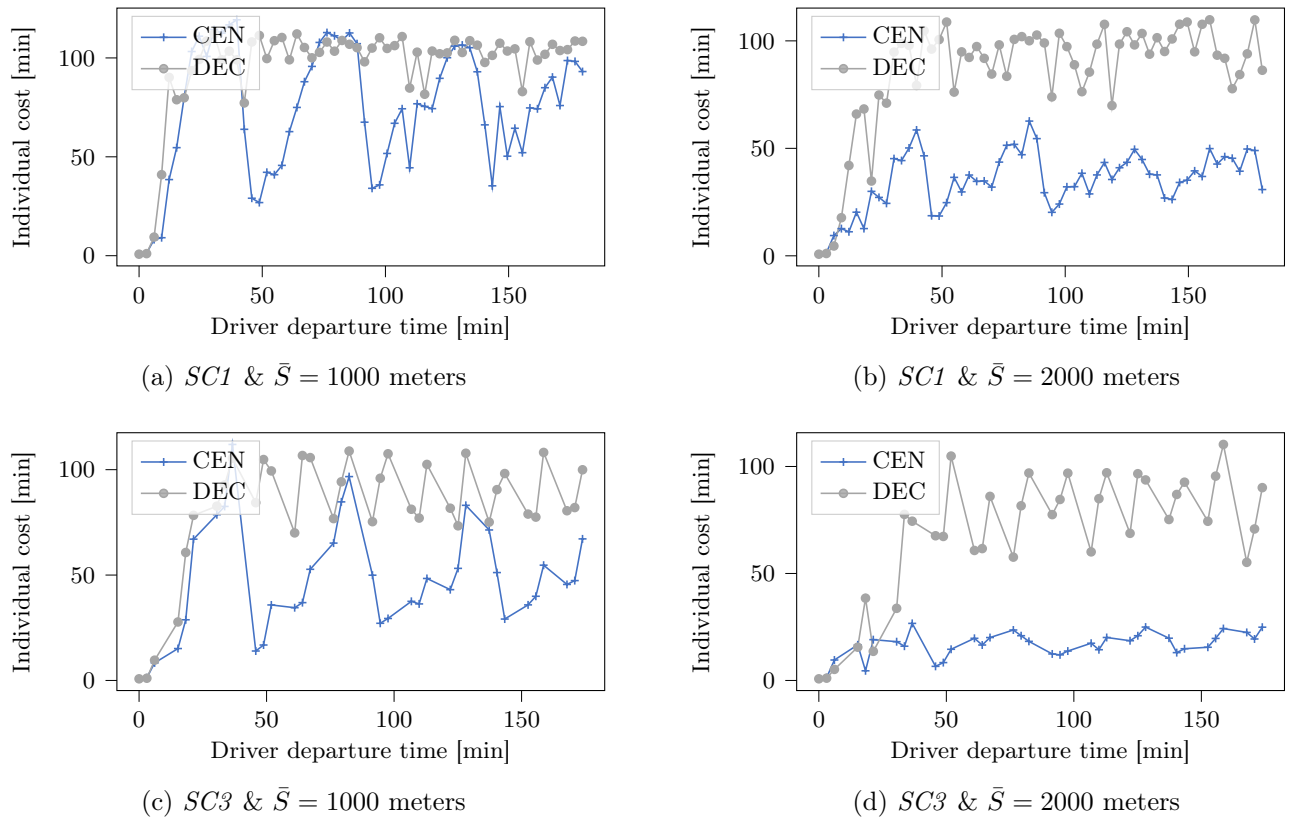
\begin{figure}[tbp]
		\begin{tabular}{x{0.527\textwidth}x{.527\textwidth}}
\begin{tikzpicture}[font=\footnotesize]

\definecolor{color0}{rgb}{0.266666666666667,0.447058823529412,0.768627450980392}

\begin{axis}[
legend cell align={left},
height=5cm,
width=8cm,
legend style={
  fill opacity=0.8,
  draw opacity=1,
  text opacity=1,
  at={(0.03,0.97)},
  anchor=north west,
  draw=white!80!black
},
tick align=outside,
tick pos=left,
x grid style={white!69.0196078431373!black},
xlabel={Driver departure time [min]},
xmin=-9, xmax=189,
xtick style={color=black},
y grid style={white!69.0196078431373!black},
ylabel={Individual cost [min]},
ymin=-5.15847227449334, ymax=125.20577779636,
ytick style={color=black}
]
\addplot [semithick, color0, mark=+, mark size=1.5, mark options={solid}]
table {%
0 0.767175455999994
3.05084745762712 1.270473888
6.10169491525424 8.12813988773333
9.15254237288136 9.0951227836
12.2033898305085 38.4895083396
15.2542372881356 54.6322744584
18.3050847457627 80.1571332469333
21.3559322033898 103.240698862933
24.4067796610169 110.983089521867
27.4576271186441 100.7671610964
30.5084745762712 114.336148767733
33.5593220338983 112.554836112
36.6101694915254 116.701754853333
39.6610169491525 119.280130065867
42.7118644067797 63.9039609050667
45.7627118644068 29.0594814474667
48.8135593220339 26.8604627170667
51.864406779661 42.1734907133333
54.9152542372881 40.9598073888
57.9661016949153 45.6617515554667
61.0169491525424 62.7339652152
64.0677966101695 74.9703589244
67.1186440677966 87.9735833418667
70.1694915254237 95.7112835636
73.2203389830509 107.832044333467
76.271186440678 112.823731095067
79.3220338983051 111.1920266176
82.3728813559322 108.7258937268
85.4237288135593 112.618960939733
88.4745762711864 107.204072544133
91.5254237288135 67.5022483153333
94.5762711864407 34.0694286386667
97.6271186440678 35.8154027816
100.677966101695 51.6956086222667
103.728813559322 66.9564072098667
106.779661016949 74.2940069788
109.830508474576 44.4546088174667
112.881355932203 76.8195015152
115.932203389831 75.492781498
118.983050847458 74.3677655933333
122.033898305085 89.6836130974667
125.084745762712 100.0981036468
128.135593220339 105.97312203
131.186440677966 106.512221064667
134.237288135593 105.168906349333
137.28813559322 92.9997631538667
140.338983050847 66.16333402
143.389830508475 35.3324283688
146.440677966102 75.3809372954667
149.491525423729 50.3191729181333
152.542372881356 64.4284522850667
155.593220338983 52.0930453648
158.64406779661 74.7221037741333
161.694915254237 74.269806316
164.745762711864 84.9420011348
167.796610169492 90.3018073238667
170.847457627119 75.9271621796
173.898305084746 98.6757168696
176.949152542373 98.2440000982667
180 93.1101870516
};
\addlegendentry{CEN}
\addplot [semithick, white!64.7058823529412!black, mark=*, mark size=1.5, mark options={solid}]
table {%
0 0.767175455999994
3.05084745762712 1.05101507959999
6.10169491525424 9.41425079279999
9.15254237288136 40.9692410334667
12.2033898305085 90.1518630621333
15.2542372881356 78.8611265572
18.3050847457627 79.84628892
21.3559322033898 93.5374265382666
24.4067796610169 99.2364124016
27.4576271186441 108.535561652667
30.5084745762712 112.1419220596
33.5593220338983 99.0677191096
36.6101694915254 103.440732749067
39.6610169491525 99.0631542446667
42.7118644067797 77.236561136
45.7627118644068 108.087186730933
48.8135593220339 111.3603936104
51.864406779661 99.6773279986667
54.9152542372881 108.709563829467
57.9661016949153 110.3341576776
61.0169491525424 99.0378472272
64.0677966101695 112.138336404267
67.1186440677966 105.265713550667
70.1694915254237 100.056150173467
73.2203389830509 102.813716104267
76.271186440678 108.049608708133
79.3220338983051 103.406160645867
82.3728813559322 108.836145031067
85.4237288135593 106.90042276
88.4745762711864 105.175493134533
91.5254237288135 98.0265237368
94.5762711864407 104.984804883333
97.6271186440678 110.1801338136
100.677966101695 104.7876056132
103.728813559322 106.226011776533
106.779661016949 110.801489623467
109.830508474576 84.7662040608
112.881355932203 102.9218997416
115.932203389831 81.6335490706667
118.983050847458 103.4800292676
122.033898305085 102.117413023067
125.084745762712 102.548064511067
128.135593220339 108.918977741867
131.186440677966 102.739593065067
134.237288135593 108.7159945444
137.28813559322 106.465853813333
140.338983050847 97.7144484172
143.389830508475 101.2797662016
146.440677966102 107.438866124667
149.491525423729 103.478863439867
152.542372881356 104.590820700267
155.593220338983 82.9741656037333
158.64406779661 108.1993282668
161.694915254237 98.9321726364
164.745762711864 101.820426541867
167.796610169492 106.8860696532
170.847457627119 103.7396212684
173.898305084746 104.193998197733
176.949152542373 108.557362472267
180 108.405724575733
};
\addlegendentry{DEC}
\end{axis}

\end{tikzpicture} &
			\hspace{-1cm} 
\begin{tikzpicture}[font=\footnotesize]

\definecolor{color0}{rgb}{0.266666666666667,0.447058823529412,0.768627450980392}

\begin{axis}[
legend cell align={left},
height=5cm,
width=8cm,
legend style={
  fill opacity=0.8,
  draw opacity=1,
  text opacity=1,
  at={(0.03,0.97)},
  anchor=north west,
  draw=white!80!black
},
tick align=outside,
tick pos=left,
x grid style={white!69.0196078431373!black},
xlabel={Driver departure time [min]},
xmin=-9, xmax=189,
xtick style={color=black},
y grid style={white!69.0196078431373!black},
ylabel={Individual cost [min]},
ymin=-4.69211181027334, ymax=115.25613355334,
ytick style={color=black}
]
\addplot [semithick, color0, mark=+, mark size=1.5, mark options={solid}]
table {%
0 0.760081160799995
3.05084745762712 1.36351732333334
6.10169491525424 9.4351414408
9.15254237288136 12.6611184474667
12.2033898305085 11.1773681933333
15.2542372881356 20.2940814644
18.3050847457627 12.6692102126667
21.3559322033898 29.9894352438667
24.4067796610169 27.1788245348
27.4576271186441 24.3988243932
30.5084745762712 45.2195458536
33.5593220338983 44.3176862170667
36.6101694915254 50.1786356954667
39.6610169491525 58.5479551676
42.7118644067797 46.5188046024
45.7627118644068 18.6530815064
48.8135593220339 18.5190062938667
51.864406779661 24.7429100590667
54.9152542372881 36.4927412708
57.9661016949153 29.7282551445333
61.0169491525424 37.5804010792
64.0677966101695 34.6340133196
67.1186440677966 34.8335651681333
70.1694915254237 31.9700822976
73.2203389830509 43.5971316618667
76.271186440678 51.418653772
79.3220338983051 51.8461754782667
82.3728813559322 47.0452339248
85.4237288135593 62.6754860694667
88.4745762711864 54.5206890572
91.5254237288135 29.3724714498667
94.5762711864407 20.2361453441333
97.6271186440678 24.0815954205333
100.677966101695 32.0984895005333
103.728813559322 32.1687762582667
106.779661016949 38.4048170078667
109.830508474576 28.8016232925333
112.881355932203 37.6303389
115.932203389831 43.4207068397333
118.983050847458 35.5119266092
122.033898305085 40.9770349022667
125.084745762712 43.5112106577333
128.135593220339 49.5352288844
131.186440677966 44.8178085198667
134.237288135593 37.9606200962667
137.28813559322 37.6426113988
140.338983050847 26.891629106
143.389830508475 26.2268164901333
146.440677966102 34.1290956170667
149.491525423729 35.1801020777333
152.542372881356 39.5191586909333
155.593220338983 36.9165769681333
158.64406779661 49.8766998126667
161.694915254237 42.6891850853333
164.745762711864 46.1354867317333
167.796610169492 45.3826126229333
170.847457627119 39.3676279710667
173.898305084746 49.6796658793333
176.949152542373 48.9817665122667
180 30.837550552
};
\addlegendentry{CEN}
\addplot [semithick, white!64.7058823529412!black, mark=*, mark size=1.5, mark options={solid}]
table {%
0 0.760243551999995
3.05084745762712 1.07076658559999
6.10169491525424 4.55796341639999
9.15254237288136 17.7250464701333
12.2033898305085 42.0372563350666
15.2542372881356 65.9591311265333
18.3050847457627 68.3207089798667
21.3559322033898 34.7853812488
24.4067796610169 74.8257518562667
27.4576271186441 71.0315237918667
30.5084745762712 94.7824764468
33.5593220338983 98.8266377018667
36.6101694915254 97.9141590745333
39.6610169491525 79.2645656416
42.7118644067797 104.7210397308
45.7627118644068 96.1785881297333
48.8135593220339 100.598885536267
51.864406779661 108.7624407688
54.9152542372881 76.1666488354667
57.9661016949153 94.8858044884
61.0169491525424 92.2976692784
64.0677966101695 97.3549522698667
67.1186440677966 91.8770162885333
70.1694915254237 84.5614408841333
73.2203389830509 98.1401809602667
76.271186440678 83.5026173872
79.3220338983051 100.6849064804
82.3728813559322 101.979046317333
85.4237288135593 100.042333577067
88.4745762711864 102.7142495744
91.5254237288135 99.026305552
94.5762711864407 73.9221853948
97.6271186440678 103.534866084533
100.677966101695 97.3132261937333
103.728813559322 88.8789798818667
106.779661016949 76.3455904504
109.830508474576 85.4689207709333
112.881355932203 98.4671111926667
115.932203389831 107.694996760267
118.983050847458 69.9035556942667
122.033898305085 98.4857758916
125.084745762712 104.255149522
128.135593220339 98.1166432116
131.186440677966 103.447277009333
134.237288135593 93.8018229322667
137.28813559322 101.466843480267
140.338983050847 95.0444228854667
143.389830508475 100.8828063244
146.440677966102 107.756171664533
149.491525423729 108.764404409733
152.542372881356 94.9006598068
155.593220338983 107.711418521867
158.64406779661 109.803940582267
161.694915254237 93.340908758
164.745762711864 91.8834324762667
167.796610169492 77.7507453364
170.847457627119 84.3182013208
173.898305084746 93.9739244669333
176.949152542373 109.7557646828
180 86.3269698741333
};
\addlegendentry{DEC}
\end{axis}

\end{tikzpicture} \\
			\small (a) \gls{sc1} \& $\bar{S}=1000$ meters & \small (b) \gls{sc1} \& $\bar{S}=2000$ meters \\
			&\\
\begin{tikzpicture}[font=\footnotesize]

\definecolor{color0}{rgb}{0.266666666666667,0.447058823529412,0.768627450980392}

\begin{axis}[
legend cell align={left},
height=5cm,
width=8cm,
legend style={
  fill opacity=0.8,
  draw opacity=1,
  text opacity=1,
  at={(0.03,0.97)},
  anchor=north west,
  draw=white!80!black
},
tick align=outside,
tick pos=left,
x grid style={white!69.0196078431373!black},
xlabel={Driver departure time [min]},
xmin=-8.69491525423729, xmax=182.593220338983,
xtick style={color=black},
y grid style={white!69.0196078431373!black},
ylabel={Individual cost [min]},
ymin=-4.79293797058001, ymax=117.52955741418,
ytick style={color=black}
]
\addplot [semithick, color0, mark=+, mark size=1.5, mark options={solid}]
table {%
0 0.767175455999994
3.05084745762712 1.270473888
6.10169491525424 8.33002393266667
15.2542372881356 15.1363453654667
18.3050847457627 28.7796829213333
21.3559322033898 67.0299408793333
30.5084745762712 78.6393858774667
33.5593220338983 82.5613735596
36.6101694915254 111.9694439876
45.7627118644068 13.9639662425333
48.8135593220339 16.8101140981333
51.864406779661 35.8191249741333
61.0169491525424 34.51393517
64.0677966101695 36.917782436
67.1186440677966 52.72710796
76.271186440678 65.1660833050667
79.3220338983051 84.7237292462667
82.3728813559322 96.7149800337333
91.5254237288135 49.9412832982667
94.5762711864407 27.1225998426667
97.6271186440678 29.408411184
106.779661016949 37.5353635785333
109.830508474576 36.3095917436
112.881355932203 48.3842848284
122.033898305085 43.0710490473333
125.084745762712 53.1771851933333
128.135593220339 83.0847788170667
137.28813559322 71.3974740569333
140.338983050847 51.1820042518667
143.389830508475 29.1305654550667
152.542372881356 35.8513789566667
155.593220338983 39.9422294645333
158.64406779661 54.6887579052
167.796610169492 45.5490934030667
170.847457627119 47.3733483296
173.898305084746 67.1604971208
};
\addlegendentry{CEN}
\addplot [semithick, white!64.7058823529412!black, mark=*, mark size=1.5, mark options={solid}]
table {%
0 0.767175455999994
3.05084745762712 1.05243521879999
6.10169491525424 9.61510426439998
15.2542372881356 27.7407573590667
18.3050847457627 60.7103784325333
21.3559322033898 78.3350735710667
30.5084745762712 82.6894550529333
33.5593220338983 92.7445743098667
36.6101694915254 105.449092780933
45.7627118644068 84.3518165808
48.8135593220339 104.811766338533
51.864406779661 99.4065091708
61.0169491525424 70.0033511721333
64.0677966101695 106.722888619733
67.1186440677966 105.731344882
76.271186440678 76.7852395654667
79.3220338983051 94.2476088710667
82.3728813559322 108.8385799276
91.5254237288135 75.3274040101333
94.5762711864407 95.9395670812
97.6271186440678 107.550231348533
106.779661016949 81.2162104765333
109.830508474576 77.0333290412
112.881355932203 102.496535773467
122.033898305085 81.7181896344
125.084745762712 73.3613966194667
128.135593220339 107.8009932308
137.28813559322 75.1207147576
140.338983050847 90.5176783390667
143.389830508475 98.1683914452
152.542372881356 78.9294616842667
155.593220338983 77.4934946301333
158.64406779661 108.182506798933
167.796610169492 80.515743842
170.847457627119 82.0170789738667
173.898305084746 99.9542546504
};
\addlegendentry{DEC}
\end{axis}

\end{tikzpicture} &
			\hspace{-1cm} 
\begin{tikzpicture}[font=\footnotesize]

\definecolor{color0}{rgb}{0.266666666666667,0.447058823529412,0.768627450980392}

\begin{axis}[
legend cell align={left},
height=5cm,
width=8cm,
legend style={
  fill opacity=0.8,
  draw opacity=1,
  text opacity=1,
  at={(0.03,0.97)},
  anchor=north west,
  draw=white!80!black
},
tick align=outside,
tick pos=left,
x grid style={white!69.0196078431373!black},
xlabel={Driver departure time [min]},
xmin=-8.69491525423729, xmax=182.593220338983,
xtick style={color=black},
y grid style={white!69.0196078431373!black},
ylabel={Individual cost [min]},
ymin=-4.71657880890667, ymax=115.77351313104,
ytick style={color=black}
]
\addplot [semithick, color0, mark=+, mark size=1.5, mark options={solid}]
table {%
0 0.760243551999995
3.05084745762712 1.32050904946667
6.10169491525424 9.55902767906667
15.2542372881356 16.6473198114667
18.3050847457627 4.51139116666667
21.3559322033898 19.0744845446667
30.5084745762712 18.1139897689333
33.5593220338983 16.0071525373333
36.6101694915254 26.671853764
45.7627118644068 6.59300973933333
48.8135593220339 8.34847687293334
51.864406779661 14.5712143462666
61.0169491525424 19.6988676809333
64.0677966101695 16.5623542630667
67.1186440677966 20.0535351456
76.271186440678 23.5843621252
79.3220338983051 20.9142954344
82.3728813559322 18.2733971432
91.5254237288135 12.3906978773333
94.5762711864407 11.9598991161333
97.6271186440678 13.7423988317333
106.779661016949 17.4021966538667
109.830508474576 14.3628587798667
112.881355932203 20.0789050557333
122.033898305085 18.5705393764
125.084745762712 20.8992402202667
128.135593220339 24.9490971118667
137.28813559322 19.7535750116
140.338983050847 13.0053512350667
143.389830508475 14.8236759706667
152.542372881356 15.5222089793333
155.593220338983 19.6453809041333
158.64406779661 24.2587809537333
167.796610169492 22.4482178402667
170.847457627119 19.3826480504
173.898305084746 24.8957560204
};
\addlegendentry{CEN}
\addplot [semithick, white!64.7058823529412!black, mark=*, mark size=1.5, mark options={solid}]
table {%
0 0.760243551999995
3.05084745762712 1.07010839426666
6.10169491525424 5.17450591653333
15.2542372881356 15.5446357197333
18.3050847457627 38.4567857628
21.3559322033898 13.6085143997333
30.5084745762712 33.7046371977333
33.5593220338983 77.5967701709333
36.6101694915254 74.4311701021333
45.7627118644068 67.6195588556
48.8135593220339 67.3010239156
51.864406779661 104.861092667467
61.0169491525424 60.7513462453333
64.0677966101695 61.6250356948
67.1186440677966 86.0252131461333
76.271186440678 57.6333613998667
79.3220338983051 81.6407799032
82.3728813559322 96.9453352496
91.5254237288135 77.5000100502667
94.5762711864407 84.5788232674667
97.6271186440678 96.8862221106667
106.779661016949 60.0887944608
109.830508474576 84.955961584
112.881355932203 97.064621106
122.033898305085 68.7540877109333
125.084745762712 96.6190158045333
128.135593220339 93.7197637593333
137.28813559322 75.2157302562667
140.338983050847 86.9212798636
143.389830508475 92.6573707086667
152.542372881356 74.3877559705333
155.593220338983 95.488887788
158.64406779661 110.296690770133
167.796610169492 55.1905379322667
170.847457627119 70.8042219710667
173.898305084746 90.1092637068
};
\addlegendentry{DEC}
\end{axis}

\end{tikzpicture} \\
			\small (c) \gls{sc3} \& $\bar{S}=1000$ meters & \small (d) \gls{sc3} \& $\bar{S}=2000$ meters \\
		\end{tabular}
		\caption{Comparison of DEC and CEN in high-availability scenarios for a three-hours planning horizon}\label{fig:indiv_cost_60}
		\fnote{\footnotesize Each subplot shows the individual realized cost $\hat{\alpha}_i$ for each considered setting, per number of drivers $N$, averaged over all values of $r^s \in \{100,300,700\}$ meters.}
	\end{figure}
	While the system performances increase for a larger search radius, in-line with short planning horizons results, Figure~\ref{fig:indiv_cost_60} shows that decreasing the requests frequency has however a larger positive impact for a smaller search radius.

	Finally, we observe that long planning horizon results reveal cyclic patterns for individual drivers cost (see~Figure~\ref{fig:indiv_cost_60}). These patterns show a decreasing amplitude for larger charging demand heterogeneity or search radii.
	This effect results from early drivers having a higher chance of reaching closely related charging stations and thus obtaining lower individual cost than succeeding drivers. When these stations are freed after $\Delta T^v_{\text{charge}}$ minutes, they get revisited first with a higher recovered probability to be available to succeeding drivers, which replicates the pattern. With a smaller overlap between drivers' searches, which can be realized either by larger search radii or larger temporal disconnect between drivers entering the system, the amplitude of these patterns decreases as a smaller overlap increases the chance for each driver to reach an unoccupied closely related station.
	%
	Figure~\ref{fig:indiv_cost_25} in Appendix~\ref{app:large-planning} shows complementary analyses for the low-availability scenario.
	Similar to short planning horizon results, the cost reduction obtained in CEN compared to DEC decreases in low-availability scenarios, especially for a small search radius with $\bar{S}=1000$ meters.

\section{Conclusion}\label{sec:conclusion}

In this paper, we studied the multi-agent charging station search problem in stochastic environments, which we define as a single-decision maker MDP. 
We showed that by constraining agents' individual policies to be executed independent of each other, we can simplify the global MDP representation to a set of single-agent MDPs.
We then introduced several online algorithms that solve centralized and decentralized decision-making settings, applicable to static and dynamic policies, and different levels of information-sharing. Specifically, we analyzed the benefits of intention-sharing, i.e., drivers sharing their planned visits, observation-sharing, i.e., drivers sharing observed occupancies of charging stations, or both.

Using a case-study for the city of Berlin, we analyzed the benefits of coordination between multiple agents' search: our results show that coordination increases the system performance while individually benefiting each driver in general.
Analyzing the performance from a system perspective, our results show that a static decentralized coordination strategy achieves a 26\% cost decrease as long as drivers share visit intentions. A centralized coordination strategy requires higher computational load but achieves only 2\%  additional cost decrease. 
We further highlight the benefit of enforcing collaboration in intention-sharing settings.
Moreover, our results show that observation-sharing performs on average worse than intention-sharing setting.
However, observation-sharing requires less data and computational resources, and may be used to derive more accurate availability probabilities, which makes it interesting for practitioners. 
We show that from a driver perspective coordination may save up to 23\% of a driver's search time, while increasing her search's success rate by 9\% on average.	
We find that coordinated searches outperform uncoordinated searches for individual best and worst case scenarios.
Finally, we observe similar effects in longer planning horizon settings.

Finally, one comment on our study is in order. We assumed a non-adversarial setting such that agents always follow their navigation device visit recommendations, which may however be challenged by drivers behavior in practice. If drivers deviate from the recommended visits, intention-sharing might become misleading. 
Analyzing competitive searches within a game-theoretical setting by relaxing the non-adversarial assumption opens a new avenue for further research in this context.


%
\singlespacing{
\bibliographystyle{model5-names}
\bibliography{main}} 
%
\onehalfspacing
\begin{appendices}
	\normalsize
	\section{Proof of Proposition \ref{prop:equal_MDP}}

To prove Proposition~\ref{prop:equal_MDP}, we seek to show that
\begin{equation}\label{eq:equivalence}
\begin{split}
\forall \pi \in \Pi^{\text{ind}} \text{  , } \alpha^{\pi} = V^{\pi}(x_0)
\end{split}
\end{equation}
holds, i.e., that the expected MDP cost defined in Section~\ref{subsec:mdp_central} for a user-independent policy $\pi$ (see Section~\ref{sec:pol_category}) can be expressed as a function of all expected single-agent MDP costs (see Section~\ref{subsec:indiv_mdp}).
We recall that a user-independent policy can be expressed as a set of single-agent policies, with $\pi = \{\pi^i, i \in \mathcal{D}\}$ and $\pi^i$ being represented by a sequence of station visits $C^i=(v^i_0,..., v^i_n)$. 

To show that Equation~\ref{eq:equivalence} holds, we first reformulate the policy-specific cost function $V^{\pi}(x)$ for any state~$x\in \mathcal{S}$. We then show by recursion that it holds for any state, which allows us to finally prove the equality relation for the initial state~$x_0$.

\begin{description}
	\item[Step 1:] Given decision epoch $t$ with deciding agent~$\acute{\lambda}$, we explicitly define for both the partially successful state~${\acute{x}}^{\text{s}}$ and the unsuccessful state~${\acute{x}}^{\text{f}}$ the policy cost function $V^{\pi}({\acute{x}}^{\text{s}})$ and $V^{\pi}({\acute{x}}^{\text{f}})$. $\mathcal{J}'_{\text{f}}$, respectively $\mathcal{J}'_{\text{s}}$ represent the set of non-terminated drivers at ${\acute{x}}^{\text{f}}$, respectively ${\acute{x}}^{\text{s}}$. 
	We consider for all drivers $i\in\mathcal{J}'_{\text{f}}$ their truncated policies, i.e., actions prescribed from their last visited stations in decision epoch $t$. We represent agent $i$'s truncated policy by the sequence of station visits $\bar{C}^i=(\bar{v}^i_0, ..., \bar{v}^i_l)$, with $\bar{v}^i_0$ being the assigned station to $i$ in $x$. With a slight abuse of notation, we refer to $\bar{v}^i_k$ as $v^i_k$ in the remainder of this proof.

	In the following, we let indices~$k$ describe the time-ordered visit events of stations included in $\bar{C}^i$ for all drivers in $\mathcal{J}'_{\text{f}}$ for both ${\acute{x}}^{\text{f}}$ and ${\acute{x}}^{\text{s}}$. Let $n(\mathcal{J}'_{\text{f}})$ be the maximum number of all possible visit events, considering all stations defined by all respective $\bar{C}^i$ such that $n=\sum_{i \in \mathcal{J}'_{\text{f}}} |\bar{C}^i| $, with $|\bar{C}^i|$ being the length of $\bar{C}^i$. 
	We define $k^i$ as the index in sequence $\bar{C}^i$ of the last station visited by $i$ in stage $k$.
	The binary variable $\delta_{i,k}$ indicates whether $i$ observes an occupied station and has enough time to select her next station to visit at stage $k$. 
	In this case, we let $t(i,k)$ indicate the travel time for $i$ from the station visited in stage $k$ to the next planned station, i.e., $t(i,k)=t_{v^i_{k^i}, v^i_{k^i + 1}}$.
%
	We let $\rho^i_{[k:l]}$ be the likelihood for $i$ to get a least one free station among $\bar{C}^i[k^i, l^i]$, with $\rho^i_{[k:l]} = 1 - \prod_{v\in \bar{C}^i[k^i:l^i]} (1 - p^i_v)$ and $p^i_v$ being the user-dependent availability probability. Binary variables $\delta'_{\text{s}}$, respectively $\delta'_{\text{f}}$, indicate whether at least one driver failed her search already, i.e., $s=\text{'t'}$, in state~${\acute{x}}^{\text{s}}$, respectively in state~${\acute{x}}^{\text{f}}$.

	We now explicit the expected cost of value function $V$ from states ${\acute{x}}^{\text{s}}$ and ${\acute{x}}^{\text{f}}$ as follows
	\begin{equation}\label{eq:explicit_cost}
	\begin{split}
	V^{\pi}({\acute{x}}^{\text{s}}) =& \sum_{k=0}^{n-1}  \sum_{i \in \mathcal{J}'_{\text{s}}} \bar{\rho}^i_{[0:k]} t(i,k) \delta_{i,k} + \sum_{i \in \mathcal{J}'_{\text{s}}} \bar{\rho}^i_{[0:n]} \bar{\beta} + (\delta'_{\text{s}} + (1- \delta'_{\text{s}}) (1 - \prod_{i \in \mathcal{J}'_{\text{s}}} \rho^i_{[0:n]})) \beta^G\; ,\\
	V^{\pi}({\acute{x}}^{\text{f}}) =& \sum_{k=0}^{n-1}  \sum_{i \in \mathcal{J}^{'}_{\text{f}}}  \bar{\rho}^i_{[0:k]} t(i,k) \delta_{i,k}  + \sum_{i \in \mathcal{J}^{'}_{\text{f}}}  \bar{\rho}^i_{[0:n]} \bar{\beta} + (\delta'_{\text{f}} + (1- \delta'_{\text{f}}) (1 - \prod_{i \in \mathcal{J}^{'}_{\text{f}}} \rho^i_{[0:n]})) \beta^G\; ,\\
	\end{split}
	\end{equation}
	with 
	\begin{equation}\label{eq:explicit_diff}
	\begin{split}
	\sum_{k=0}^{n-1} \sum_{i \in \mathcal{J}^{'}_{\text{f}}}  \bar{\rho}^i_{[0:k]} t(i,k) \delta_{i,k}   - \sum_{k=0}^{n-1} \sum_{i \in \mathcal{J}^{'}_{\text{s}}}  \bar{\rho}^i_{[0:k]} t(i,k) \delta_{i,k}   \stackrel{(*)}{=} \sum_{k=0}^{n-1} \bar{\rho}^{\acute{\lambda}}_{[0:k]} t(\acute{\lambda}, k)  \delta_{{\acute{\lambda}},k} \;  , \\
	\end{split}
	\end{equation}
	using in $(*)$ that the subsequent visit decisions for all drivers but $\acute{\lambda}$ do not depend on whether $\acute{\lambda}$ observes an occupied station (${\acute{x}}^{\text{s}}$) or not (${\acute{x}}^{\text{f}}$) in decision epoch $t$, because all $\pi^i$ are user-independent.
	
	For brevity in notation, we show the case for $\gamma^i_v=0 \; \forall i \in \mathcal{D}, v\in \mathcal{V}$, w.l.o.g., as one can apply the following transformation: (i) $t_{v^i_{k},v^i_{k +1}} \gets t_{v^i_{k},v^i_{k +1}} + p_{v^i_{k +1}} \cdot \gamma^i_{v^i_{k +1}}$ and (ii) $V^{\pi}({\acute{x}}^{\text{s}}) \gets V^{\pi}({\acute{x}}^{\text{s}}) + \gamma^{\acute{\lambda}}_v$. 
	
	\item[Step 2:] Equation~\ref{eq:explicit_cost} holds for the last decision epoch in state~$\acute{x}$ before the global termination state, in which only one non-terminated driver, for ${\acute{x}}^{\text{f}}$ and no driver for ${\acute{x}}^{\text{s}}$ ($n=1$) remains. We now show that if Equation~\ref{eq:explicit_cost} holds for $t$, then it also holds for $t-1$ and thus by recursion for any possible decision epoch, e.g., in the initial state with $t=0$.
	
	Let us consider all states $x$ that can lead to ${\acute{x}}^{\text{s}}$ or ${\acute{x}}^{\text{f}}$ following policy $\pi$, such that $p(\acute{x} \in \{ {\acute{x}}^{\text{f}}, {\acute{x}}^{\text{s}}\} |  x,\pi(x) ) = 1$. Let $\lambda$ be the deciding agent for these states and $t-1$ be the decision epoch accordingly. For all these states, $\acute{\lambda}$ is unterminated, as it is either deciding in ${\acute{x}}^{\text{f}}$ or successfully finished in ${\acute{x}}^{\text{s}}$.
	Without loss of generality, let us shift indexes such that $k:=k+1$ in the summations of $V({\acute{x}}^{\text{s}})$ and $V({\acute{x}}^{\text{f}})$. The first station visit from  ${\acute{x}}^{\text{f}}$ or ${\acute{x}}^{\text{s}}$ in epoch $t$ becomes the second station visit from states~${x}^{\text{f}}$ or ${x}^{\text{s}}$ in preceding epoch $t-1$.
	
	
	For brevity, we show the case in which $\acute{\lambda}$ \textit{is not terminated}, i.e., $\acute{\lambda} \in \mathcal{J}'_{\text{f}}$ and $\mathcal{J}'_{\text{f}} = \mathcal{J}'_{\text{s}} \bigcup \acute{\lambda}$. In this case, $\delta'_{\text{s}} = \delta'_{\text{f}} = \delta_{\text{f}}$. If at least one agent terminated in ${\acute{x}}^{\text{s}}$, at least one agent must have terminated in  ${\acute{x}}^{\text{f}}$ and vice-versa since $\acute{\lambda}$ did not terminate. Similarly, if no agent terminated in ${\acute{x}}^{\text{s}}$, then no agent terminated in ${\acute{x}}^{\text{f}}$ since $\acute{\lambda}$ did not terminate and vice-versa. In all cases, the termination condition holds for the previous unsuccessful state~$x^{\text{f}}$.
	
	\begin{equation}\label{eq:demo_cost_K}
	\begin{split}
	V^{\pi}({x}^f) \stackrel{(*)}{=}&  t(\lambda,0) +p_{v^{\acute{\lambda}}} V({\acute{x}}^{\text{s}})  + (1- p_{v^{\acute{\lambda}}} ) V({\acute{x}}^{\text{f}})\\
	=&  t(\lambda,0) + p_{v^{\acute{\lambda}}} ( \sum_{k=1}^{n}  \sum_{i \in \mathcal{J}'_{\text{s}}} \bar{\rho}^i_{[1:k]} t(i,k) \delta_{i,k}  + \sum_{i \in \mathcal{J}'_{\text{s}}} \bar{\rho}^i_{[1:n+1]} \bar{\beta})  + \\
	&  (1 - p_{v^{\acute{\lambda}}} ) ( \sum_{k=1}^{n}  \sum_{i \in \mathcal{J}^{'}_{\text{f}}}  \bar{\rho}^i_{[1:k]} t(i,k) \delta_{i,k}  + \sum_{i \in \mathcal{J}^{'}_{\text{f}}}  \bar{\rho}^i_{[1:n+1]} \bar{\beta}) \\
	& + p_{v^{\acute{\lambda}}}  ( \delta'_{\text{s}}  + (1-\delta'_{\text{s}}) (1 - \prod_{i \in \mathcal{J}'_{\text{s}}} \rho^i_{[1:n+1]})) \beta^G + (1 - p_{v^{\acute{\lambda}}} )( \delta'_{\text{f}}  + (1-\delta'_{\text{f}}) (1 - \prod_{i \in \mathcal{J}^{'}_{\text{f}}} \rho^i_{[1:n+1]})) \beta^G \\
	\stackrel{(**)}{=}&  \sum_{k=0}^{n}  \sum_{i \in \mathcal{J}'_{\text{s}}} \bar{\rho}^i_{[0:k]} t(i,k) \delta_{i,k}  + \sum_{i \in \mathcal{J}'_{\text{s}}} \bar{\rho}^i_{[0:n+1]} \bar{\beta} + (1 - p_{v^{\acute{\lambda}}} ) ( \sum_{k=1}^{n}  \bar{\rho}^{\acute{\lambda}}_{[1:k]} t(\acute{\lambda}, k) \delta_{\acute{\lambda},l}  + \bar{\rho}^{\acute{\lambda}}_{[1:n+1]} \bar{\beta}) + \\
	& + p_{v^{\acute{\lambda}}} (1- \delta'_{\text{s}}) (1 - \prod_{i \in \mathcal{J}^{'}_{\text{s}}} \rho^i_{[1:n+1]}) \beta^G + (1 - p_{v^{\acute{\lambda}}}) (1-\delta'_{\text{f}}) (1 - \prod_{i \in \mathcal{J}'_{\text{f}}} \rho^i_{[1:n+1]}) \beta^G \\
	&+ ((1 - p_{v^{\acute{\lambda}}}) \delta'_{\text{f}}  + p_{v^{\acute{\lambda}}} \delta'_{\text{s}})  \beta^G\\
	%
	=&  \sum_{k=0}^{n}  \sum_{i \in \mathcal{J}^{'}_{\text{f}}} \bar{\rho}^i_{[0:k]} t(i,k) \delta_{i,k} + \sum_{i \in \mathcal{J}^{'}_{\text{f}}}  \bar{\rho}^i_{[0:k]} \bar{\beta} + \\
	&  (1-\delta_{\text{f}}) (p_{v^{\acute{\lambda}}}  (1 - \prod_{i \in \mathcal{J}^{'}_{\text{s}}} \rho^i_{[1:n+1]}) + (1 - p_{v^{\acute{\lambda}}}) (1 - \prod_{i \in \mathcal{J}^{'}_{\text{f}}} \rho^i_{[1:n+1]})) \beta^G  +  \delta_{\text{f}} \beta^G\\
	\stackrel{(***)}{=}  &  \sum_{k=0}^{n}  \sum_{i \in \mathcal{J}_{\text{f}}} \bar{\rho}^i_{[0:k]} t(i,k) \delta_{i,k}  + \sum_{i \in \mathcal{J}_{\text{f}}}  \bar{\rho}^i_{[0:n+1]} \bar{\beta} + (1-\delta_{\text{f}})(1 - \prod_{i \in \mathcal{J}_{\text{f}}} \rho^i_{[0:n+1]}) \beta^G + \delta_{\text{f}} \beta^G\; .\\
	\end{split}
	\end{equation}

	We use in $(*)$ the recursive definition of $V^\pi$ and in $(**)$ Equation~\ref{eq:explicit_diff}. In $(***)$, we exploit that $\mathcal{J}^{'}_{\text{f}} = \mathcal{J}_{\text{f}}$, which follows from the following reasoning: if $\acute{\lambda}$ belongs to $\mathcal{J}^{'}_{\text{f}}$, then it is not terminated in $x^{\text{f}}$ and accordingly belongs to $\mathcal{J}_{\text{f}}$; as $x^{\text{f}}$ is a partially unsuccessful state for $\lambda$, it follows that $\lambda$ belongs to both $\mathcal{J}_{\text{f}}$ and $\mathcal{J}^{'}_{\text{f}}$.
	We show that $(***)$ holds, given the following equality
	\begin{equation}\label{eq:circ}
	\begin{split}
	&p_{v^{\acute{\lambda}}} (1 - \prod_{i \in \mathcal{J}^{'}_{\text{s}}} \rho^i_{[1:n+1]}) + (1 - p_{v^{\acute{\lambda}}} ) (1 - \prod_{i \in \mathcal{J}^{'}_{\text{f}}} \rho^i_{[1:n+1]}) \\
	=& p_{v^{\acute{\lambda}}}  (1 - \prod_{i \in \mathcal{J}^{'}_{\text{s}}} \rho^i_{[1:n+1]}) + (1 - p_{v^{\acute{\lambda}}} ) (1 - \rho^{\acute{\lambda}}_{[1:n+1]}. \prod_{i \in \mathcal{J}^{'}_{\text{s}}} \rho^i_{[1:n+1]} ) \\
	=& 1  - (\prod_{i \in \mathcal{J}^{'}_{\text{s}}} \rho^i_{[1:n+1]} ) (\rho^{\acute{\lambda}}_{[1:n+1]} +  p_{v^{\acute{\lambda}}}   - \rho^{\acute{\lambda}}_{[1:n+1]}  p_{v^{\acute{\lambda}}})\\
	\stackrel{(****)}{=}& 1  - (\prod_{i \in \mathcal{J}^{'}_{\text{s}}} \rho^i_{[0:n+1]} ) (\rho^{\acute{\lambda}}_{[0:n+1]})\\
	=& 1  - (\prod_{i \in \mathcal{J}^{'}_{\text{f}}} \rho^i_{[0:n+1]} ),\\ 
	\end{split}
	\end{equation}
	using in $(****)$ that
	$\forall i \in \mathcal{J}^{'}_{\text{s}} \text{ }  \rho^i_{[0:n+1]} =  \rho^i_{[1:n+1]}$
	and that 
	$$ (1 - \rho^{\acute{\lambda}}_{[0:n+1]} ) =  (1 - \rho^{\acute{\lambda}}_{[1:n+1]}) (1 -  p_{v^{\acute{\lambda}}})  \Leftrightarrow \rho^{\acute{\lambda}}_{[0:n+1]} =  \rho^{\acute{\lambda}}_{[1:n+1]} +  p_{v^{\acute{\lambda}}}   - \rho^{\acute{\lambda}}_{[1:n+1]} p_{v^{\acute{\lambda}}} \; .$$
	Finally, we have $n(\mathcal{J}_{\text{f}}) := n(\mathcal{J}'_{\text{f}}) + 1 = n+1$.

	We can analogously show that \ref{eq:explicit_cost} holds for $V^{\pi}({x}^{\text{f}})$ when $\acute{\lambda}$ is terminated in ${\acute{x}}_{\text{f}}$. In this case, $\mathcal{J}'_{\text{s}}=\mathcal{J}'_{\text{f}}$, $\mathcal{J}'_{\text{f}} = \mathcal{J}_{\text{f}} \bigcup \acute{\lambda}$, $\delta'_{\text{f}} = 1$, and  $\delta'_{\text{s}} = \delta_{\text{f}}$. Given that $\prod_{i \in \mathcal{J}_{\text{f}}} \rho^i_{[0:n+1]} = p_{v^{\acute{\lambda}}}.\prod_{i \in \mathcal{J}^{'}_{\text{f}}} \rho^i_{[1:n+1]} $, we can verify by substitution that
	\begin{equation}\label{eq:circ}
	\begin{split}
	p_{v^{\acute{\lambda}}} (1- \delta'_{\text{s}}) (1 - \prod_{i \in \mathcal{J}^{'}_{\text{s}}} \rho^i_{[1:n+1]}) \beta^G + (1 - p_{v^{\acute{\lambda}}}) (1-\delta'_{\text{f}}) (1 - \prod_{i \in \mathcal{J}'_{\text{f}}} \rho^i_{[1:n+1]}) \beta^G + ((1 - p_{v^{\acute{\lambda}}}) \delta'_{\text{f}}  + p_{v^{\acute{\lambda}}} \delta'_{\text{s}})  \beta^G &=\\
	(\delta_{\text{f}} + (1- \delta_{\text{f}}) (1 - \prod_{i \in \mathcal{J}_{\text{f}}} \rho^i_{[0:n]})) \beta^G& \; .\\
	\end{split}
	\end{equation}
	
	We analogously show that \ref{eq:explicit_cost} holds for $V^{\pi}({x}^s)$. 
	\item[Step 3:] Finally, we have
	\begin{equation}\label{eq:demo_cost_K}
	\begin{split}
	V^{\pi}(x_0) \stackrel{(*)}{=}& \sum_{k=0}^{n -1}  \sum_{i \in \mathcal{J}_{0}}  \bar{\rho}^i_{[0:l]} t(i,k) \delta_{i,k}  + \sum_{i \in \mathcal{J}_{0}}  \bar{\rho}^i_{[0:n]} \bar{\beta} + (1 - \prod_{i \in \mathcal{J}_{0}} \rho^i_{[0:n+1]} ) \beta^G \\
	\stackrel{(**)}{=}& \sum_{i \in \mathcal{J}_{0}}(\alpha^i) + (1 - \prod_{i \in \mathcal{J}_{0}} \rho^i_{[0:n]}) \beta^G \; , \\
	\end{split}
	\end{equation}
	with $n=n(\mathcal{J}_{0})$. In the initial state~$x_0$, $\delta=0$ holds, because no driver has terminated in the initial state by assumption. We use in $(*)$ Equation~\ref{eq:explicit_cost} and in $(**)$ the definition of $\alpha^i$.
	
	Finally, we justified that for $\pi \in \Pi^{\text{ind}}$, with $x_0$ as initial state at the decision epoch $t=0$ and $\mathcal{J}_{0} = \mathcal{D}$,  that
	\begin{equation}\label{eq:final}
	\begin{split}
	V^\pi(x_0) &= \sum_{i \in \mathcal{D}}(\alpha^i)  + (1 - \prod_{i \in \mathcal{D}} \rho^i_{[0:n]}) \beta^G \\
	&= \sum_{i \in \mathcal{D}} {F^i}^{\pi^i}(x^i_0)  + (1 - \prod_{i \in \mathcal{D}} \rho^i_{[0:n]}) \beta^G. \\
	\end{split}
	\end{equation}
	
\end{description}

\noindent This concludes the proof.

\section{Pseudo code of the LH algorithm}\label{app:lh_algo}
\begin{figure}[bp]
	\caption{LH algorithm.}
	\label{alg:lh_algo}
	\begin{algorithmic}[1]
		\State $\mathcal{L}^a\gets\{L^i_0\}$, $L^*\gets L^i_0$
		\While{$\mathcal{L}^a\neq \emptyset$}
		\State $L\gets \texttt{costMinimumLabel(}\mathcal{L}^a\texttt{)}$
		\State $\mathcal{L}^a\gets \mathcal{L}^a\setminus\{L\}$
		\For{$(v,\acute{v}) \in \delta^+(L)$}
		\State $L' \gets \mathcal{F}_{v\acute{v}}(L)$
		\If{\texttt{isNotDominated($L',\mathcal{ L}^a$)}}
		\State \texttt{dominanceCheck($\mathcal{L}^a,L'$)}
		\State $\mathcal{L}^a\gets\mathcal{L}^a\cup\{L'\}$
		\If{$ (\delta^+(L') = \emptyset) \land \alpha(L') < \alpha(L^*)$}
		\State $L^*\gets L'$
		\EndIf
		\EndIf
		\EndFor
		\EndWhile
		\State \Return $L^*$
	\end{algorithmic}
	\vspace{-0.5cm}
\end{figure}
Figure~\ref{alg:lh_algo} shows the pseudo-code of the multi-label setting algorithm introduced in \citet{GuilletHiermannEtAl2021}. 
We introduce the set of active labels $\mathcal{L}^a$ and let $L^i_0$ be the initial label corresponding to an agent's start location. We let $\mathcal{F}_{v,\acute{v}}(L)$  be the function that returns the propagated label $L'$ (with a partial policy ending at $\acute{v}$) of $L$ (with partial policy ending at $v$).
The cost $\alpha(L)$ is the cost associated with label~$L$. Function $\delta^{+}(L)$ returns a set of tuples $(v,\acute{v})$ which denotes all feasible physical successor locations $\acute{v}\in \mathcal{V}$ for a label~$L$ whose partial policy ends at $v\in\mathcal{V}$.

\section{Additional numerical results}\label{app:num_res}
In this section, we substantiate our results discussion with additional analyses.

\subsection{Sensitivity analysis}\label{app:beta_sensitivity}
In the following, we analyze the sensitivity of the algorithmic settings DEC-I, DEC-IO, and CEN to the value of parameter $\beta^G$, and recall that DEC, DEC-O, and DEC-O-d are insensitive to this parameter.

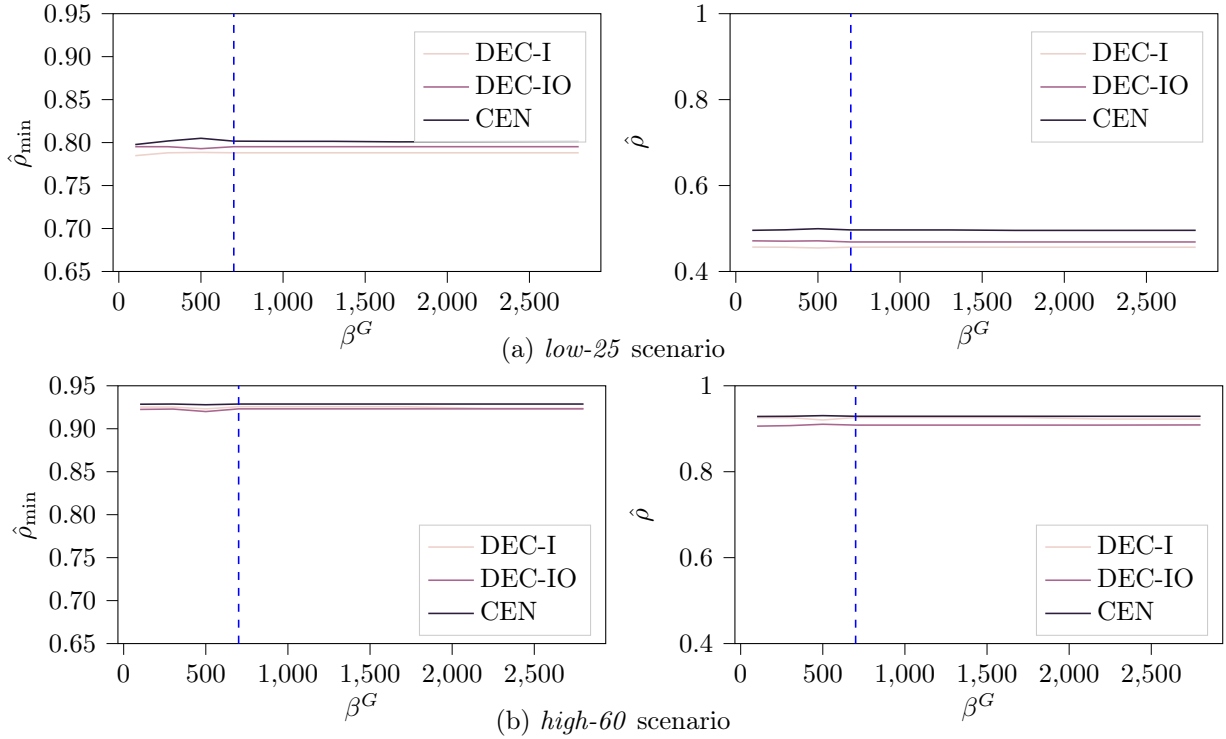
\begin{figure}[bp]
		\scalebox{1}{\begin{tabular}{c @{\qquad}}
				\vspace{-0.5cm}
\begin{tikzpicture}[font=\small]

\definecolor{color0}{rgb}{0.931269222332537,0.820192179608212,0.797148097466359}
\definecolor{color1}{rgb}{0.662652747046639,0.402798937278128,0.559929395373557}
\definecolor{color2}{rgb}{0.175086564895221,0.118400233069168,0.242159891378365}

\begin{groupplot}[group style={group size=2 by 1, horizontal sep=1.7cm}]
\nextgroupplot[
legend cell align={left},
legend style={
	fill opacity=0.8,
	draw opacity=1,
	text opacity=1,
	at={(0.97,0.5)},
	anchor=south east,
	draw=white!80!black
},
tick align=outside,
width=8cm,
height=5cm,
tick pos=left,
x grid style={white!69.0196078431373!black},
xmin=-35, xmax=2935,
xtick style={color=black},
xlabel={$\beta^G$},
y grid style={white!69.0196078431373!black},
ylabel={$\hat{\rho}_{\text{min}}$},
ymin=0.4, ymax=1,
ytick style={color=black},
ytick={0.4,0.5,0.6,0.7,0.8,0.9,1},
yticklabels={0.65,0.70,0.75,0.80,0.85,0.90,0.95}
]
\addplot [semithick, color0]
table {%
100 0.669583333333333
300 0.67625
500 0.677083333333333
700 0.67625
1000 0.67625
1300 0.67625
1700 0.67625
2200 0.67625
2800 0.67625
};
\addlegendentry{DEC-I}
\addplot [semithick, color1]
table {%
100 0.690416666666667
300 0.690416666666667
500 0.685833333333333
700 0.690416666666667
1000 0.690416666666667
1300 0.690416666666667
1700 0.690416666666667
2200 0.690416666666667
2800 0.690416666666667
};
\addlegendentry{DEC-IO}
\addplot [semithick, color2]
table {%
100 0.695416666666667
300 0.70375
500 0.71
700 0.703333333333333
1000 0.702916666666667
1300 0.702916666666667
1700 0.701666666666667
2200 0.702083333333333
2800 0.702916666666667
};
\addlegendentry{CEN}
\addplot [semithick, blue, dashed]
table {%
700 0.4
700 1
};

\nextgroupplot[
legend cell align={left},
legend style={
	fill opacity=0.8,
	draw opacity=1,
	text opacity=1,
	at={(0.97,0.5)},
	anchor=south east,
	draw=white!80!black
},
tick align=outside,
tick pos=left,
width=8cm,
height=5cm,
x grid style={white!69.0196078431373!black},
xmin=-35, xmax=2935,
xtick style={color=black},
xlabel={$\beta^G$},
y grid style={white!69.0196078431373!black},
ylabel={$\hat{\rho}$},
ymin=0.4, ymax=1,
ytick style={color=black}
]
\addplot [semithick, color0]
table {%
	100 0.456775
	300 0.456526875
	500 0.454701875
	700 0.456621875
	1000 0.456621875
	1300 0.456621875
	1700 0.456474375
	2200 0.456474375
	2800 0.456474375
};
\addlegendentry{DEC-I}
\addplot [semithick, color1]
table {%
	100 0.47135625
	300 0.470543125
	500 0.47131375
	700 0.46870375
	1000 0.46870375
	1300 0.46870375
	1700 0.46870375
	2200 0.46870375
	2800 0.46870375
};
\addlegendentry{DEC-IO}
\addplot [semithick, color2]
table {%
	100 0.495810625
	300 0.49691375
	500 0.49956375
	700 0.496735
	1000 0.49667
	1300 0.49667
	1700 0.495619375
	2200 0.495650625
	2800 0.495780625
};
\addlegendentry{CEN}
\addplot [semithick, blue, dashed]
table {%
	700 0.4
	700 1
};
\end{groupplot}

\end{tikzpicture} \\
				 \small (a) \gls{low_avail} scenario \\ 
				\vspace{-0.5cm}
\begin{tikzpicture}[font=\small]

\definecolor{color0}{rgb}{0.931269222332537,0.820192179608212,0.797148097466359}
\definecolor{color1}{rgb}{0.662652747046639,0.402798937278128,0.559929395373557}
\definecolor{color2}{rgb}{0.175086564895221,0.118400233069168,0.242159891378365}

\begin{groupplot}[group style={group size=2 by 1, horizontal sep=1.7cm}]
\nextgroupplot[
legend cell align={left},
legend style={
	fill opacity=0.8,
	draw opacity=1,
	text opacity=1,
	at={(0.97,0.03)},
	anchor=south east,
	draw=white!80!black
},
tick align=outside,
tick pos=left,
width=8cm,
height=5cm,
x grid style={white!69.0196078431373!black},
xlabel={$\beta^G$},
xmin=-35, xmax=2935,
xtick style={color=black},
y grid style={white!69.0196078431373!black},
ylabel={$\hat{\rho}_{\text{min}}$},
ymin=0.4, ymax=1,
ytick style={color=black},
ytick={0.4,0.5,0.6,0.7,0.8,0.9,1},
yticklabels={0.65,0.70,0.75,0.80,0.85,0.90,0.95}
]
\addplot [semithick, color0]
table {%
100 0.950416666666667
300 0.950416666666667
500 0.94625
700 0.95125
1000 0.95125
1300 0.95125
1700 0.95125
2200 0.9475
2800 0.9475
};
\addlegendentry{DEC-I}
\addplot [semithick, color1]
table {%
100 0.945
300 0.945833333333333
500 0.94
700 0.94625
1000 0.94625
1300 0.94625
1700 0.94625
2200 0.94625
2800 0.94625
};
\addlegendentry{DEC-IO}
\addplot [semithick, color2]
table {%
100 0.957083333333333
300 0.9575
500 0.955833333333333
700 0.9575
1000 0.9575
1300 0.9575
1700 0.9575
2200 0.9575
2800 0.9575
};
\addlegendentry{CEN}
\addplot [semithick, blue, dashed]
table {%
700 0.4
700 1
};

\nextgroupplot[
legend cell align={left},
legend style={
	fill opacity=0.8,
	draw opacity=1,
	text opacity=1,
	at={(0.97,0.03)},
	anchor=south east,
	draw=white!80!black
},
tick align=outside,
tick pos=left,
width=8cm,
height=5cm,
x grid style={white!69.0196078431373!black},
xlabel={$\beta^G$},
xmin=-35, xmax=2935,
xtick style={color=black},
y grid style={white!69.0196078431373!black},
ylabel={$\hat{\rho}$},
ymin=0.4, ymax=1,
ytick style={color=black}
]
\addplot [semithick, color0]
table {%
	100 0.92587
	300 0.92587
	500 0.9203575
	700 0.9262925
	1000 0.9262925
	1300 0.9262925
	1700 0.9262925
	2200 0.9225675
	2800 0.9225675
};
\addlegendentry{DEC-I}
\addplot [semithick, color1]
table {%
	100 0.906064375
	300 0.907295
	500 0.9104225
	700 0.908474375
	1000 0.908474375
	1300 0.908474375
	1700 0.908474375
	2200 0.908474375
	2800 0.908866875
};
\addlegendentry{DEC-IO}
\addplot [semithick, color2]
table {%
	100 0.928583125
	300 0.928996875
	500 0.930536875
	700 0.928996875
	1000 0.928996875
	1300 0.928996875
	1700 0.928996875
	2200 0.928996875
	2800 0.928996875
};
\addlegendentry{CEN}
\addplot [semithick, blue, dashed]
table {%
	700 0.4
	700 1
};
\end{groupplot}
\end{tikzpicture}\\
				\small (b) \gls{high_avail} scenario \\
		\end{tabular}}
	\caption{Impact of $\beta^G$ on system and worst success rates \label{fig:sensitivity_p}}
	\fnote{\footnotesize We compute $\rho_{\text{min}}$ as follows: $\hat{\rho}_{\text{min}} = \min_{i \in \mathcal{D}} \hat{\rho}^i$ with $\mathcal{D}$ being the set of drivers considered for each instance. We average values over all instances corresponding to a \gls{low_avail} or \gls{high_avail} scenario that correspond to $t^s=300$ m.}
\end{figure}

\begin{figure}[tbp]
		\scalebox{1}{\begin{tabular}{c @{\qquad}}
				\vspace{-0.5cm}
\begin{tikzpicture}[font=\small]

\definecolor{color0}{rgb}{0.931269222332537,0.820192179608212,0.797148097466359}
\definecolor{color1}{rgb}{0.662652747046639,0.402798937278128,0.559929395373557}
\definecolor{color2}{rgb}{0.175086564895221,0.118400233069168,0.242159891378365}

\begin{groupplot}[group style={group size=2 by 1, horizontal sep=1.7cm }]
\nextgroupplot[
tick align=outside,
legend cell align={left},
legend style={
	fill opacity=0.8,
	draw opacity=1,
	text opacity=1,
	at={(0.97,0.03)},
	anchor=south east,
	draw=white!80!black
},
tick align=outside,
tick pos=left,
width=8cm,
height=5cm,
xlabel={$\beta^G$},
x grid style={white!69.0196078431373!black},
xmin=-35, xmax=2935,
xtick style={color=black},
y grid style={white!69.0196078431373!black},
ylabel={$\hat{t}_{\text{max}}$ [min]},
ymin=1.7957125, ymax=2.63328,
ytick style={color=black}
]
\addplot [semithick, color0]
table {%
100 2.18340125
300 2.22409875
500 2.23691
700 2.23884125
1000 2.23884125
1300 2.23884125
1700 2.235656875
2200 2.235656875
2800 2.235656875
};
\addlegendentry{DEC-I}
\addplot [semithick, color1]
table {%
100 2.593371875
300 2.59520875
500 2.331515
700 2.59520875
1000 2.59520875
1300 2.59520875
1700 2.59520875
2200 2.59520875
2800 2.59520875
};
\addlegendentry{DEC-IO}
\addplot [semithick, color2]
table {%
100 2.2517825
300 2.235113125
500 2.212065
700 2.23321875
1000 2.231770625
1300 2.231770625
1700 2.229654375
2200 2.228014375
2800 2.229140625
};
\addlegendentry{CEN}
\addplot [semithick, blue, dashed]
table {%
700 1.7957125
700 2.63328
};

\nextgroupplot[
tick align=outside,
legend cell align={left},
legend style={
	fill opacity=0.8,
	draw opacity=1,
	text opacity=1,
	at={(0.97,0.03)},
	anchor=south east,
	draw=white!80!black
},
tick pos=left,
width=8cm,
height=5cm,
xlabel={$\beta^G$},
x grid style={white!69.0196078431373!black},
xmin=-35, xmax=2935,
xtick style={color=black},
y grid style={white!69.0196078431373!black},
ylabel={$\hat{t}$ [min]},
ymin=1.27944703125, ymax=1.97578484375,
ytick style={color=black}
]
\addplot [semithick, color0]
table {%
	100 1.776385
	300 1.790693125
	500 1.783615
	700 1.803538125
	1000 1.803538125
	1300 1.803538125
	1700 1.800451875
	2200 1.800451875
	2800 1.800451875
};
\addlegendentry{DEC-I}
\addplot [semithick, color1]
table {%
	100 1.941306875
	300 1.94278375
	500 1.847630625
	700 1.944133125
	1000 1.943973125
	1300 1.943973125
	1700 1.943973125
	2200 1.943973125
	2800 1.943973125
};
\addlegendentry{DEC-IO}
\addplot [semithick, color2]
table {%
	100 1.803266875
	300 1.814365625
	500 1.810785
	700 1.81389
	1000 1.81283
	1300 1.81283
	1700 1.812265
	2200 1.81283625
	2800 1.813723125
};
\addlegendentry{CEN}
\addplot [semithick, blue, dashed]
table {%
	700 1.27944703125
	700 1.97578484375
};
\end{groupplot}

\end{tikzpicture} \\
				\small (a) \gls{low_avail} scenario \\ 
				\vspace{-0.5cm}
\begin{tikzpicture}[font=\small]

\definecolor{color0}{rgb}{0.931269222332537,0.820192179608212,0.797148097466359}
\definecolor{color1}{rgb}{0.662652747046639,0.402798937278128,0.559929395373557}
\definecolor{color2}{rgb}{0.175086564895221,0.118400233069168,0.242159891378365}

\begin{groupplot}[group style={group size=2 by 1, horizontal sep=1.7cm}]
\nextgroupplot[
legend cell align={left},
legend style={
	fill opacity=0.8,
	draw opacity=1,
	text opacity=1,
	at={(0.97,0.5)},
	anchor=south east,
	draw=white!80!black
},
tick align=outside,
tick pos=left,
width=8cm,
height=5cm,
x grid style={white!69.0196078431373!black},
xlabel={$\beta^G$},
xmin=-35, xmax=2935,
xtick style={color=black},
y grid style={white!69.0196078431373!black},
ylabel={$\hat{t}_{\text{max}}$ [min]},
ymin=1.7957125, ymax=2.63328,
ytick style={color=black}
]
\addplot [semithick, color0]
table {%
100 1.852035625
300 1.852035625
500 1.884851875
700 1.83378375
1000 1.86500375
1300 1.86500375
1700 1.86500375
2200 1.872575625
2800 1.872575625
};
\addlegendentry{DEC-I}
\addplot [semithick, color1]
table {%
100 1.925405
300 1.917549375
500 1.90165
700 1.947354375
1000 1.947354375
1300 1.947354375
1700 1.948998125
2200 1.948998125
2800 1.948998125
};
\addlegendentry{DEC-IO}
\addplot [semithick, color2]
table {%
100 1.918785625
300 1.9135575
500 1.92048
700 1.92035875
1000 1.92035875
1300 1.92035875
1700 1.92035875
2200 1.928585625
2800 1.928585625
};
\addlegendentry{CEN}
\addplot [semithick, blue, dashed]
table {%
700 1.7957125
700 2.63328
};

\nextgroupplot[
tick align=outside,
legend cell align={left},
legend style={
	fill opacity=0.8,
	draw opacity=1,
	text opacity=1,
	at={(0.97,0.5)},
	anchor=south east,
	draw=white!80!black
},
tick align=outside,
tick pos=left,
width=8cm,
height=5cm,
x grid style={white!69.0196078431373!black},
xlabel={$\beta^G$},
xmin=-35, xmax=2935,
xtick style={color=black},
y grid style={white!69.0196078431373!black},
ylabel={$\hat{t}$ [min]},
ymin=1.27944703125, ymax=1.97578484375,
ytick style={color=black}
]
\addplot [semithick, color0]
table {%
	100 1.3153375
	300 1.3153375
	500 1.315239375
	700 1.31109875
	1000 1.334961875
	1300 1.334961875
	1700 1.334961875
	2200 1.33732
	2800 1.33732
};
\addlegendentry{DEC-I}
\addplot [semithick, color1]
table {%
	100 1.3285325
	300 1.331719375
	500 1.3587825
	700 1.33912
	1000 1.33912
	1300 1.33912
	1700 1.339668125
	2200 1.339668125
	2800 1.3432175
};
\addlegendentry{DEC-IO}
\addplot [semithick, color2]
table {%
	100 1.32664125
	300 1.325280625
	500 1.325408125
	700 1.3288375
	1000 1.3288375
	1300 1.3288375
	1700 1.328681875
	2200 1.3331275
	2800 1.3331275
};
\addlegendentry{CEN}
\addplot [semithick, blue, dashed]
table {%
	700 1.27944703125
	700 1.97578484375
};
\end{groupplot}

\end{tikzpicture}\\
				\small (b) \gls{high_avail} scenario \\
		\end{tabular}}
	\caption{Impact of $\beta^G$ on on average and worst search times \label{fig:sensitivity_t}}
	\fnote{\footnotesize We compute $\hat{t}_{\text{max}}$ as follows: $\hat{t}_{\text{max}} = \max_{i \in \mathcal{D}} \hat{t}^i$ with $\mathcal{D}$ being the set of drivers considered for each instance. We average values over all instances corresponding to a \gls{low_avail} or \gls{high_avail} scenario and that correspond to $t^s=300$ m.}
\end{figure}
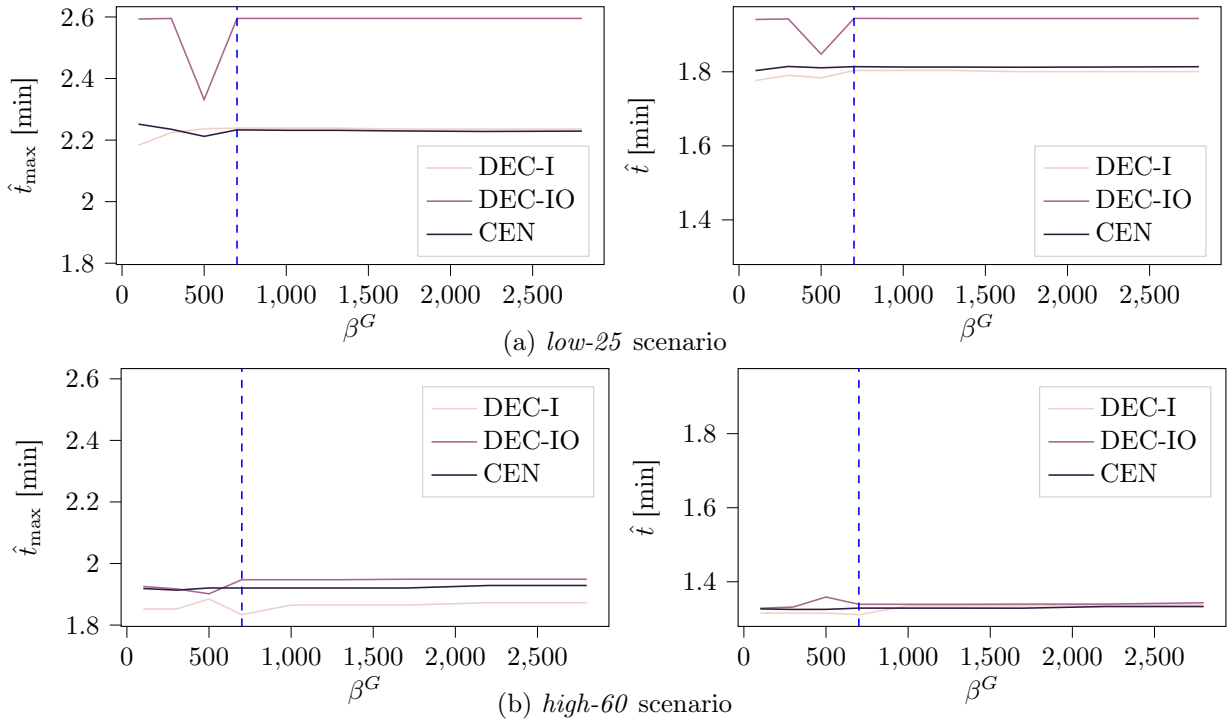

As the value of cost~$\hat{\alpha}$ depends on the value of $\beta^G$, we analyze the impact on $\beta^G$-independent metrics: 
Figure~\ref{fig:sensitivity_p} evaluates the impact of $\beta^G$ on the system success rate $\hat{\rho}$ and the worst individual success rate $\hat{\rho}_{\text{min}}$ that a driver may obtain, while Figure~\ref{fig:sensitivity_t} evaluates the impact of $\beta^G$ on the average search time $\hat{t}$ and the largest search time $\hat{t}_{\text{max}}$ that a driver may obtain in a successful search.

As can be seen, the impact of $\beta^G$ on both the success rate and the search time is low. In both cases, results show opposite effects in low- and high-availability scenarios.
While a value of $\beta^G$ lower than $700$~min yields lower search times (worst and average case) in a low-availability scenario for DEC-IO, it also yields larger search times in the worst and average case.
We note that values of $\beta^G$ larger than $700$ have only a marginal impact for all settings and set $\beta^G=700$~min.

\subsection{Collaboration in intention-sharing settings}

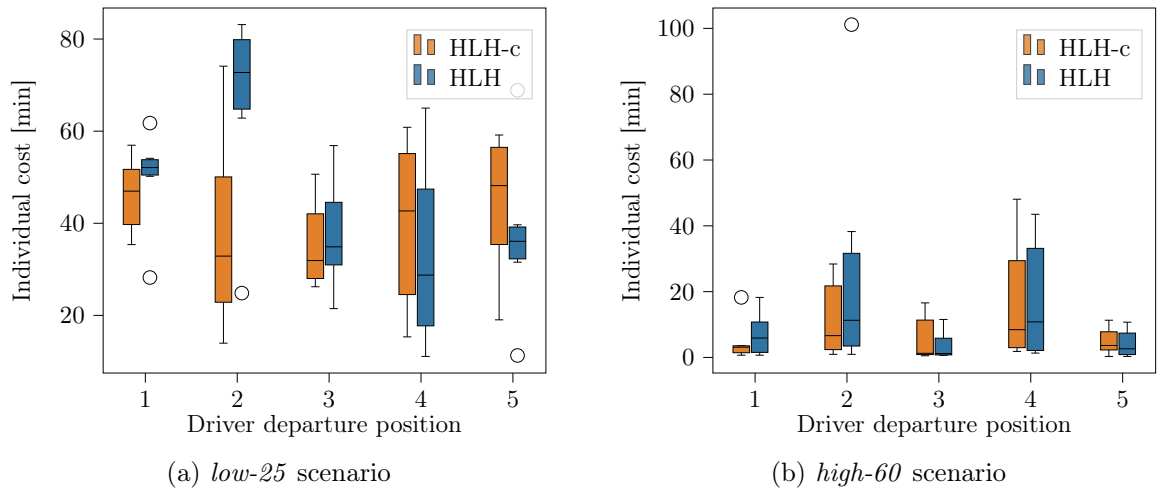
\begin{figure}[tbp]
	\begin{tabular}{c @{\qquad} c @{\qquad}}
		\scalebox{0.85}{
\begin{tikzpicture}

\definecolor{color0}{rgb}{0.194607843137255,0.453431372549019,0.632843137254902}
\definecolor{color1}{rgb}{0.881862745098039,0.505392156862745,0.173039215686275}

\begin{axis}[
legend cell align={left},
legend style={fill opacity=0.8, draw opacity=1, text opacity=1, draw=white!80!black, at={(0.97,0.75)}, anchor=south east, legend columns=1, /tikz/every even column/.append style={column sep=0.5cm}, /tikz/every odd column/.append style={column sep=0.15cm}},
tick align=outside,
tick pos=left,
x grid style={white!69.0196078431373!black},
xlabel={Driver departure position},
xmin=-0.459, xmax=4.359,
xticklabels={0,1,2,3,4,5},
xtick style={color=black},
y grid style={white!69.0196078431373!black},
ylabel={Individual cost [min]},
ymin=7.48378911959999, ymax=86.7401733684,
ytick style={color=black}
]

\draw[draw=white!23.921568627451!black,fill=color1,line width=0.3pt] (axis cs:0,0) rectangle (axis cs:0,0);
\addlegendimage{ybar,ybar legend,draw=white!23.921568627451!black,fill=color1,line width=0.3pt};
\addlegendentry{\gls{HLHC}}

\draw[draw=white!23.921568627451!black,fill=color0,line width=0.3pt] (axis cs:0,0) rectangle (axis cs:0,0);
\addlegendimage{ybar,ybar legend,draw=white!23.921568627451!black,fill=color0,line width=0.3pt};
\addlegendentry{\gls{HLH}}

\addplot [black, forget plot]
table {%
-0.15 39.721726127
-0.15 35.370203424
};
\addplot [black, forget plot]
table {%
-0.15 51.702499932
-0.15 56.944986884
};
\addplot [black, forget plot]
table {%
-0.195 35.370203424
-0.105 35.370203424
};
\addplot [black, forget plot]
table {%
-0.195 56.944986884
-0.105 56.944986884
};
\addplot [black, forget plot]
table {%
0.85 22.854823203
0.85 13.965418088
};
\addplot [black, forget plot]
table {%
0.85 50.06207271
0.85 74.1117078
};
\addplot [black, forget plot]
table {%
0.805 13.965418088
0.895 13.965418088
};
\addplot [black, forget plot]
table {%
0.805 74.1117078
0.895 74.1117078
};
\addplot [black, forget plot]
table {%
1.85 28.008389394
1.85 26.217625236
};
\addplot [black, forget plot]
table {%
1.85 42.045738813
1.85 50.65040832
};
\addplot [black, forget plot]
table {%
1.805 26.217625236
1.895 26.217625236
};
\addplot [black, forget plot]
table {%
1.805 50.65040832
1.895 50.65040832
};
\addplot [black, forget plot]
table {%
2.85 24.51907806
2.85 15.333383388
};
\addplot [black, forget plot]
table {%
2.85 55.135524213
2.85 60.838335732
};
\addplot [black, forget plot]
table {%
2.805 15.333383388
2.895 15.333383388
};
\addplot [black, forget plot]
table {%
2.805 60.838335732
2.895 60.838335732
};
\addplot [black, forget plot]
table {%
3.85 35.388330103
3.85 19.032978984
};
\addplot [black, forget plot]
table {%
3.85 56.484676893
3.85 59.162943092
};
\addplot [black, forget plot]
table {%
3.805 19.032978984
3.895 19.032978984
};
\addplot [black, forget plot]
table {%
3.805 59.162943092
3.895 59.162943092
};
\addplot [black, forget plot]
table {%
0.05 50.469011103
0.05 50.19846702
};
\addplot [black, forget plot]
table {%
0.05 53.802499932
0.05 54.102499932
};
\addplot [black, forget plot]
table {%
0.00500000000000002 50.19846702
0.095 50.19846702
};
\addplot [black, forget plot]
table {%
0.00500000000000002 54.102499932
0.095 54.102499932
};
\addplot [black, mark=o, mark size=3, mark options={solid,fill opacity=0}, only marks, forget plot]
table {%
0.05 28.220343376
0.05 61.753617212
};
\addplot [black, forget plot]
table {%
1.05 64.782635831
1.05 62.83332348
};
\addplot [black, forget plot]
table {%
1.05 79.863366678
1.05 83.137610448
};
\addplot [black, forget plot]
table {%
1.005 62.83332348
1.095 62.83332348
};
\addplot [black, forget plot]
table {%
1.005 83.137610448
1.095 83.137610448
};
\addplot [black, mark=o, mark size=3, mark options={solid,fill opacity=0}, only marks, forget plot]
table {%
1.05 24.836127372
};
\addplot [black, forget plot]
table {%
2.05 30.960720216
2.05 21.46643454
};
\addplot [black, forget plot]
table {%
2.05 44.544787869
2.05 56.862784968
};
\addplot [black, forget plot]
table {%
2.005 21.46643454
2.095 21.46643454
};
\addplot [black, forget plot]
table {%
2.005 56.862784968
2.095 56.862784968
};
\addplot [black, forget plot]
table {%
3.05 17.726061468
3.05 11.08635204
};
\addplot [black, forget plot]
table {%
3.05 47.419102187
3.05 65.010853092
};
\addplot [black, forget plot]
table {%
3.005 11.08635204
3.095 11.08635204
};
\addplot [black, forget plot]
table {%
3.005 65.010853092
3.095 65.010853092
};
\addplot [black, forget plot]
table {%
4.05 32.266983983
4.05 31.554260616
};
\addplot [black, forget plot]
table {%
4.05 39.188146971
4.05 39.664524396
};
\addplot [black, forget plot]
table {%
4.005 31.554260616
4.095 31.554260616
};
\addplot [black, forget plot]
table {%
4.005 39.664524396
4.095 39.664524396
};
\addplot [black, mark=o, mark size=3, mark options={solid,fill opacity=0}, only marks, forget plot]
table {%
4.05 11.293389904
4.05 68.86680646
};
\path [draw=black, fill=color1]
(axis cs:-0.24,39.721726127)
--(axis cs:-0.06,39.721726127)
--(axis cs:-0.06,51.702499932)
--(axis cs:-0.24,51.702499932)
--(axis cs:-0.24,39.721726127)
--cycle;
\path [draw=black, fill=color1]
(axis cs:0.76,22.854823203)
--(axis cs:0.94,22.854823203)
--(axis cs:0.94,50.06207271)
--(axis cs:0.76,50.06207271)
--(axis cs:0.76,22.854823203)
--cycle;
\path [draw=black, fill=color1]
(axis cs:1.76,28.008389394)
--(axis cs:1.94,28.008389394)
--(axis cs:1.94,42.045738813)
--(axis cs:1.76,42.045738813)
--(axis cs:1.76,28.008389394)
--cycle;
\path [draw=black, fill=color1]
(axis cs:2.76,24.51907806)
--(axis cs:2.94,24.51907806)
--(axis cs:2.94,55.135524213)
--(axis cs:2.76,55.135524213)
--(axis cs:2.76,24.51907806)
--cycle;
\path [draw=black, fill=color1]
(axis cs:3.76,35.388330103)
--(axis cs:3.94,35.388330103)
--(axis cs:3.94,56.484676893)
--(axis cs:3.76,56.484676893)
--(axis cs:3.76,35.388330103)
--cycle;
\path [draw=black, fill=color0]
(axis cs:-0.04,50.469011103)
--(axis cs:0.14,50.469011103)
--(axis cs:0.14,53.802499932)
--(axis cs:-0.04,53.802499932)
--(axis cs:-0.04,50.469011103)
--cycle;
\path [draw=black, fill=color0]
(axis cs:0.96,64.782635831)
--(axis cs:1.14,64.782635831)
--(axis cs:1.14,79.863366678)
--(axis cs:0.96,79.863366678)
--(axis cs:0.96,64.782635831)
--cycle;
\path [draw=black, fill=color0]
(axis cs:1.96,30.960720216)
--(axis cs:2.14,30.960720216)
--(axis cs:2.14,44.544787869)
--(axis cs:1.96,44.544787869)
--(axis cs:1.96,30.960720216)
--cycle;
\path [draw=black, fill=color0]
(axis cs:2.96,17.726061468)
--(axis cs:3.14,17.726061468)
--(axis cs:3.14,47.419102187)
--(axis cs:2.96,47.419102187)
--(axis cs:2.96,17.726061468)
--cycle;
\path [draw=black, fill=color0]
(axis cs:3.96,32.266983983)
--(axis cs:4.14,32.266983983)
--(axis cs:4.14,39.188146971)
--(axis cs:3.96,39.188146971)
--(axis cs:3.96,32.266983983)
--cycle;
\addplot [black, forget plot]
table {%
-0.24 46.973026366
-0.06 46.973026366
};
\addplot [black, forget plot]
table {%
0.76 32.869578204
0.94 32.869578204
};
\addplot [black, forget plot]
table {%
1.76 31.90253829
1.94 31.90253829
};
\addplot [black, forget plot]
table {%
2.76 42.67879158
2.94 42.67879158
};
\addplot [black, forget plot]
table {%
3.76 48.177371474
3.94 48.177371474
};
\addplot [black, forget plot]
table {%
-0.04 52.091571642
0.14 52.091571642
};
\addplot [black, forget plot]
table {%
0.96 72.73552123
1.14 72.73552123
};
\addplot [black, forget plot]
table {%
1.96 34.886677446
2.14 34.886677446
};
\addplot [black, forget plot]
table {%
2.96 28.744649602
3.14 28.744649602
};
\addplot [black, forget plot]
table {%
3.96 36.08208439
4.14 36.08208439
};
\end{axis}

\end{tikzpicture}} &
		\scalebox{0.85}{
\begin{tikzpicture}

\definecolor{color0}{rgb}{0.194607843137255,0.453431372549019,0.632843137254902}
\definecolor{color1}{rgb}{0.881862745098039,0.505392156862745,0.173039215686275}

\begin{axis}[
legend cell align={left},
legend style={fill opacity=0.8, draw opacity=1, text opacity=1, draw=white!80!black, at={(0.97,0.75)}, anchor=south east, legend columns=1, /tikz/every even column/.append style={column sep=0.5cm}, /tikz/every odd column/.append style={column sep=0.15cm}},
tick align=outside,
tick pos=left,
x grid style={white!69.0196078431373!black},
xlabel={Driver departure position},
xmin=-0.459, xmax=4.359,
xticklabels={0,1,2,3,4,5},
xtick style={color=black},
y grid style={white!69.0196078431373!black},
ylabel={Individual cost [min]},
ymin=-4.7349904128, ymax=106.1945202528,
ytick style={color=black}
]

\draw[draw=white!23.921568627451!black,fill=color1,line width=0.3pt] (axis cs:0,0) rectangle (axis cs:0,0);
\addlegendimage{ybar,ybar legend,draw=white!23.921568627451!black,fill=color1,line width=0.3pt};
\addlegendentry{\gls{HLHC}}

\draw[draw=white!23.921568627451!black,fill=color0,line width=0.3pt] (axis cs:0,0) rectangle (axis cs:0,0);
\addlegendimage{ybar,ybar legend,draw=white!23.921568627451!black,fill=color0,line width=0.3pt};
\addlegendentry{\gls{HLH}}

\addplot [black, forget plot]
table {%
-0.15 1.43925332
-0.15 0.678967379999999
};
\addplot [black, forget plot]
table {%
-0.15 3.523557128
-0.15 3.589925828
};
\addplot [black, forget plot]
table {%
-0.195 0.678967379999999
-0.105 0.678967379999999
};
\addplot [black, forget plot]
table {%
-0.195 3.589925828
-0.105 3.589925828
};
\addplot [black, mark=o, mark size=3, mark options={solid,fill opacity=0}, only marks, forget plot]
table {%
-0.15 18.244960064
};
\addplot [black, forget plot]
table {%
0.85 2.38047958
0.85 0.945826472000001
};
\addplot [black, forget plot]
table {%
0.85 21.738721119
0.85 28.400866452
};
\addplot [black, forget plot]
table {%
0.805 0.945826472000001
0.895 0.945826472000001
};
\addplot [black, forget plot]
table {%
0.805 28.400866452
0.895 28.400866452
};
\addplot [black, forget plot]
table {%
1.85 0.872951620999999
1.85 0.515378579999999
};
\addplot [black, forget plot]
table {%
1.85 11.34975541
1.85 16.581999756
};
\addplot [black, forget plot]
table {%
1.805 0.515378579999999
1.895 0.515378579999999
};
\addplot [black, forget plot]
table {%
1.805 16.581999756
1.895 16.581999756
};
\addplot [black, forget plot]
table {%
2.85 2.973032832
2.85 1.827763584
};
\addplot [black, forget plot]
table {%
2.85 29.405761695
2.85 48.089987892
};
\addplot [black, forget plot]
table {%
2.805 1.827763584
2.895 1.827763584
};
\addplot [black, forget plot]
table {%
2.805 48.089987892
2.895 48.089987892
};
\addplot [black, forget plot]
table {%
3.85 2.274614841
3.85 0.307260072
};
\addplot [black, forget plot]
table {%
3.85 7.797919372
3.85 11.314391276
};
\addplot [black, forget plot]
table {%
3.805 0.307260072
3.895 0.307260072
};
\addplot [black, forget plot]
table {%
3.805 11.314391276
3.895 11.314391276
};
\addplot [black, forget plot]
table {%
0.05 1.539444392
0.05 0.678967379999999
};
\addplot [black, forget plot]
table {%
0.05 10.766884609
0.05 18.244960064
};
\addplot [black, forget plot]
table {%
0.00500000000000002 0.678967379999999
0.095 0.678967379999999
};
\addplot [black, forget plot]
table {%
0.00500000000000002 18.244960064
0.095 18.244960064
};
\addplot [black, forget plot]
table {%
1.05 3.473502474
1.05 0.945826472000001
};
\addplot [black, forget plot]
table {%
1.05 31.641828615
1.05 38.255429472
};
\addplot [black, forget plot]
table {%
1.005 0.945826472000001
1.095 0.945826472000001
};
\addplot [black, forget plot]
table {%
1.005 38.255429472
1.095 38.255429472
};
\addplot [black, mark=o, mark size=3, mark options={solid,fill opacity=0}, only marks, forget plot]
table {%
1.05 101.152269768
};
\addplot [black, forget plot]
table {%
2.05 0.763540782
2.05 0.636955219999999
};
\addplot [black, forget plot]
table {%
2.05 5.83951055900001
2.05 11.546649436
};
\addplot [black, forget plot]
table {%
2.005 0.636955219999999
2.095 0.636955219999999
};
\addplot [black, forget plot]
table {%
2.005 11.546649436
2.095 11.546649436
};
\addplot [black, forget plot]
table {%
3.05 2.122250727
3.05 1.31772932
};
\addplot [black, forget plot]
table {%
3.05 33.151708853
3.05 43.51009194
};
\addplot [black, forget plot]
table {%
3.005 1.31772932
3.095 1.31772932
};
\addplot [black, forget plot]
table {%
3.005 43.51009194
3.095 43.51009194
};
\addplot [black, forget plot]
table {%
4.05 0.869383083
4.05 0.307260072
};
\addplot [black, forget plot]
table {%
4.05 7.39518082800001
4.05 10.722217012
};
\addplot [black, forget plot]
table {%
4.005 0.307260072
4.095 0.307260072
};
\addplot [black, forget plot]
table {%
4.005 10.722217012
4.095 10.722217012
};
\path [draw=black, fill=color1]
(axis cs:-0.24,1.43925332)
--(axis cs:-0.06,1.43925332)
--(axis cs:-0.06,3.523557128)
--(axis cs:-0.24,3.523557128)
--(axis cs:-0.24,1.43925332)
--cycle;
\path [draw=black, fill=color1]
(axis cs:0.76,2.38047958)
--(axis cs:0.94,2.38047958)
--(axis cs:0.94,21.738721119)
--(axis cs:0.76,21.738721119)
--(axis cs:0.76,2.38047958)
--cycle;
\path [draw=black, fill=color1]
(axis cs:1.76,0.872951620999999)
--(axis cs:1.94,0.872951620999999)
--(axis cs:1.94,11.34975541)
--(axis cs:1.76,11.34975541)
--(axis cs:1.76,0.872951620999999)
--cycle;
\path [draw=black, fill=color1]
(axis cs:2.76,2.973032832)
--(axis cs:2.94,2.973032832)
--(axis cs:2.94,29.405761695)
--(axis cs:2.76,29.405761695)
--(axis cs:2.76,2.973032832)
--cycle;
\path [draw=black, fill=color1]
(axis cs:3.76,2.274614841)
--(axis cs:3.94,2.274614841)
--(axis cs:3.94,7.797919372)
--(axis cs:3.76,7.797919372)
--(axis cs:3.76,2.274614841)
--cycle;
\path [draw=black, fill=color0]
(axis cs:-0.04,1.539444392)
--(axis cs:0.14,1.539444392)
--(axis cs:0.14,10.766884609)
--(axis cs:-0.04,10.766884609)
--(axis cs:-0.04,1.539444392)
--cycle;
\path [draw=black, fill=color0]
(axis cs:0.96,3.473502474)
--(axis cs:1.14,3.473502474)
--(axis cs:1.14,31.641828615)
--(axis cs:0.96,31.641828615)
--(axis cs:0.96,3.473502474)
--cycle;
\path [draw=black, fill=color0]
(axis cs:1.96,0.763540782)
--(axis cs:2.14,0.763540782)
--(axis cs:2.14,5.83951055900001)
--(axis cs:1.96,5.83951055900001)
--(axis cs:1.96,0.763540782)
--cycle;
\path [draw=black, fill=color0]
(axis cs:2.96,2.122250727)
--(axis cs:3.14,2.122250727)
--(axis cs:3.14,33.151708853)
--(axis cs:2.96,33.151708853)
--(axis cs:2.96,2.122250727)
--cycle;
\path [draw=black, fill=color0]
(axis cs:3.96,0.869383083)
--(axis cs:4.14,0.869383083)
--(axis cs:4.14,7.39518082800001)
--(axis cs:3.96,7.39518082800001)
--(axis cs:3.96,0.869383083)
--cycle;
\addplot [black, forget plot]
table {%
-0.24 3.124068884
-0.06 3.124068884
};
\addplot [black, forget plot]
table {%
0.76 6.623855858
0.94 6.623855858
};
\addplot [black, forget plot]
table {%
1.76 1.25826786
1.94 1.25826786
};
\addplot [black, forget plot]
table {%
2.76 8.440394838
2.94 8.440394838
};
\addplot [black, forget plot]
table {%
3.76 3.641236238
3.94 3.641236238
};
\addplot [black, forget plot]
table {%
-0.04 5.894917716
0.14 5.894917716
};
\addplot [black, forget plot]
table {%
0.96 11.291702646
1.14 11.291702646
};
\addplot [black, forget plot]
table {%
1.96 1.200414394
2.14 1.200414394
};
\addplot [black, forget plot]
table {%
2.96 10.804837456
3.14 10.804837456
};
\addplot [black, forget plot]
table {%
3.96 2.633500872
4.14 2.633500872
};
\end{axis}

\end{tikzpicture}} \\
		\small (a) \gls{low_avail} scenario & \small(b) \gls{high_avail} scenario \\
	\end{tabular}
	\caption{Comparison of individual cost $\hat{\alpha}^i$ obtained with \gls{HLH} and \gls{HLHC} in the \gls{2} setting for $t^s=1$~min \label{fig:dec_quality_1}}
	\fnote{\footnotesize Each plot shows for each driver $i$, depending on her departure position, the distribution of the realized individual cost $\hat{\alpha}^i$ over all test instances that correspond to $t^s=1$~min, $r^s\in \{100,300,700\}$ m, $\bar{S} \in \{1000,2000\}$ m, and for $N=5$ drivers.}
\end{figure}

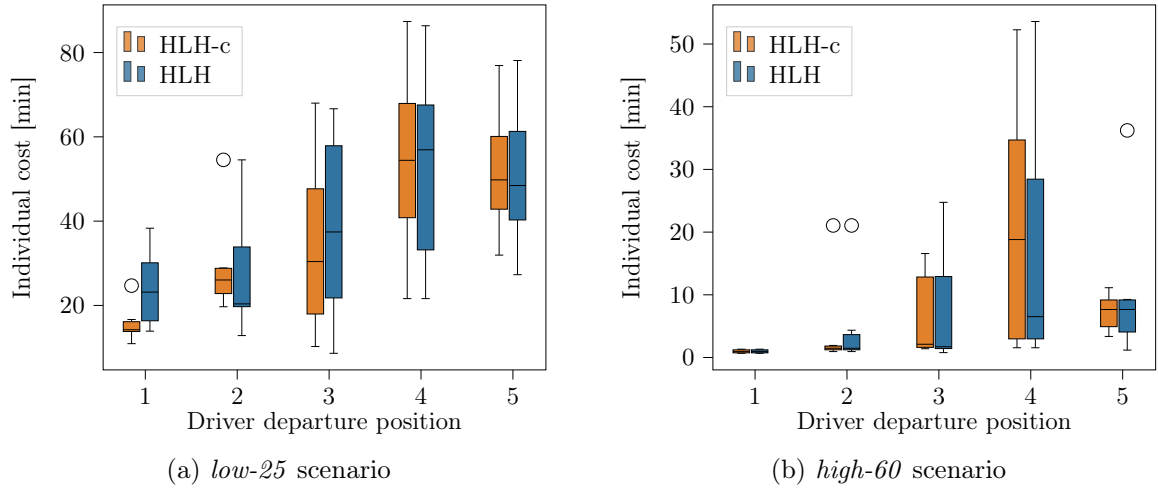
\begin{figure}[tp]
		\begin{tabular}{c @{\qquad} c @{\qquad}}
			\scalebox{0.85}{
\begin{tikzpicture}

\definecolor{color0}{rgb}{0.194607843137255,0.453431372549019,0.632843137254902}
\definecolor{color1}{rgb}{0.881862745098039,0.505392156862745,0.173039215686275}

\begin{axis}[
legend cell align={left},
legend style={fill opacity=0.8, draw opacity=1, text opacity=1, draw=white!80!black, at={(0.03,0.75)}, anchor=south west, legend columns=1, /tikz/every even column/.append style={column sep=0.5cm}, /tikz/every odd column/.append style={column sep=0.15cm}},
tick align=outside,
tick pos=left,
x grid style={white!69.0196078431373!black},
xlabel={Driver departure position},
xmin=-0.459, xmax=4.359,
xticklabels={0,1,2,3,4,5},
xtick style={color=black},
y grid style={white!69.0196078431373!black},
ylabel={Individual cost [min]},
ymin=4.70366824400001, ymax=91.307596908,
ytick style={color=black}
]

\draw[draw=white!23.921568627451!black,fill=color1,line width=0.3pt] (axis cs:0,0) rectangle (axis cs:0,0);
\addlegendimage{ybar,ybar legend,draw=white!23.921568627451!black,fill=color1,line width=0.3pt};
\addlegendentry{\gls{HLHC}}

\draw[draw=white!23.921568627451!black,fill=color0,line width=0.3pt] (axis cs:0,0) rectangle (axis cs:0,0);
\addlegendimage{ybar,ybar legend,draw=white!23.921568627451!black,fill=color0,line width=0.3pt};
\addlegendentry{\gls{HLH}}

\addplot [black, forget plot]
table {%
-0.15 13.819957122
-0.15 10.928076444
};
\addplot [black, forget plot]
table {%
-0.15 16.141464642
-0.15 16.664307164
};
\addplot [black, forget plot]
table {%
-0.195 10.928076444
-0.105 10.928076444
};
\addplot [black, forget plot]
table {%
-0.195 16.664307164
-0.105 16.664307164
};
\addplot [black, mark=o, mark size=3, mark options={solid,fill opacity=0}, only marks, forget plot]
table {%
-0.15 24.698829852
};
\addplot [black, forget plot]
table {%
0.85 22.814639949
0.85 19.688657936
};
\addplot [black, forget plot]
table {%
0.85 28.809219915
0.85 28.898118852
};
\addplot [black, forget plot]
table {%
0.805 19.688657936
0.895 19.688657936
};
\addplot [black, forget plot]
table {%
0.805 28.898118852
0.895 28.898118852
};
\addplot [black, mark=o, mark size=3, mark options={solid,fill opacity=0}, only marks, forget plot]
table {%
0.85 54.536631192
};
\addplot [black, forget plot]
table {%
1.85 17.966161709
1.85 10.24991322
};
\addplot [black, forget plot]
table {%
1.85 47.671566567
1.85 68.00150634
};
\addplot [black, forget plot]
table {%
1.805 10.24991322
1.895 10.24991322
};
\addplot [black, forget plot]
table {%
1.805 68.00150634
1.895 68.00150634
};
\addplot [black, forget plot]
table {%
2.85 40.812362618
2.85 21.619602396
};
\addplot [black, forget plot]
table {%
2.85 67.923636975
2.85 87.371054696
};
\addplot [black, forget plot]
table {%
2.805 21.619602396
2.895 21.619602396
};
\addplot [black, forget plot]
table {%
2.805 87.371054696
2.895 87.371054696
};
\addplot [black, forget plot]
table {%
3.85 42.834517777
3.85 31.937355884
};
\addplot [black, forget plot]
table {%
3.85 60.097179495
3.85 76.919150772
};
\addplot [black, forget plot]
table {%
3.805 31.937355884
3.895 31.937355884
};
\addplot [black, forget plot]
table {%
3.805 76.919150772
3.895 76.919150772
};
\addplot [black, forget plot]
table {%
0.05 16.356850182
0.05 13.898829852
};
\addplot [black, forget plot]
table {%
0.05 30.110284432
0.05 38.323002752
};
\addplot [black, forget plot]
table {%
0.00500000000000002 13.898829852
0.095 13.898829852
};
\addplot [black, forget plot]
table {%
0.00500000000000002 38.323002752
0.095 38.323002752
};
\addplot [black, forget plot]
table {%
1.05 19.707117996
1.05 12.845467836
};
\addplot [black, forget plot]
table {%
1.05 33.874924461
1.05 54.536631192
};
\addplot [black, forget plot]
table {%
1.005 12.845467836
1.095 12.845467836
};
\addplot [black, forget plot]
table {%
1.005 54.536631192
1.095 54.536631192
};
\addplot [black, forget plot]
table {%
2.05 21.782957865
2.05 8.64021045600001
};
\addplot [black, forget plot]
table {%
2.05 57.885811116
2.05 66.663056144
};
\addplot [black, forget plot]
table {%
2.005 8.64021045600001
2.095 8.64021045600001
};
\addplot [black, forget plot]
table {%
2.005 66.663056144
2.095 66.663056144
};
\addplot [black, forget plot]
table {%
3.05 33.179699724
3.05 21.619602396
};
\addplot [black, forget plot]
table {%
3.05 67.55888718
3.05 86.358293196
};
\addplot [black, forget plot]
table {%
3.005 21.619602396
3.095 21.619602396
};
\addplot [black, forget plot]
table {%
3.005 86.358293196
3.095 86.358293196
};
\addplot [black, forget plot]
table {%
4.05 40.281901365
4.05 27.288708072
};
\addplot [black, forget plot]
table {%
4.05 61.29338364
4.05 78.115307424
};
\addplot [black, forget plot]
table {%
4.005 27.288708072
4.095 27.288708072
};
\addplot [black, forget plot]
table {%
4.005 78.115307424
4.095 78.115307424
};
\path [draw=black, fill=color1]
(axis cs:-0.24,13.819957122)
--(axis cs:-0.06,13.819957122)
--(axis cs:-0.06,16.141464642)
--(axis cs:-0.24,16.141464642)
--(axis cs:-0.24,13.819957122)
--cycle;
\path [draw=black, fill=color1]
(axis cs:0.76,22.814639949)
--(axis cs:0.94,22.814639949)
--(axis cs:0.94,28.809219915)
--(axis cs:0.76,28.809219915)
--(axis cs:0.76,22.814639949)
--cycle;
\path [draw=black, fill=color1]
(axis cs:1.76,17.966161709)
--(axis cs:1.94,17.966161709)
--(axis cs:1.94,47.671566567)
--(axis cs:1.76,47.671566567)
--(axis cs:1.76,17.966161709)
--cycle;
\path [draw=black, fill=color1]
(axis cs:2.76,40.812362618)
--(axis cs:2.94,40.812362618)
--(axis cs:2.94,67.923636975)
--(axis cs:2.76,67.923636975)
--(axis cs:2.76,40.812362618)
--cycle;
\path [draw=black, fill=color1]
(axis cs:3.76,42.834517777)
--(axis cs:3.94,42.834517777)
--(axis cs:3.94,60.097179495)
--(axis cs:3.76,60.097179495)
--(axis cs:3.76,42.834517777)
--cycle;
\path [draw=black, fill=color0]
(axis cs:-0.04,16.356850182)
--(axis cs:0.14,16.356850182)
--(axis cs:0.14,30.110284432)
--(axis cs:-0.04,30.110284432)
--(axis cs:-0.04,16.356850182)
--cycle;
\path [draw=black, fill=color0]
(axis cs:0.96,19.707117996)
--(axis cs:1.14,19.707117996)
--(axis cs:1.14,33.874924461)
--(axis cs:0.96,33.874924461)
--(axis cs:0.96,19.707117996)
--cycle;
\path [draw=black, fill=color0]
(axis cs:1.96,21.782957865)
--(axis cs:2.14,21.782957865)
--(axis cs:2.14,57.885811116)
--(axis cs:1.96,57.885811116)
--(axis cs:1.96,21.782957865)
--cycle;
\path [draw=black, fill=color0]
(axis cs:2.96,33.179699724)
--(axis cs:3.14,33.179699724)
--(axis cs:3.14,67.55888718)
--(axis cs:2.96,67.55888718)
--(axis cs:2.96,33.179699724)
--cycle;
\path [draw=black, fill=color0]
(axis cs:3.96,40.281901365)
--(axis cs:4.14,40.281901365)
--(axis cs:4.14,61.29338364)
--(axis cs:3.96,61.29338364)
--(axis cs:3.96,40.281901365)
--cycle;
\addplot [black, forget plot]
table {%
-0.24 14.232426654
-0.06 14.232426654
};
\addplot [black, forget plot]
table {%
0.76 26.036692956
0.94 26.036692956
};
\addplot [black, forget plot]
table {%
1.76 30.408349252
1.94 30.408349252
};
\addplot [black, forget plot]
table {%
2.76 54.430586326
2.94 54.430586326
};
\addplot [black, forget plot]
table {%
3.76 49.789883726
3.94 49.789883726
};
\addplot [black, forget plot]
table {%
-0.04 23.164466438
0.14 23.164466438
};
\addplot [black, forget plot]
table {%
0.96 20.367443562
1.14 20.367443562
};
\addplot [black, forget plot]
table {%
1.96 37.436362842
2.14 37.436362842
};
\addplot [black, forget plot]
table {%
2.96 56.932603212
3.14 56.932603212
};
\addplot [black, forget plot]
table {%
3.96 48.453389412
4.14 48.453389412
};
\end{axis}

\end{tikzpicture}} &
			\scalebox{0.85}{
\begin{tikzpicture}

\definecolor{color0}{rgb}{0.194607843137255,0.453431372549019,0.632843137254902}
\definecolor{color1}{rgb}{0.881862745098039,0.505392156862745,0.173039215686275}

\begin{axis}[
legend cell align={left},
legend style={fill opacity=0.8, draw opacity=1, text opacity=1, draw=white!80!black, at={(0.03,0.75)}, anchor=south west, legend columns=1, /tikz/every even column/.append style={column sep=0.5cm}, /tikz/every odd column/.append style={column sep=0.15cm}},
tick align=outside,
tick pos=left,
x grid style={white!69.0196078431373!black},
xlabel={Driver departure position},
xmin=-0.459, xmax=4.359,
xticklabels={0,1,2,3,4,5},
xtick style={color=black},
y grid style={white!69.0196078431373!black},
ylabel={Individual cost [min]},
ymin=-1.9776892912, ymax=56.2400046432,
ytick style={color=black}
]

\draw[draw=white!23.921568627451!black,fill=color1,line width=0.3pt] (axis cs:0,0) rectangle (axis cs:0,0);
\addlegendimage{ybar,ybar legend,draw=white!23.921568627451!black,fill=color1,line width=0.3pt};
\addlegendentry{\gls{HLHC}}

\draw[draw=white!23.921568627451!black,fill=color0,line width=0.3pt] (axis cs:0,0) rectangle (axis cs:0,0);
\addlegendimage{ybar,ybar legend,draw=white!23.921568627451!black,fill=color0,line width=0.3pt};
\addlegendentry{\gls{HLH}}

\addplot [black, forget plot]
table {%
-0.15 0.742736615999999
-0.15 0.668569523999999
};
\addplot [black, forget plot]
table {%
-0.15 1.223859339
-0.15 1.326697352
};
\addplot [black, forget plot]
table {%
-0.195 0.668569523999999
-0.105 0.668569523999999
};
\addplot [black, forget plot]
table {%
-0.195 1.326697352
-0.105 1.326697352
};
\addplot [black, forget plot]
table {%
0.85 1.228564986
0.85 0.945826472000001
};
\addplot [black, forget plot]
table {%
0.85 1.830629286
0.85 1.926178512
};
\addplot [black, forget plot]
table {%
0.805 0.945826472000001
0.895 0.945826472000001
};
\addplot [black, forget plot]
table {%
0.805 1.926178512
0.895 1.926178512
};
\addplot [black, mark=o, mark size=3, mark options={solid,fill opacity=0}, only marks, forget plot]
table {%
0.85 21.060212724
};
\addplot [black, forget plot]
table {%
1.85 1.61455549
1.85 1.39525946
};
\addplot [black, forget plot]
table {%
1.85 12.838292128
1.85 16.581999756
};
\addplot [black, forget plot]
table {%
1.805 1.39525946
1.895 1.39525946
};
\addplot [black, forget plot]
table {%
1.805 16.581999756
1.895 16.581999756
};
\addplot [black, forget plot]
table {%
2.85 2.973032832
2.85 1.554725484
};
\addplot [black, forget plot]
table {%
2.85 34.695612552
2.85 52.269605928
};
\addplot [black, forget plot]
table {%
2.805 1.554725484
2.895 1.554725484
};
\addplot [black, forget plot]
table {%
2.805 52.269605928
2.895 52.269605928
};
\addplot [black, forget plot]
table {%
3.85 4.916946348
3.85 3.3594273
};
\addplot [black, forget plot]
table {%
3.85 9.179808713
3.85 11.131506212
};
\addplot [black, forget plot]
table {%
3.805 3.3594273
3.895 3.3594273
};
\addplot [black, forget plot]
table {%
3.805 11.131506212
3.895 11.131506212
};
\addplot [black, forget plot]
table {%
0.05 0.742736615999999
0.05 0.668569523999999
};
\addplot [black, forget plot]
table {%
0.05 1.223859339
0.05 1.326697352
};
\addplot [black, forget plot]
table {%
0.00500000000000002 0.668569523999999
0.095 0.668569523999999
};
\addplot [black, forget plot]
table {%
0.00500000000000002 1.326697352
0.095 1.326697352
};
\addplot [black, forget plot]
table {%
1.05 1.228564986
1.05 0.945826472000001
};
\addplot [black, forget plot]
table {%
1.05 3.648864717
1.05 4.35049242
};
\addplot [black, forget plot]
table {%
1.005 0.945826472000001
1.095 0.945826472000001
};
\addplot [black, forget plot]
table {%
1.005 4.35049242
1.095 4.35049242
};
\addplot [black, mark=o, mark size=3, mark options={solid,fill opacity=0}, only marks, forget plot]
table {%
1.05 21.060212724
};
\addplot [black, forget plot]
table {%
2.05 1.425263333
2.05 0.779554931999999
};
\addplot [black, forget plot]
table {%
2.05 12.914599093
2.05 24.749404116
};
\addplot [black, forget plot]
table {%
2.005 0.779554931999999
2.095 0.779554931999999
};
\addplot [black, forget plot]
table {%
2.005 24.749404116
2.095 24.749404116
};
\addplot [black, forget plot]
table {%
3.05 2.973032832
3.05 1.554725484
};
\addplot [black, forget plot]
table {%
3.05 28.444759883
3.05 53.593745828
};
\addplot [black, forget plot]
table {%
3.005 1.554725484
3.095 1.554725484
};
\addplot [black, forget plot]
table {%
3.005 53.593745828
3.095 53.593745828
};
\addplot [black, forget plot]
table {%
4.05 4.067466477
4.05 1.179237368
};
\addplot [black, forget plot]
table {%
4.05 9.179808713
4.05 9.26595611199999
};
\addplot [black, forget plot]
table {%
4.005 1.179237368
4.095 1.179237368
};
\addplot [black, forget plot]
table {%
4.005 9.26595611199999
4.095 9.26595611199999
};
\addplot [black, mark=o, mark size=3, mark options={solid,fill opacity=0}, only marks, forget plot]
table {%
4.05 36.219555484
};
\path [draw=black, fill=color1]
(axis cs:-0.24,0.742736615999999)
--(axis cs:-0.06,0.742736615999999)
--(axis cs:-0.06,1.223859339)
--(axis cs:-0.24,1.223859339)
--(axis cs:-0.24,0.742736615999999)
--cycle;
\path [draw=black, fill=color1]
(axis cs:0.76,1.228564986)
--(axis cs:0.94,1.228564986)
--(axis cs:0.94,1.830629286)
--(axis cs:0.76,1.830629286)
--(axis cs:0.76,1.228564986)
--cycle;
\path [draw=black, fill=color1]
(axis cs:1.76,1.61455549)
--(axis cs:1.94,1.61455549)
--(axis cs:1.94,12.838292128)
--(axis cs:1.76,12.838292128)
--(axis cs:1.76,1.61455549)
--cycle;
\path [draw=black, fill=color1]
(axis cs:2.76,2.973032832)
--(axis cs:2.94,2.973032832)
--(axis cs:2.94,34.695612552)
--(axis cs:2.76,34.695612552)
--(axis cs:2.76,2.973032832)
--cycle;
\path [draw=black, fill=color1]
(axis cs:3.76,4.916946348)
--(axis cs:3.94,4.916946348)
--(axis cs:3.94,9.179808713)
--(axis cs:3.76,9.179808713)
--(axis cs:3.76,4.916946348)
--cycle;
\path [draw=black, fill=color0]
(axis cs:-0.04,0.742736615999999)
--(axis cs:0.14,0.742736615999999)
--(axis cs:0.14,1.223859339)
--(axis cs:-0.04,1.223859339)
--(axis cs:-0.04,0.742736615999999)
--cycle;
\path [draw=black, fill=color0]
(axis cs:0.96,1.228564986)
--(axis cs:1.14,1.228564986)
--(axis cs:1.14,3.648864717)
--(axis cs:0.96,3.648864717)
--(axis cs:0.96,1.228564986)
--cycle;
\path [draw=black, fill=color0]
(axis cs:1.96,1.425263333)
--(axis cs:2.14,1.425263333)
--(axis cs:2.14,12.914599093)
--(axis cs:1.96,12.914599093)
--(axis cs:1.96,1.425263333)
--cycle;
\path [draw=black, fill=color0]
(axis cs:2.96,2.973032832)
--(axis cs:3.14,2.973032832)
--(axis cs:3.14,28.444759883)
--(axis cs:2.96,28.444759883)
--(axis cs:2.96,2.973032832)
--cycle;
\path [draw=black, fill=color0]
(axis cs:3.96,4.067466477)
--(axis cs:4.14,4.067466477)
--(axis cs:4.14,9.179808713)
--(axis cs:3.96,9.179808713)
--(axis cs:3.96,4.067466477)
--cycle;
\addplot [black, forget plot]
table {%
-0.24 0.939243251999998
-0.06 0.939243251999998
};
\addplot [black, forget plot]
table {%
0.76 1.47289824
0.94 1.47289824
};
\addplot [black, forget plot]
table {%
1.76 2.11135642
1.94 2.11135642
};
\addplot [black, forget plot]
table {%
2.76 18.812762472
2.94 18.812762472
};
\addplot [black, forget plot]
table {%
3.76 7.653209044
3.94 7.653209044
};
\addplot [black, forget plot]
table {%
-0.04 0.939243251999998
0.14 0.939243251999998
};
\addplot [black, forget plot]
table {%
0.96 1.47289824
1.14 1.47289824
};
\addplot [black, forget plot]
table {%
1.96 1.713836028
2.14 1.713836028
};
\addplot [black, forget plot]
table {%
2.96 6.518391214
3.14 6.518391214
};
\addplot [black, forget plot]
table {%
3.96 7.653209044
4.14 7.653209044
};
\end{axis}

\end{tikzpicture}} \\
			\small (a) \gls{low_avail} scenario & \small(b) \gls{high_avail} scenario \\
		\end{tabular}
	\caption{Comparison of individual cost $\hat{\alpha}^i$ obtained with \gls{HLH} and \gls{HLHC} in the \gls{2} setting for $t^s=5$~min \label{fig:dec_quality_5}}
	\fnote{\footnotesize Each plot shows for each driver $i$, depending on her departure position, the distribution of the realized individual cost $\hat{\alpha}^i$ over all test instances that correspond to $t^s=5$~min, $r^s\in \{100,300,700\}$ m, $\bar{S} \in \{1000,2000\}$ m, and for $N=5$ drivers.}
\end{figure}

\begin{figure}[tp]
		\begin{tabular}{c @{\qquad} c @{\qquad}}
			\scalebox{0.85}{
\begin{tikzpicture}

\definecolor{color0}{rgb}{0.194607843137255,0.453431372549019,0.632843137254902}
\definecolor{color1}{rgb}{0.881862745098039,0.505392156862745,0.173039215686275}

\begin{axis}[
legend cell align={left},
legend style={fill opacity=0.8, draw opacity=1, text opacity=1, draw=white!80!black, at={(0.03,0.75)}, anchor=south west, legend columns=1, /tikz/every even column/.append style={column sep=0.5cm}, /tikz/every odd column/.append style={column sep=0.15cm}},
tick align=outside,
tick pos=left,
x grid style={white!69.0196078431373!black},
xlabel={Driver departure position},
xmin=-0.459, xmax=4.359,
xtick style={color=black},
xticklabels={0,1,2,3,4,5},
y grid style={white!69.0196078431373!black},
ylabel={Individual cost [min]},
ymin=0.679017274400003, ymax=98.8048590136,
ytick style={color=black}
]

\draw[draw=white!23.921568627451!black,fill=color1,line width=0.3pt] (axis cs:0,0) rectangle (axis cs:0,0);
\addlegendimage{ybar,ybar legend,draw=white!23.921568627451!black,fill=color1,line width=0.3pt};
\addlegendentry{\gls{HLHC}}

\draw[draw=white!23.921568627451!black,fill=color0,line width=0.3pt] (axis cs:0,0) rectangle (axis cs:0,0);
\addlegendimage{ybar,ybar legend,draw=white!23.921568627451!black,fill=color0,line width=0.3pt};
\addlegendentry{\gls{HLH}}

\addplot [black, forget plot]
table {%
-0.15 5.289710448
-0.15 5.139282808
};
\addplot [black, forget plot]
table {%
-0.15 7.850046768
-0.15 8.49884602800001
};
\addplot [black, forget plot]
table {%
-0.195 5.139282808
-0.105 5.139282808
};
\addplot [black, forget plot]
table {%
-0.195 8.49884602800001
-0.105 8.49884602800001
};
\addplot [black, forget plot]
table {%
0.85 18.420335237
0.85 14.996013696
};
\addplot [black, forget plot]
table {%
0.85 29.130949953
0.85 37.4536192
};
\addplot [black, forget plot]
table {%
0.805 14.996013696
0.895 14.996013696
};
\addplot [black, forget plot]
table {%
0.805 37.4536192
0.895 37.4536192
};
\addplot [black, forget plot]
table {%
1.85 24.821758314
1.85 19.353478056
};
\addplot [black, forget plot]
table {%
1.85 41.856709314
1.85 49.935222732
};
\addplot [black, forget plot]
table {%
1.805 19.353478056
1.895 19.353478056
};
\addplot [black, forget plot]
table {%
1.805 49.935222732
1.895 49.935222732
};
\addplot [black, forget plot]
table {%
2.85 39.299545665
2.85 21.619602396
};
\addplot [black, forget plot]
table {%
2.85 79.437215466
2.85 91.684394748
};
\addplot [black, forget plot]
table {%
2.805 21.619602396
2.895 21.619602396
};
\addplot [black, forget plot]
table {%
2.805 91.684394748
2.895 91.684394748
};
\addplot [black, forget plot]
table {%
3.85 57.354654378
3.85 41.955006812
};
\addplot [black, forget plot]
table {%
3.85 81.276476574
3.85 94.34459348
};
\addplot [black, forget plot]
table {%
3.805 41.955006812
3.895 41.955006812
};
\addplot [black, forget plot]
table {%
3.805 94.34459348
3.895 94.34459348
};
\addplot [black, forget plot]
table {%
0.05 5.289710448
0.05 5.139282808
};
\addplot [black, forget plot]
table {%
0.05 7.850046768
0.05 8.49884602800001
};
\addplot [black, forget plot]
table {%
0.00500000000000002 5.139282808
0.095 5.139282808
};
\addplot [black, forget plot]
table {%
0.00500000000000002 8.49884602800001
0.095 8.49884602800001
};
\addplot [black, forget plot]
table {%
1.05 18.420335237
1.05 14.996013696
};
\addplot [black, forget plot]
table {%
1.05 34.318690959
1.05 54.536631192
};
\addplot [black, forget plot]
table {%
1.005 14.996013696
1.095 14.996013696
};
\addplot [black, forget plot]
table {%
1.005 54.536631192
1.095 54.536631192
};
\addplot [black, forget plot]
table {%
2.05 20.570848416
2.05 17.116102804
};
\addplot [black, forget plot]
table {%
2.05 41.856709314
2.05 49.935222732
};
\addplot [black, forget plot]
table {%
2.005 17.116102804
2.095 17.116102804
};
\addplot [black, forget plot]
table {%
2.005 49.935222732
2.095 49.935222732
};
\addplot [black, forget plot]
table {%
3.05 40.199545665
3.05 21.619602396
};
\addplot [black, forget plot]
table {%
3.05 80.387910576
3.05 91.684394748
};
\addplot [black, forget plot]
table {%
3.005 21.619602396
3.095 21.619602396
};
\addplot [black, forget plot]
table {%
3.005 91.684394748
3.095 91.684394748
};
\addplot [black, forget plot]
table {%
4.05 53.135189826
4.05 41.955006812
};
\addplot [black, forget plot]
table {%
4.05 73.325800662
4.05 94.34459348
};
\addplot [black, forget plot]
table {%
4.005 41.955006812
4.095 41.955006812
};
\addplot [black, forget plot]
table {%
4.005 94.34459348
4.095 94.34459348
};
\path [draw=black, fill=color1]
(axis cs:-0.24,5.289710448)
--(axis cs:-0.06,5.289710448)
--(axis cs:-0.06,7.850046768)
--(axis cs:-0.24,7.850046768)
--(axis cs:-0.24,5.289710448)
--cycle;
\path [draw=black, fill=color1]
(axis cs:0.76,18.420335237)
--(axis cs:0.94,18.420335237)
--(axis cs:0.94,29.130949953)
--(axis cs:0.76,29.130949953)
--(axis cs:0.76,18.420335237)
--cycle;
\path [draw=black, fill=color1]
(axis cs:1.76,24.821758314)
--(axis cs:1.94,24.821758314)
--(axis cs:1.94,41.856709314)
--(axis cs:1.76,41.856709314)
--(axis cs:1.76,24.821758314)
--cycle;
\path [draw=black, fill=color1]
(axis cs:2.76,39.299545665)
--(axis cs:2.94,39.299545665)
--(axis cs:2.94,79.437215466)
--(axis cs:2.76,79.437215466)
--(axis cs:2.76,39.299545665)
--cycle;
\path [draw=black, fill=color1]
(axis cs:3.76,57.354654378)
--(axis cs:3.94,57.354654378)
--(axis cs:3.94,81.276476574)
--(axis cs:3.76,81.276476574)
--(axis cs:3.76,57.354654378)
--cycle;
\path [draw=black, fill=color0]
(axis cs:-0.04,5.289710448)
--(axis cs:0.14,5.289710448)
--(axis cs:0.14,7.850046768)
--(axis cs:-0.04,7.850046768)
--(axis cs:-0.04,5.289710448)
--cycle;
\path [draw=black, fill=color0]
(axis cs:0.96,18.420335237)
--(axis cs:1.14,18.420335237)
--(axis cs:1.14,34.318690959)
--(axis cs:0.96,34.318690959)
--(axis cs:0.96,18.420335237)
--cycle;
\path [draw=black, fill=color0]
(axis cs:1.96,20.570848416)
--(axis cs:2.14,20.570848416)
--(axis cs:2.14,41.856709314)
--(axis cs:1.96,41.856709314)
--(axis cs:1.96,20.570848416)
--cycle;
\path [draw=black, fill=color0]
(axis cs:2.96,40.199545665)
--(axis cs:3.14,40.199545665)
--(axis cs:3.14,80.387910576)
--(axis cs:2.96,80.387910576)
--(axis cs:2.96,40.199545665)
--cycle;
\path [draw=black, fill=color0]
(axis cs:3.96,53.135189826)
--(axis cs:4.14,53.135189826)
--(axis cs:4.14,73.325800662)
--(axis cs:3.96,73.325800662)
--(axis cs:3.96,53.135189826)
--cycle;
\addplot [black, forget plot]
table {%
-0.24 6.30009825
-0.06 6.30009825
};
\addplot [black, forget plot]
table {%
0.76 22.656525826
0.94 22.656525826
};
\addplot [black, forget plot]
table {%
1.76 27.76840221
1.94 27.76840221
};
\addplot [black, forget plot]
table {%
2.76 52.821230478
2.94 52.821230478
};
\addplot [black, forget plot]
table {%
3.76 69.038260266
3.94 69.038260266
};
\addplot [black, forget plot]
table {%
-0.04 6.30009825
0.14 6.30009825
};
\addplot [black, forget plot]
table {%
0.96 22.656525826
1.14 22.656525826
};
\addplot [black, forget plot]
table {%
1.96 26.570804574
2.14 26.570804574
};
\addplot [black, forget plot]
table {%
2.96 54.621230478
3.14 54.621230478
};
\addplot [black, forget plot]
table {%
3.96 66.46088304
4.14 66.46088304
};
\end{axis}

\end{tikzpicture}} &
			\scalebox{0.85}{
\begin{tikzpicture}

\definecolor{color0}{rgb}{0.194607843137255,0.453431372549019,0.632843137254902}
\definecolor{color1}{rgb}{0.881862745098039,0.505392156862745,0.173039215686275}

\begin{axis}[
legend cell align={left},
legend style={fill opacity=0.8, draw opacity=1, text opacity=1, draw=white!80!black, at={(0.03,0.75)}, anchor=south west, legend columns=1, /tikz/every even column/.append style={column sep=0.5cm}, /tikz/every odd column/.append style={column sep=0.15cm}},
tick align=outside,
tick pos=left,
x grid style={white!69.0196078431373!black},
xlabel={Driver departure position},
xmin=-0.459, xmax=4.359,
xticklabels={0,1,2,3,4,5},
xtick style={color=black},
y grid style={white!69.0196078431373!black},
ylabel={Individual cost [min]},
ymin=-1.8778429638, ymax=54.1432317678,
ytick style={color=black}
]

\draw[draw=white!23.921568627451!black,fill=color1,line width=0.3pt] (axis cs:0,0) rectangle (axis cs:0,0);
\addlegendimage{ybar,ybar legend,draw=white!23.921568627451!black,fill=color1,line width=0.3pt};
\addlegendentry{\gls{HLHC}}

\draw[draw=white!23.921568627451!black,fill=color0,line width=0.3pt] (axis cs:0,0) rectangle (axis cs:0,0);
\addlegendimage{ybar,ybar legend,draw=white!23.921568627451!black,fill=color0,line width=0.3pt};
\addlegendentry{\gls{HLH}}

\addplot [black, forget plot]
table {%
-0.15 0.742736615999999
-0.15 0.668569523999999
};
\addplot [black, forget plot]
table {%
-0.15 1.223859339
-0.15 1.326697352
};
\addplot [black, forget plot]
table {%
-0.195 0.668569523999999
-0.105 0.668569523999999
};
\addplot [black, forget plot]
table {%
-0.195 1.326697352
-0.105 1.326697352
};
\addplot [black, forget plot]
table {%
0.85 1.228564986
0.85 0.945826472000001
};
\addplot [black, forget plot]
table {%
0.85 1.732114629
0.85 1.794825636
};
\addplot [black, forget plot]
table {%
0.805 0.945826472000001
0.895 0.945826472000001
};
\addplot [black, forget plot]
table {%
0.805 1.794825636
0.895 1.794825636
};
\addplot [black, mark=o, mark size=3, mark options={solid,fill opacity=0}, only marks, forget plot]
table {%
0.85 9.06021272400001
};
\addplot [black, forget plot]
table {%
1.85 1.078528118
1.85 0.694942819999999
};
\addplot [black, forget plot]
table {%
1.85 3.630629693
1.85 4.479152076
};
\addplot [black, forget plot]
table {%
1.805 0.694942819999999
1.895 0.694942819999999
};
\addplot [black, forget plot]
table {%
1.805 4.479152076
1.895 4.479152076
};
\addplot [black, forget plot]
table {%
2.85 2.973032832
2.85 1.554725484
};
\addplot [black, forget plot]
table {%
2.85 13.674012276
2.85 13.957659792
};
\addplot [black, forget plot]
table {%
2.805 1.554725484
2.895 1.554725484
};
\addplot [black, forget plot]
table {%
2.805 13.957659792
2.895 13.957659792
};
\addplot [black, mark=o, mark size=3, mark options={solid,fill opacity=0}, only marks, forget plot]
table {%
2.85 51.59681928
};
\addplot [black, forget plot]
table {%
3.85 3.416588727
3.85 1.179237368
};
\addplot [black, forget plot]
table {%
3.85 22.142661495
3.85 34.237952292
};
\addplot [black, forget plot]
table {%
3.805 1.179237368
3.895 1.179237368
};
\addplot [black, forget plot]
table {%
3.805 34.237952292
3.895 34.237952292
};
\addplot [black, forget plot]
table {%
0.05 0.742736615999999
0.05 0.668569523999999
};
\addplot [black, forget plot]
table {%
0.05 1.223859339
0.05 1.326697352
};
\addplot [black, forget plot]
table {%
0.00500000000000002 0.668569523999999
0.095 0.668569523999999
};
\addplot [black, forget plot]
table {%
0.00500000000000002 1.326697352
0.095 1.326697352
};
\addplot [black, forget plot]
table {%
1.05 1.228564986
1.05 0.945826472000001
};
\addplot [black, forget plot]
table {%
1.05 1.732114629
1.05 1.794825636
};
\addplot [black, forget plot]
table {%
1.005 0.945826472000001
1.095 0.945826472000001
};
\addplot [black, forget plot]
table {%
1.005 1.794825636
1.095 1.794825636
};
\addplot [black, mark=o, mark size=3, mark options={solid,fill opacity=0}, only marks, forget plot]
table {%
1.05 21.060212724
};
\addplot [black, forget plot]
table {%
2.05 1.078528118
2.05 0.842284343999999
};
\addplot [black, forget plot]
table {%
2.05 3.630629693
2.05 4.479152076
};
\addplot [black, forget plot]
table {%
2.005 0.842284343999999
2.095 0.842284343999999
};
\addplot [black, forget plot]
table {%
2.005 4.479152076
2.095 4.479152076
};
\addplot [black, forget plot]
table {%
3.05 2.973032832
3.05 1.554725484
};
\addplot [black, forget plot]
table {%
3.05 18.174012276
3.05 19.957659792
};
\addplot [black, forget plot]
table {%
3.005 1.554725484
3.095 1.554725484
};
\addplot [black, forget plot]
table {%
3.005 19.957659792
3.095 19.957659792
};
\addplot [black, mark=o, mark size=3, mark options={solid,fill opacity=0}, only marks, forget plot]
table {%
3.05 50.933552604
};
\addplot [black, forget plot]
table {%
4.05 3.416588727
4.05 1.179237368
};
\addplot [black, forget plot]
table {%
4.05 21.826291775
4.05 30.115925824
};
\addplot [black, forget plot]
table {%
4.005 1.179237368
4.095 1.179237368
};
\addplot [black, forget plot]
table {%
4.005 30.115925824
4.095 30.115925824
};
\path [draw=black, fill=color1]
(axis cs:-0.24,0.742736615999999)
--(axis cs:-0.06,0.742736615999999)
--(axis cs:-0.06,1.223859339)
--(axis cs:-0.24,1.223859339)
--(axis cs:-0.24,0.742736615999999)
--cycle;
\path [draw=black, fill=color1]
(axis cs:0.76,1.228564986)
--(axis cs:0.94,1.228564986)
--(axis cs:0.94,1.732114629)
--(axis cs:0.76,1.732114629)
--(axis cs:0.76,1.228564986)
--cycle;
\path [draw=black, fill=color1]
(axis cs:1.76,1.078528118)
--(axis cs:1.94,1.078528118)
--(axis cs:1.94,3.630629693)
--(axis cs:1.76,3.630629693)
--(axis cs:1.76,1.078528118)
--cycle;
\path [draw=black, fill=color1]
(axis cs:2.76,2.973032832)
--(axis cs:2.94,2.973032832)
--(axis cs:2.94,13.674012276)
--(axis cs:2.76,13.674012276)
--(axis cs:2.76,2.973032832)
--cycle;
\path [draw=black, fill=color1]
(axis cs:3.76,3.416588727)
--(axis cs:3.94,3.416588727)
--(axis cs:3.94,22.142661495)
--(axis cs:3.76,22.142661495)
--(axis cs:3.76,3.416588727)
--cycle;
\path [draw=black, fill=color0]
(axis cs:-0.04,0.742736615999999)
--(axis cs:0.14,0.742736615999999)
--(axis cs:0.14,1.223859339)
--(axis cs:-0.04,1.223859339)
--(axis cs:-0.04,0.742736615999999)
--cycle;
\path [draw=black, fill=color0]
(axis cs:0.96,1.228564986)
--(axis cs:1.14,1.228564986)
--(axis cs:1.14,1.732114629)
--(axis cs:0.96,1.732114629)
--(axis cs:0.96,1.228564986)
--cycle;
\path [draw=black, fill=color0]
(axis cs:1.96,1.078528118)
--(axis cs:2.14,1.078528118)
--(axis cs:2.14,3.630629693)
--(axis cs:1.96,3.630629693)
--(axis cs:1.96,1.078528118)
--cycle;
\path [draw=black, fill=color0]
(axis cs:2.96,2.973032832)
--(axis cs:3.14,2.973032832)
--(axis cs:3.14,18.174012276)
--(axis cs:2.96,18.174012276)
--(axis cs:2.96,2.973032832)
--cycle;
\path [draw=black, fill=color0]
(axis cs:3.96,3.416588727)
--(axis cs:4.14,3.416588727)
--(axis cs:4.14,21.826291775)
--(axis cs:3.96,21.826291775)
--(axis cs:3.96,3.416588727)
--cycle;
\addplot [black, forget plot]
table {%
-0.24 0.939243251999998
-0.06 0.939243251999998
};
\addplot [black, forget plot]
table {%
0.76 1.47289824
0.94 1.47289824
};
\addplot [black, forget plot]
table {%
1.76 1.455267206
1.94 1.455267206
};
\addplot [black, forget plot]
table {%
2.76 7.92832458
2.94 7.92832458
};
\addplot [black, forget plot]
table {%
3.76 9.13060131
3.94 9.13060131
};
\addplot [black, forget plot]
table {%
-0.04 0.939243251999998
0.14 0.939243251999998
};
\addplot [black, forget plot]
table {%
0.96 1.47289824
1.14 1.47289824
};
\addplot [black, forget plot]
table {%
1.96 1.455267206
2.14 1.455267206
};
\addplot [black, forget plot]
table {%
2.96 7.92832458
3.14 7.92832458
};
\addplot [black, forget plot]
table {%
3.96 8.49786187
4.14 8.49786187
};
\end{axis}

\end{tikzpicture}} \\
			\small (a) \gls{low_avail} scenario & \small(b) \gls{high_avail} scenario \\
		\end{tabular}
	\caption{Comparison of individual cost $\hat{\alpha}^i$ obtained with \gls{HLH} and \gls{HLHC} in the \gls{2} setting for $t^s=15$~min \label{fig:dec_quality_15}}
	\fnote{\footnotesize Each plot shows for each driver $i$, depending on her departure position, the distribution of the realized individual cost $\hat{\alpha}^i$ over all test instances that correspond to $t^s=15$~min, $r^s\in \{100,300,700\}$ m, $\bar{S} \in \{1000,2000\}$ m, and for $N=5$ drivers.}
\end{figure}
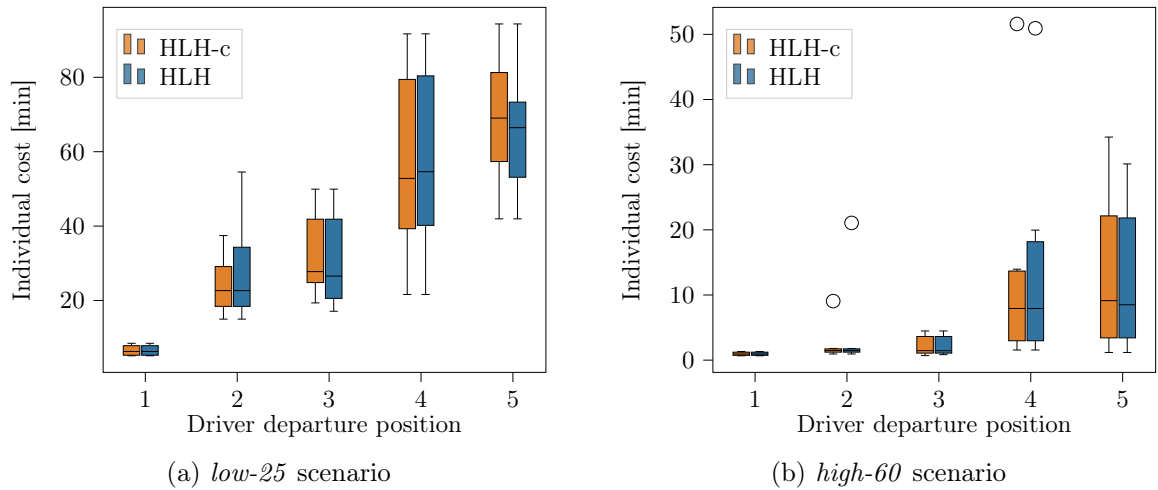

Similar to Figure~\ref{fig:dec_quality}, Figures \ref{fig:dec_quality_1},\ref{fig:dec_quality_5}\&\ref{fig:dec_quality_15} compare the individual driver's costs depending on their departure order, obtained with the collaborative procedure (\gls{HLHC}) and the non-collaborative procedure (\gls{HLH}) for $t^s=1$~min, respectively $t^s=5$~min, and $t^s=15$~min, in low and high-availability scenarios for instances with $N=5$ drivers.

As can be seen, collaboration significantly increases the solution's fairness for $t^s=1$~min, in particular in low-availability scenarios. However for a larger $t^s$, collaboration affects marginally the individual solution that a driver may obtain, independent of her departure position.

\subsection{General performance evaluation}
\begin{figure}[tbp]
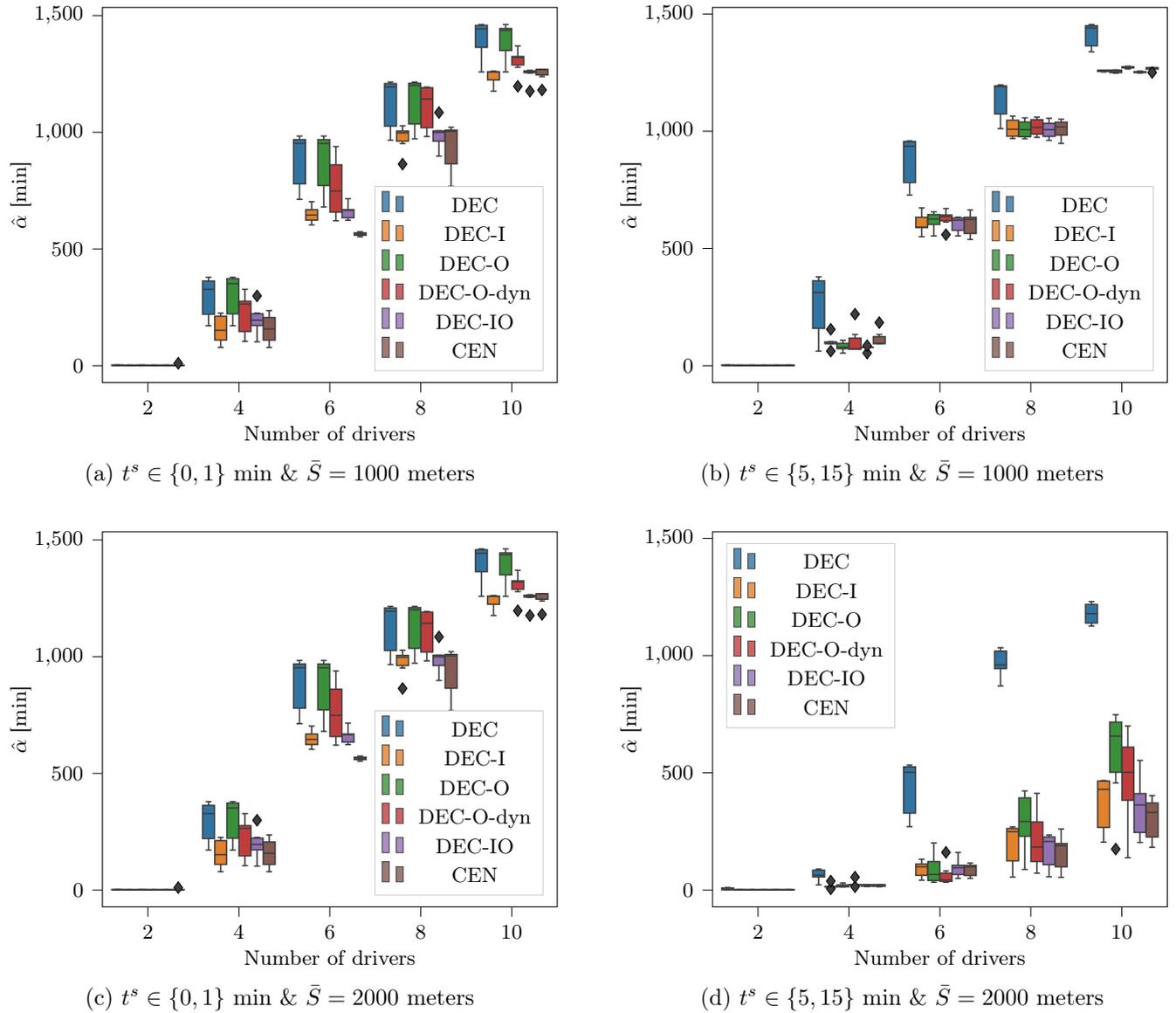

	\begin{tabular}{x{0.527\textwidth}x{.527\textwidth}}
			\input{chapters/figures/st_0_1_60_1000} &
			\hspace{-0.5cm} \input{chapters/figures/st_5_15_60_1000} \\
			\small (a) $t^s\in \{0,1\}$~min \& $\bar{S}=1000$ meters & \small (b) $t^s \in \{5,15\} $~min \& $\bar{S}=1000$ meters \\
			& \\
			\input{chapters/figures/st_0_1_60_2000} &
			\hspace{-0.5cm} \input{chapters/figures/st_5_15_60_2000} \\
			\small (c) $t^s\in \{0,1\}$~min \& $\bar{S}=2000$ meters & \small (d) $t^s \in \{5,15\} $~min \& $\bar{S}=2000$ meters \\
	\end{tabular}
	\caption{Comparison of decentralized and centralized decision-making in high-availability scenarios\label{fig:dec_comp_high}}
	\fnote{\footnotesize Each subplot shows the distribution of the realized system cost $\hat{\alpha}$ for each considered setting, per number of drivers $N$, and search radius $\bar{S}$.}
\end{figure}

Analogously to Figure~\ref{fig:dec_comp_low}, Figure~\ref{fig:dec_comp_high} compares all decentralized settings (\gls{1}, \gls{2}, \gls{3}, \gls{4}, \gls{9}) with the centralized setting (\gls{5}) with respect to cost $\hat{\alpha}$ in high-availability scenarios. We split results between short departure time horizons ($t^s\in\{0,1\} $~min) and large departure time horizons ($t^s\in\{5,15\} $~min), and between a small search radius ($\bar{S} = 1000 $ m) and a large search radius ($\bar{S} = 2000 $ meters).
Overall we observe similar result trends as for low-availability scenarios. 
However, the difference between the improvement obtained with coordination in small search areas and the improvement obtained in large search areas is significantly higher than for low-availability scenarios.
Here, the benefit of deviating from overlapping selfish solutions increases due the higher number of alternative available candidate stations compared to a low-availability scenario.
We additionally note that the amplitude of the cost distribution obtained in the \gls{9} setting for large search radius instances (see Figure~\ref{fig:dec_comp_high}a) is larger than in low-availability scenarios.

\FloatBarrier
\subsection{Coordination with a long planning horizon:}\label{app:large-planning}

\begin{figure}[H]
			\begin{tabular}{x{0.527\textwidth}x{.527\textwidth}}
\begin{tikzpicture}[font=\footnotesize]

\definecolor{color0}{rgb}{0.266666666666667,0.447058823529412,0.768627450980392}

\begin{axis}[
legend cell align={left},
height=5cm,
width=8cm,
legend style={
  fill opacity=0.8,
  draw opacity=1,
  text opacity=1,
  at={(0.03,0.97)},
  anchor=north west,
  draw=white!80!black
},
tick align=outside,
tick pos=left,
x grid style={white!69.0196078431373!black},
xlabel={Driver departure time [min]},
xmin=-9, xmax=189,
xtick style={color=black},
y grid style={white!69.0196078431373!black},
ylabel={Individual cost [min]},
ymin=1.06548674262668, ymax=119.36928847684,
ytick style={color=black}
]
\addplot [semithick, color0, mark=+, mark size=1.5, mark options={solid}]
table {%
0 10.7342157050667
3.05084745762712 35.555108674
6.10169491525424 53.6479107701333
9.15254237288136 58.5696786774667
12.2033898305085 90.9923518172
15.2542372881356 96.3740343481333
18.3050847457627 87.1377945910667
21.3559322033898 106.5965672152
24.4067796610169 105.493018306933
27.4576271186441 98.7936115333333
30.5084745762712 110.358194950133
33.5593220338983 109.727742685067
36.6101694915254 110.743100739733
39.6610169491525 113.991842943467
42.7118644067797 106.5268006176
45.7627118644068 61.1614880224
48.8135593220339 65.2737812913333
51.864406779661 78.8373884589333
54.9152542372881 71.5776162072
57.9661016949153 82.4042553701333
61.0169491525424 96.6895023796
64.0677966101695 95.1989399718667
67.1186440677966 93.4643282128
70.1694915254237 99.2170327810667
73.2203389830509 103.908661962667
76.271186440678 104.9502593008
79.3220338983051 106.304941671333
82.3728813559322 105.9099437144
85.4237288135593 107.006287043067
88.4745762711864 105.9557194852
91.5254237288135 72.4750903137333
94.5762711864407 75.4212847749333
97.6271186440678 86.4082974058667
100.677966101695 86.8552963450667
103.728813559322 94.9129266285333
106.779661016949 98.0270545258667
109.830508474576 77.7928212566667
112.881355932203 96.5459579250667
115.932203389831 96.1595378808
118.983050847458 91.0495892728
122.033898305085 90.4839438001333
125.084745762712 96.8971429705334
128.135593220339 100.956334518667
131.186440677966 102.486792032
134.237288135593 103.767172481733
137.28813559322 92.1687170976
140.338983050847 84.3101508797333
143.389830508475 84.0534037301333
146.440677966102 90.333521042
149.491525423729 91.1454159606667
152.542372881356 90.3396966034667
155.593220338983 95.5346998537333
158.64406779661 91.0659307154667
161.694915254237 90.7149421608
164.745762711864 93.0838044414667
167.796610169492 95.9746509533333
170.847457627119 95.1832951753333
173.898305084746 97.1623393606667
176.949152542373 101.894951254267
180 91.7536051093333
};
\addlegendentry{CEN}
\addplot [semithick, white!64.7058823529412!black, mark=*, mark size=1.5, mark options={solid}]
table {%
0 6.44293227600001
3.05084745762712 31.3134426990667
6.10169491525424 74.6015123822667
9.15254237288136 68.5867457669333
12.2033898305085 98.6697206889333
15.2542372881356 92.8110015918667
18.3050847457627 83.0886039632
21.3559322033898 103.254500089333
24.4067796610169 100.117803343733
27.4576271186441 102.275255126133
30.5084745762712 107.9831081544
33.5593220338983 103.146685430667
36.6101694915254 102.8192572056
39.6610169491525 107.044714676133
42.7118644067797 108.8993089504
45.7627118644068 104.243108391467
48.8135593220339 101.896459068533
51.864406779661 97.7628929281333
54.9152542372881 103.272167973333
57.9661016949153 104.436962267733
61.0169491525424 105.666454980533
64.0677966101695 104.512606612267
67.1186440677966 105.344203504
70.1694915254237 104.073819714267
73.2203389830509 108.802439329733
76.271186440678 102.170438569467
79.3220338983051 106.200086358667
82.3728813559322 106.5158938124
85.4237288135593 104.789095046267
88.4745762711864 105.885309664533
91.5254237288135 97.2690517790667
94.5762711864407 103.3097003172
97.6271186440678 105.5681428816
100.677966101695 101.069576902
103.728813559322 108.300663977333
106.779661016949 109.872272396533
109.830508474576 92.8922999490667
112.881355932203 101.256307732933
115.932203389831 107.862867652933
118.983050847458 103.7425145632
122.033898305085 107.976506472
125.084745762712 91.444603952
128.135593220339 104.760721082133
131.186440677966 103.029621039467
134.237288135593 106.3501800384
137.28813559322 105.275622641733
140.338983050847 95.2710445414667
143.389830508475 103.435251118
146.440677966102 103.607813776667
149.491525423729 101.9893656196
152.542372881356 105.5849009928
155.593220338983 111.032017345867
158.64406779661 102.5321826456
161.694915254237 102.2345721872
164.745762711864 104.628876077733
167.796610169492 107.196744535333
170.847457627119 98.1328119716
173.898305084746 104.235487150133
176.949152542373 103.219309299733
180 104.175419484133
};
\addlegendentry{DEC}
\end{axis}

\end{tikzpicture} &
				\hspace{-0.5cm} 
\begin{tikzpicture}[font=\footnotesize]

\definecolor{color0}{rgb}{0.266666666666667,0.447058823529412,0.768627450980392}

\begin{axis}[
legend cell align={left},
height=5cm,
width=8cm,
legend style={
  fill opacity=0.8,
  draw opacity=1,
  text opacity=1,
  at={(0.03,0.97)},
  anchor=north west,
  draw=white!80!black
},
tick align=outside,
tick pos=left,
x grid style={white!69.0196078431373!black},
xlabel={Driver departure time [min]},
xmin=-9, xmax=189,
xtick style={color=black},
y grid style={white!69.0196078431373!black},
ylabel={Individual cost [min]},
ymin=0.048842907600001, ymax=113.819860569733,
ytick style={color=black}
]
\addplot [semithick, color0, mark=+, mark size=1.5, mark options={solid}]
table {%
0 10.0253933722667
3.05084745762712 25.2483730769333
6.10169491525424 44.1311913494667
9.15254237288136 71.1137378189333
12.2033898305085 76.7439855272
15.2542372881356 91.0508397289333
18.3050847457627 88.5167773612
21.3559322033898 96.249103318
24.4067796610169 99.9260637364
27.4576271186441 100.214853329067
30.5084745762712 98.1874660457333
33.5593220338983 90.2514062470667
36.6101694915254 87.0209077890667
39.6610169491525 100.3498472012
42.7118644067797 84.1608394365333
45.7627118644068 72.6534343105333
48.8135593220339 62.2489595982667
51.864406779661 65.7338575244
54.9152542372881 67.3084339442667
57.9661016949153 80.6093176568
61.0169491525424 89.7491860885333
64.0677966101695 92.5449804396
67.1186440677966 90.9881412042667
70.1694915254237 88.881983028
73.2203389830509 94.5107101681333
76.271186440678 93.5140327578667
79.3220338983051 94.4224798298667
82.3728813559322 93.0179377873333
85.4237288135593 98.1143353448
88.4745762711864 95.3159403226667
91.5254237288135 78.1523681714667
94.5762711864407 68.5809450505333
97.6271186440678 78.3105207816
100.677966101695 79.6092871492
103.728813559322 70.36640768
106.779661016949 86.3993009342667
109.830508474576 61.9713499693333
112.881355932203 86.8089225418667
115.932203389831 92.7781026858667
118.983050847458 82.6744942362667
122.033898305085 93.7124767761333
125.084745762712 91.3122781082667
128.135593220339 93.2991976829333
131.186440677966 86.7891652150667
134.237288135593 92.7934503470667
137.28813559322 86.1805869781333
140.338983050847 74.4199792541333
143.389830508475 75.2430003062667
146.440677966102 83.1000578373333
149.491525423729 82.6207719412
152.542372881356 85.8381330926667
155.593220338983 82.8337228697333
158.64406779661 88.8036726266667
161.694915254237 88.63516288
164.745762711864 86.1010989108
167.796610169492 89.6958588638667
170.847457627119 67.5940663025333
173.898305084746 88.8216182809333
176.949152542373 90.6784226436
180 94.419388706
};
\addlegendentry{CEN}
\addplot [semithick, white!64.7058823529412!black, mark=*, mark size=1.5, mark options={solid}]
table {%
0 5.22025280133333
3.05084745762712 18.8232467866667
6.10169491525424 56.0519240937333
9.15254237288136 74.5600380641333
12.2033898305085 99.3593107458667
15.2542372881356 97.7617347409333
18.3050847457627 95.6449346209333
21.3559322033898 85.3921934012
24.4067796610169 100.4997775852
27.4576271186441 105.510272779333
30.5084745762712 101.462859548133
33.5593220338983 104.085583964133
36.6101694915254 102.547039658933
39.6610169491525 94.8725063904
42.7118644067797 105.169424832
45.7627118644068 106.390892763467
48.8135593220339 99.708422108
51.864406779661 102.6249801364
54.9152542372881 104.004869428667
57.9661016949153 104.532385401467
61.0169491525424 105.252552210667
64.0677966101695 104.560158471733
67.1186440677966 102.412970925333
70.1694915254237 99.6069255582667
73.2203389830509 104.494786347067
76.271186440678 92.7499645274667
79.3220338983051 102.5944298204
82.3728813559322 104.193768217333
85.4237288135593 106.731979851867
88.4745762711864 105.722727321733
91.5254237288135 95.2021210370667
94.5762711864407 103.991304063067
97.6271186440678 106.447782335333
100.677966101695 103.524239733333
103.728813559322 96.2189972285333
106.779661016949 108.648450676
109.830508474576 71.6856998772
112.881355932203 95.3922579870667
115.932203389831 107.303431344133
118.983050847458 100.742966488267
122.033898305085 108.458414518267
125.084745762712 102.876341054133
128.135593220339 105.219902731867
131.186440677966 99.4772052825333
134.237288135593 103.5799479836
137.28813559322 101.346350531333
140.338983050847 99.7169517194667
143.389830508475 101.887793399733
146.440677966102 105.277528004267
149.491525423729 104.431481727067
152.542372881356 102.407556528
155.593220338983 104.801658167733
158.64406779661 103.282010589867
161.694915254237 98.5296567937333
164.745762711864 96.3654675036
167.796610169492 105.389332146267
170.847457627119 73.8089377185333
173.898305084746 99.6401519077333
176.949152542373 101.502868730533
180 105.435751078267
};
\addlegendentry{DEC}
\end{axis}

\end{tikzpicture} \\
				\small (a) \gls{sc1} \& $\bar{S}=1000$ meters & \small (b) \gls{sc1} \& $\bar{S}=2000$ meters \\
				&\\
\begin{tikzpicture}[font=\footnotesize]

\definecolor{color0}{rgb}{0.266666666666667,0.447058823529412,0.768627450980392}

\begin{axis}[
legend cell align={left},
height=5cm,
width=8cm,
legend style={
  fill opacity=0.8,
  draw opacity=1,
  text opacity=1,
  at={(0.03,0.97)},
  anchor=north west,
  draw=white!80!black
},
tick align=outside,
tick pos=left,
x grid style={white!69.0196078431373!black},
xlabel={Driver departure time [min]},
xmin=-8.69491525423729, xmax=182.593220338983,
xtick style={color=black},
y grid style={white!69.0196078431373!black},
ylabel={Individual cost [min]},
ymin=1.43882269804668, ymax=111.52923341302,
ytick style={color=black}
]
\addplot [semithick, color0, mark=+, mark size=1.5, mark options={solid}]
table {%
0 10.6921100932
3.05084745762712 35.9673484821333
6.10169491525424 58.7766012782667
15.2542372881356 72.3772529174667
18.3050847457627 66.0456873306667
21.3559322033898 88.6692707393333
30.5084745762712 86.4139666172
33.5593220338983 88.1668820236
36.6101694915254 104.599239279867
45.7627118644068 50.5177223897333
48.8135593220339 55.3204263604
51.864406779661 76.9553552054667
61.0169491525424 68.6727930296
64.0677966101695 74.3943060716
67.1186440677966 82.7670347946667
76.271186440678 79.6367022504
79.3220338983051 92.3272434148
82.3728813559322 96.5451360270667
91.5254237288135 56.8260210634667
94.5762711864407 63.3717640802667
97.6271186440678 77.8018472961333
106.779661016949 73.3030886086667
109.830508474576 67.0630247926667
112.881355932203 84.2281990916
122.033898305085 67.2380473418667
125.084745762712 79.518476286
128.135593220339 89.1883099917333
137.28813559322 73.0502668745333
140.338983050847 68.0148541266667
143.389830508475 76.2802167681333
152.542372881356 68.6851713492
155.593220338983 78.1719507157333
158.64406779661 79.9003103438667
167.796610169492 69.7678797210667
170.847457627119 74.9770537493333
173.898305084746 87.4024969501333
};
\addlegendentry{CEN}
\addplot [semithick, white!64.7058823529412!black, mark=*, mark size=1.5, mark options={solid}]
table {%
0 6.44293227600001
3.05084745762712 30.9159053310667
6.10169491525424 57.0332814318667
15.2542372881356 67.9411924741333
18.3050847457627 69.9025581192
21.3559322033898 96.9760584305333
30.5084745762712 78.4919588744
33.5593220338983 93.4979351278667
36.6101694915254 103.030807063333
45.7627118644068 78.8390024265333
48.8135593220339 96.0535870225333
51.864406779661 95.5239951886667
61.0169491525424 76.0667502569333
64.0677966101695 98.4938539550667
67.1186440677966 105.995243492133
76.271186440678 85.1250366464
79.3220338983051 94.6938317304
82.3728813559322 106.525123835067
91.5254237288135 75.2634165797333
94.5762711864407 92.0577730634667
97.6271186440678 104.030384639067
106.779661016949 90.1338524572
109.830508474576 78.9739447745333
112.881355932203 100.741940594667
122.033898305085 79.9624958465333
125.084745762712 88.1301690122667
128.135593220339 90.0549579032
137.28813559322 78.2192690725333
140.338983050847 90.0533435584
143.389830508475 102.146208545733
152.542372881356 76.8984675329333
155.593220338983 100.977895850933
158.64406779661 102.225124798133
167.796610169492 83.1701159869333
170.847457627119 86.5437136172
173.898305084746 99.6914511210667
};
\addlegendentry{DEC}
\end{axis}

\end{tikzpicture} &
				\hspace{-0.5cm} 
\begin{tikzpicture}[font=\footnotesize]

\definecolor{color0}{rgb}{0.266666666666667,0.447058823529412,0.768627450980392}

\begin{axis}[
legend cell align={left},
height=5cm,
width=8cm,
legend style={
  fill opacity=0.8,
  draw opacity=1,
  text opacity=1,
  at={(0.03,0.97)},
  anchor=north west,
  draw=white!80!black
},
tick align=outside,
tick pos=left,
x grid style={white!69.0196078431373!black},
xlabel={Driver departure time [min]},
xmin=-8.69491525423729, xmax=182.593220338983,
xtick style={color=black},
y grid style={white!69.0196078431373!black},
ylabel={Individual cost [min]},
ymin=0.325176906693327, ymax=108.832941810107,
ytick style={color=black}
]
\addplot [semithick, color0, mark=+, mark size=1.5, mark options={solid}]
table {%
0 9.94388886039999
3.05084745762712 25.0472746984
6.10169491525424 44.2854484586667
15.2542372881356 72.3338891016
18.3050847457627 62.0044110789333
21.3559322033898 86.0337170238667
30.5084745762712 87.3410845449333
33.5593220338983 81.8337493330666
36.6101694915254 81.3194952202667
45.7627118644068 56.1218851028
48.8135593220339 54.7353218078667
51.864406779661 63.3436030886667
61.0169491525424 62.9511464437333
64.0677966101695 69.4104814954667
67.1186440677966 72.6989301068
76.271186440678 74.0556116173333
79.3220338983051 77.2461100228
82.3728813559322 86.3810344142667
91.5254237288135 67.9084821202667
94.5762711864407 58.4278087974667
97.6271186440678 69.4123546308
106.779661016949 63.0748340998667
109.830508474576 51.8506422094667
112.881355932203 69.7270444554667
122.033898305085 65.9222376705333
125.084745762712 74.134631694
128.135593220339 80.7961228889333
137.28813559322 69.1417004142667
140.338983050847 62.7540756052
143.389830508475 67.7080980412
152.542372881356 68.2250837538667
155.593220338983 62.1089163649333
158.64406779661 76.9694308336
167.796610169492 71.3533911374667
170.847457627119 53.9397163812
173.898305084746 74.5657065918667
};
\addlegendentry{CEN}
\addplot [semithick, white!64.7058823529412!black, mark=*, mark size=1.5, mark options={solid}]
table {%
0 5.25734803866666
3.05084745762712 19.8679568344
6.10169491525424 57.2520593933333
15.2542372881356 58.8204093669333
18.3050847457627 82.0187310436
21.3559322033898 81.3488372918667
30.5084745762712 75.8506738418667
33.5593220338983 96.4553670412
36.6101694915254 97.4355775601333
45.7627118644068 80.3154763997333
48.8135593220339 92.0968373656
51.864406779661 99.5736968897333
61.0169491525424 65.1202802716
64.0677966101695 95.9045196144
67.1186440677966 103.0603783604
76.271186440678 77.1053287310667
79.3220338983051 96.2577171885333
82.3728813559322 101.865624554133
91.5254237288135 70.0045202397333
94.5762711864407 93.117558556
97.6271186440678 102.813555163067
106.779661016949 77.8081729968
109.830508474576 62.5020857677333
112.881355932203 94.0710775957333
122.033898305085 82.6338266069333
125.084745762712 88.9941171750667
128.135593220339 103.068854830533
137.28813559322 71.6295163504
140.338983050847 85.9708307312
143.389830508475 99.2594470576
152.542372881356 79.7554645541333
155.593220338983 97.3600156232
158.64406779661 103.900770678133
167.796610169492 71.6747712962667
170.847457627119 68.4123658293333
173.898305084746 95.1869063277333
};
\addlegendentry{DEC}
\end{axis}

\end{tikzpicture} \\
				\small (c) \gls{sc3} \& $\bar{S}=1000$ meters & \small (d) \gls{sc3} \& $\bar{S}=2000$ meters \\
			\end{tabular}
		\caption{Comparison of DEC and CEN in high-availability scenarios for a three-hours planning horizon\label{fig:indiv_cost_25}}
		\fnote{\footnotesize Each subplot shows the individual realized cost $\hat{\alpha}_i$ for each considered setting, per number of drivers $N$, averaged over all values of $r^s \in \{100,300,700\}$ meters.}
	\end{figure}
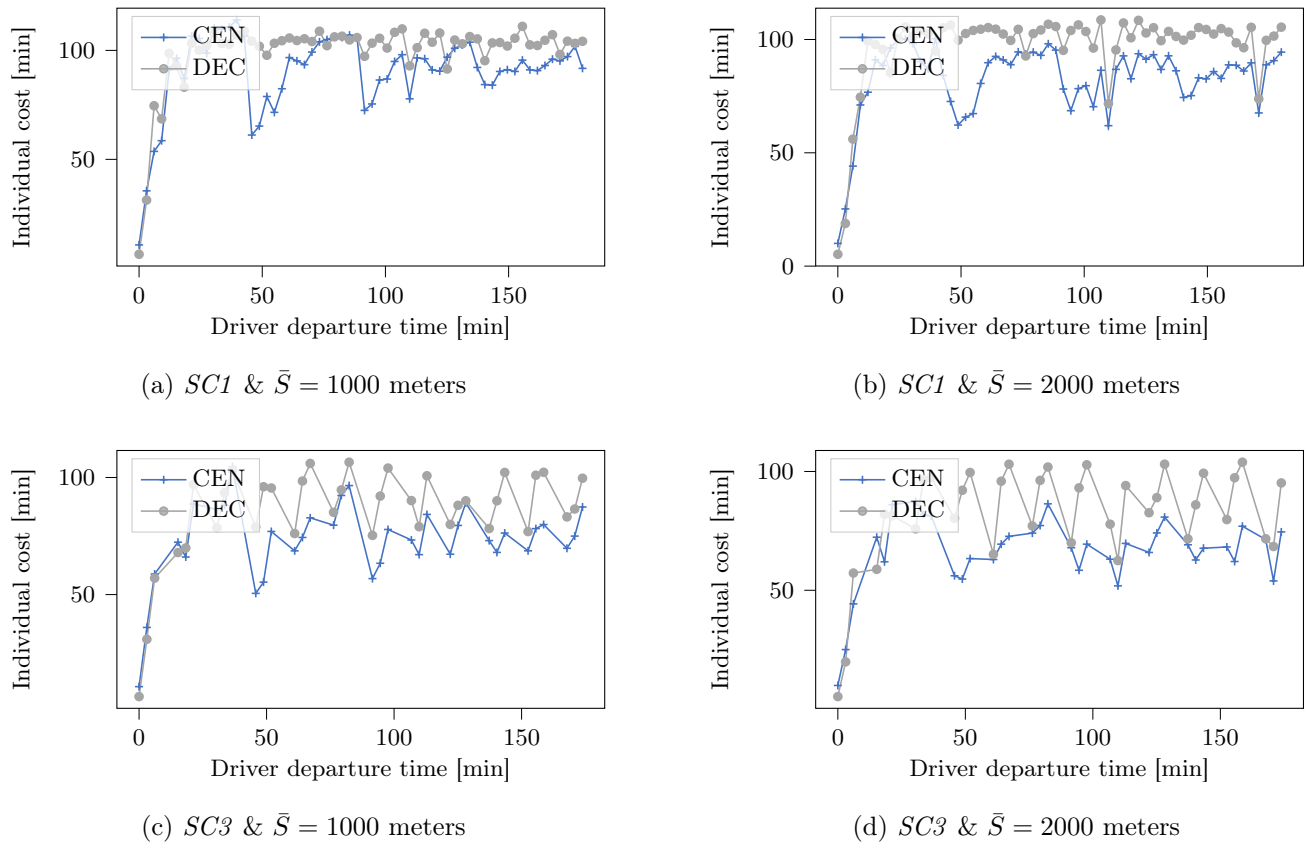

	Similar to Figure~\ref{fig:indiv_cost_60}, Figure~\ref{fig:indiv_cost_25} shows individual drivers' cost depending on their departure time, but in a low-availability scenario. Coordination generally improves a non-coordinated system, but with a less visible impact than in the high-availability scenario case, in-line with the main results discussion (cf.~Figure~\ref{fig:indiv_cost}). In Figure~\ref{fig:indiv_cost_25}a, a lower search radius $\bar{S}=1000$ meters combined with \gls{low_avail} increases the number of drivers that compete for a very limited number of expectedly available stations, which significantly limits the overall coordination potential. 
	In contrast to high-availability scenarios, results show a visible system transition from the initial state to congested states.
	
\end{appendices}

\end{document}